\documentclass[12pt]{article}
\usepackage{amssymb}
\newtheorem{tver}{Assertion}

\def\bt{\begin{tver}}
\def\et{\end{tver}}
\newtheorem{lema}{Lema}

\def\bl{\begin{lema}}
\def\el{\end{lema}}

\def\be{\begin{equation}}
\def\ee{\end{equation}}

\author{R.Z.~Zhdanov \thanks {On leave of absence from the
Institute of Mathematics of NAN of Ukraine, 3 Tereshchenkivska
Street, 01601 Kyiv-4, Ukraine} \\ \small Complutense University,
28040 Madrid, Spain \\ \small E-mail: renat@ciruelo.fis.ucm.es
\and V.I.~Lahno \\ \small State Pedagogical University, 2
Ostrogradskogo Street, 36000 Poltava, Ukraine \\ \small E-mail:
laggo@poltava.bank.gov.ua}

\title{Symmetry and Exact Solutions of the Maxwell and
$SU(2)$ Yang-Mills Equations}

\date{}

\begin{document}
\maketitle

\begin{abstract}
We give the overview of solution techniques for the general
con\-for\-ma\-lly-invariant linear and nonlinear wave equations
centered around the idea of dimensional reductions by their
symmetry groups. The efficiency of these techniques is
demonstrated on the examples of the $SU(2)$ Yang-Mills and the
vacuum Maxwell equations. For the Yang-Mills equations we have
derived the most general form of the conformally-invariant
solution and construct a number of their new analytical
non-Abelian solutions in explicit form. We have completely solved
the problem of symmetry reduction of the Maxwell equations by
subgroups of the conformal group. This yields twelve
multi-parameter families of their exact solutions, a majority of
which are new and might be of considerable interest for
applications.
\end{abstract}

\begin{center}
{\bf Plan.}
\end{center}
\begin{enumerate}
\item {\footnotesize Introduction}
\item {\footnotesize Conformally-invariant ansatzes for an
arbitrary vector field}
\item {\footnotesize Exact solutions of the Yang-Mills equations}
\item {\footnotesize Conditional symmetry and new solutions of the
Yang-Mills equations}
\item {\footnotesize Exact solutions of the Maxwell equations}
\item {\footnotesize Concluding remarks}
\end{enumerate}

\section{Introduction.}
\setcounter{equation}{0}
\setcounter{tver}{0}
\setcounter{lema}{0}

The famous paper \cite{m1} written by Yang and Mills is a
milestone of the modern quantum physics, where the role played by
the equations introduced in the paper (called now the $SU(2)$
Yang-Mills equations) can be compared only to that of the
Klein-Gordon-Fock, Schr\"odinger, Maxwell and Dirac equations.
However, the real importance of the Yang-Mills equations has been
understood only in the late sixties, when the concept of the gauge
fields as of those responsible for all the four fundamental
physical interactions (gravitational, electro-magnetic, weak and
strong interactions) has become widely spread.

The simplest example of the gauge theory in the
$(1+3)$-dimensional space is the system of the Maxwell equations
for the four-component vector-potential of the electro-magnetic
field, whose gauge group is the one-pa\-ra\-me\-ter group $U(1)$. The
simplest example of the non-Abelian gauge group is the group
$SU(2)$. This very group is realized as the symmetry group
admitted by the Yang-Mills equations describing the triplet of the
gauge fields (called in the sequel the Yang-Mills field) ${\bf
A}_\mu (x)= (A^a_\mu (x), a =1,2,3)$, where $\mu =1,2,3,4,\ x=
(x_1, x_2, x_3, x_4)$ for the case of the four-dimensional Euclid
space and $\mu = 0,1,2,3$,\ $x= (x_0 =t, x_1, x_2, x_3)=$ $ (t,
{\bf x})$ for the case of the Minkowski space. The matrix vector
field $A_\mu = A_\mu (x)$ is defined as follows: $$ A_\mu = e
\frac{\sigma_a}{2i}A^a_\mu. $$ Here $\sigma_a$, ($a=1,2,3$) are
the Pauli matrices $$ \sigma_1 = \pmatrix{0&1\cr 1&0 \cr}, \quad
\sigma_2 = \pmatrix{0&-i\cr i&0 \cr}, \quad \sigma_3 =
\pmatrix{1&0\cr 0&-1 \cr} $$ and $e$ is the real constant called
the gauge coupling constant.

Using the matrix gauge potentials one constructs the matrix-valued
field
$$
F_{\mu \nu} \equiv \partial^\mu A_\nu - \partial^\nu A_\mu
+[A_\mu,\ A_\nu], \quad \mu, \nu = 0,1,2,3.
$$

Writing the above expressions component-wise yields
$$
F_{\mu \nu} \equiv e \frac{\sigma_a}{2i} F^a_{\mu \nu},\quad
F^a_{\mu \nu} = \partial^\mu A^a_\nu -\partial^\nu A^a_\mu + e
f^a_{bc} A^b_\mu A^c_\nu,
$$
where $\mu, \nu =0,1,2,3$, $a=1,2,3$ and the symbols
$f^a_{bc}$,\ ($a,b,c=1,2,3$) stand for the structure constants
determining the Lie algebra of the gauge group (note that for the
case of the group $SU(2),\ f^a_{bc} = \varepsilon_{abc}$,
$\varepsilon_{abc}$ being the anti-symmetric tensor with
$\varepsilon_{123} =1$,\ $a,b,c = 1,2,3$).

Hereafter we use the following designations:
$$
\partial_\mu =\partial_{x_\mu} = \frac{\partial}{\partial x_\mu}.
$$
Furthermore, lowering and rising the indices $\mu, \nu$ is
performed with the help of the metric tensor of the space of the
variables $x_\mu$ and the summation over the repeated indices is
carried out.

The $SU(2)$ Yang-Mills equations are obtained from the Lagrangian
$$
{\cal L} = -\frac{1}{4} F_{\mu \nu} F^{\mu \nu}
$$
and are of the form
\be \label{1.1}
\partial_\mu F_{\mu \nu} +[A^\mu, F_{\mu \nu}] = [D_\mu, F_{\mu
\nu}] =0,
\ee
where $D_\mu = \partial_\mu + A^\mu$ is the covariant derivative,
$\mu, \nu = 0,1,2,3$.

One of the most popular and exciting parts of the general theory of
the Yang-Mills equations is that devoted to constructing their
exact analytical solutions. There is a vast literature devoted
solely to constructing and analyzing exact solutions of
(\ref{1.1}) (see, the review by Actor \cite{m2} and the monograph
\cite{m3} for the extensive list of references). The majority of
the results is obtained for the case of the Yang-Mills equations
in the Euclidean space. The principal reason for this is the fact
that equations (\ref{1.1}) in the Euclidean space have the
monopole and instanton solutions \cite{m3,m4}, that admit numerous
physical interpretations and have highly non-trivial geometrical
and algebraic properties. Note that these and some other classes
of exact solutions of system (\ref{1.1}) can be also obtained
by solving the so-called self-dual Yang-Mills equations
\be
\label{1.2} F_{\mu \nu} = *F_{\mu \nu},\quad \mu, \nu = 0,1,2,3.
\ee
Evidently, each solution of (\ref{1.2}) satisfies (\ref{1.1}),
while the reverse assertion does not hold.

Provided we consider the Euclidean case, $*F_{\mu \nu} = $
$\frac{1}{2} \varepsilon_{\mu \nu\lambda \rho}F_{\lambda \rho},\
(\mu, \nu,\lambda, \rho=1, 2, 3, 4)$, where $\varepsilon_{\mu
\nu\lambda \rho}$ is the completely anti-symmetric tensor, and
equations (\ref{1.2}) form the system of four real first-order
partial differential equations.

It was the self-duality property of the instanton solutions of
(\ref{1.1}) in the Euclidean space, that had enabled using the
ansatz, suggested by t'Hooft \cite{m5}, Corrigan and Fairlie
\cite{m6}, Wilczek \cite{m7} and Witten \cite{m8}, in order to
construct these solutions. Furthermore, the well-known monopole
solution by Prasad and Sommerfield \cite{m9}, as well as, the
solutions obtainable via the Atyah-Hitchin-Drinfeld-Manin method
\cite{m10} exploit explicitly the self-duality condition.

One more important property of the self-dual Yang-Mills equations
is that they are equivalent to the compatibility conditions of
some over-determined system of linear partial differential
equations \cite{m11,m12}. In other words, the self-dual Yang-Mills
equations admit the Lax representation and, in this sense, are
integrable. By this very reason it is possible to reduce equations
(\ref{1.2}) to the well-studied solitonic equations, such as the
Euler-Arnold, Burgers and Devy-Stuardson equations (Chakravarty et
al, \cite{m13,m14}) and Liouville and sine-Gordon equations
(Tafel, \cite{m15}) by the use of the symmetry reduction method.

For the case, when the Yang-Mills field are defined in the
Minkowski space, we have, $*F_{\mu \nu}$ $ =\frac{i}{2}$
$\varepsilon_{\mu \nu \lambda \rho}$ $F^{\lambda \rho},\ (\mu,
\nu, \lambda, \rho=0,1,2,3)$. Consequently, equations (\ref{1.2})
form the system of complex first-order differential equations. In
view of this fact, exploitation of the above mentioned methods and
results for study of the $SU(2)$ Yang-Mills equations (\ref{1.1})
in the Minkowski space yields complex-valued solutions. That is
why, the above mentioned methods for solving equations (\ref{1.1})
fail to be efficient for the case of the Minkowski space.
Consequently, there is a need for developing the new methods, that
do not rely on the self-duality condition. This problem has been
addressed by one of the creators of the inverse scattering
technique V.E.Zaharov, who wrote in the foreword to the Russian
translation of the monograph by Calogero and Degasperis
\cite{m16}, that a number of important problems of the nonlinear
ma\-the\-ma\-ti\-cal physics (including the Yang-Mills equations in
the Minkowski space) are still waiting for new efficient solution
techniques to appear.

On the other hand, it is known \cite{m17} (see, also, \cite{m18})
that equations (\ref{1.1}) have rich symmetry. Namely, their
maximal (in the Lie sense) symmetry group is the group $G \otimes
SU(2)$, where $G$ is
\begin{itemize}
\item the conformal group $C(1, 3)$, if the Yang-Mills equations
are defined in the Minkowski space;
\item the conformal group $C(4)$, if the Yang-Mills equations are
defined in the Euclidean space;
\item the conformal group $C(2,2)$, if the Yang-Mills equations
are defined in the pseudo-Euclidean space having the metric tensor
with the signature $(-,\ -,\ +,\ +)$.
\end{itemize}
Note that the maximal symmetry groups admitted by the self-dual
Yang-Mills equations (\ref{1.2}) coincide with the symmetry groups
of the corresponding equations (\ref{1.1}).

The rich symmetry of equations (\ref{1.1}), (\ref{1.2}) enables
efficient exploitation of the symmetry reduction routine for the
sake of dimensional reduction of the Yang-Mills equations either
to ordinary differential equations integrable by quadratures or to
integrable solitonic equations in two or three independent
variables \cite{m19}--\cite{m21}. In particular, some subgroups of
the generalized Poincar\'e group $P(2,2)$, which is the subgroup
of the conformal group $C(2,2)$, were used in order to reduce the
self-dual Yang-Mills equations, defined in the pseudo-Euclidean
space having the metric tensor with the signature $(-,$\ $-,$\
$+,$\ $+)$, to a number of known integrable systems, like, the
Ernst, cubic Schr\"odinger and Euler-Calogero-Moser equations
(see, \cite{m22} and the references therein). Legar\'e et al have
carried out systematic investigation of the problem of symmetry
reduction of system (\ref{1.2}) in the Euclidean space by
subgroups of the Euclid group $E(4)\in C(4)$ \cite{m23,m24}. What
is more, some of the known analytical solutions of equations
(\ref{1.1}) in the Euclidean space (namely, the non self-dual
meron solution, obtained by Alfaro, Fubini and Furlan \cite{m25},
and the instanton solution, constructed by Belavin, Polyakov,
Schwartz and Tyupkin \cite{m26}) can also be obtained within the
framework of the symmetry reduction approach (see, e.g.,
\cite{m21}).

To the best of our knowledge, the first paper devoted to symmetry
reduction of the $SU(2)$ Yang-Mills equations in the Minkowski
space has been published by Fushchych and Shtelen \cite{m27} (see,
also, \cite{m21}). They use two confor\-mal\-ly-invariant
ansatzes in order to perform reduction of equations (\ref{1.1}) to
systems of ordinary differential equations. Integrating the
latter yields several exact solutions of the Yang-Mills equations
(\ref{1.1}).

Let us note that the full solution of the problem of symmetry
reduction of fundamental equations of relativistic physics, whose
symmetry groups are subgroups of the conformal group $C(1,3)$, has
been obtained for the scalar wave equation only (see, for further
details, \cite{m21}, \cite{m28}--\cite{m30}). This fact is
explained by the extreme cumbersomity of the calculations needed
to perform a systematic symmetry reduction of systems of partial
differential equations by all inequivalent subgroups of the
conformal group $C(1,3)$. The complete solution of the problem
symmetry reduction to systems of ordinary differential equations
has been obtained for the conformally-invariant non\-li\-ne\-ar
spinor equations \cite{m31}--\cite{m33}, that generalize the
Dirac equation for an electron. In our recent publications we have
carried out symmetry reduction of the Yang-Mills equations
(\ref{1.1}), (\ref{1.2}) by subgroups of the Poincar\'e group and
have constructed a number of their exact solutions
\cite{m44}--\cite{m53}.

The principal aim of the present paper is two-fold. Firstly, we
will review the already known ideas, methods and results, centered
around the solution techniques, that are based on the symmetry
reduction method for the Yang-Mills equations (\ref{1.1}),
(\ref{1.2}) in the Minkowski space. Secondly, we will expose the
general reduction routine, developed by us recently, that enables
the unified treatment of both the classical and non-classical
symmetry reduction approaches for an arbitrary
relativistically-invariant system of partial differential
equations. As a by-product, this approach yields exhaustive
solution of the problem of symmetry reduction of the vacuum
Maxwell equations
\be \label{1.3}
\matrix{{\rm rot}\, {\bf E}= -\frac{\textstyle{\partial {\bf
H}}}{\textstyle{\partial t}}, & {\rm div}\, {\bf H} =0,  \cr & \cr
{\rm rot}\, {\bf H}= \frac{\textstyle{\partial {\bf
E}}}{\textstyle{\partial t}}, & \ {\rm div}\, {\bf E} =0. \cr}
\ee

The history of the study of symmetry properties of equations
(\ref{1.3}) goes back to the beginning of the century.
Invariance properties of the Maxwell equations have been studied
by Lorentz \cite{m34} and Poincar\'e \cite{m35,m36}. They have
proved that equa\-ti\-ons (\ref{1.3}) are invariant with respect
to the transformation group named by the Poincar\'e's suggestion
the Lorentz group. Furthermore, Larmor \cite{m37} and Rai\-nich
\cite{m38} have found that equations (\ref{1.3}) are invariant
with respect the one-parameter transformation group
\be \label{1.4} {\bf E}\to {\bf E}\cos \theta + {\bf H} \sin
\theta, \quad
 {\bf H}\to {\bf H}\cos \theta - {\bf E} \sin \theta
\ee
called now the Heviside-Larmor-Rainich group. Later on,
Bateman \cite{m39} and Cunningham \cite{m40} showed that the
Maxwell equations are invariant with respect to the conformal
group.

Much later, Ibragimov \cite{m41} have proved that the group
$C(1,3) \otimes H$, where $C(1,3)$ is the group of conformal
transformations of the Minkowski space and $H$ is the
Heviside-Larmor-Rainich group (\ref{1.4}), is the maximal in Lie's
sense invariance group of equations (\ref{1.3}). Note that this
result coincides with that obtained earlier without explicit use
of the infinitesimal Lie algorithm \cite{m19,m20}. A further
progress in study of symmetries of the Maxwell equations has
become possible, when Fushchych and Nikitin suggested the non-Lie
approach to investigating symmetry properties of linear systems of
partial differential equations (see, for more details,
\cite{m42}).

The present review is based mainly on our publications
\cite{m33}--\cite{m45}, \cite{m47,m48,m52,m53},
\cite{m46}--\cite{m54} and has the following structure. In the
second section we give the detailed description of the general
reduction routine for an arbitrary relativistically-invariant
systems of partial differential equations. The results of this
section are used in the third one in order to solve the problem of
symmetry reduction of the Yang-Mills equations (\ref{1.1}) by
subgroups of the Pojncar\'e group $P(1,3)$ and to construct their
exact (non-Abelian) solutions. In the next section we review the
techniques for non-classical reductions of the $SU(2)$ Yang-Mills
equations, that are based on their conditional symmetry. These
techniques enable obtaining the principally new classes of exact
solutions of (\ref{1.1}), that are not derivable within the
framework of the standard symmetry reduction technique. In the
fifth section we give an overview of the known invariant solutions
of the Maxwell equations and construct multi-parameter families of
new ones.

\section{Conformally-invariant ansatzes for an arbitrary vector
field}
\setcounter{equation}{0}
\setcounter{tver}{0}
\setcounter{lema}{0}

In this section we describe the general approach to constructing
conforma\-lly-invariant ansatzes applicable to any (linear or
non-linear) system of partial differential equations, on whose
solution set a linear covariant representation of the conformal
group $C(1,3)$ is realized. Since the majority of the equations of
the relativistic physics, including the Klein-Gordon-Fock,
Maxwell, massless Dirac and Yang-Mills equations respect this
requirement, they can be handled within the framework of this
approach.

Note that all our subsequent considerations are local and the
functions involved are supposed to be as many times continuously
differentiable, as it is necessary for performing the
corresponding mathematical operations.

\subsection{On the linear form of invariant ansatzes}

Consider the system of partial differential equations (we denote
it as $S$)
\be
\label{2.1} S: \mathop F \limits_A ({\bf x}, {\bf u}, \mathop {\bf
u} \limits_1, \ldots, \mathop {\bf u} \limits_r) =0,\quad A=1,
\ldots, m,
\ee
defined on the open subset $M \subset X \times U \simeq R^q \times
R^p$ of the space of $p$ independent and $q$ dependent variables.
In (\ref{2.1}) we use the notations, ${\bf x} = (x_1, \ldots, x_p)
\in X, \ {\bf u} = (u^1, \ldots, u^q) \in U$, $\mathop {\bf u}
\limits_l = \Bigl \{ \frac{\textstyle{\partial^l
u^k}}{\textstyle{\partial x_1^{\alpha_1} \partial x_2^{\alpha_2}
\ldots \partial x_p^{\alpha_p}}}$, $ 0 \leq \alpha_i \leq l, \
\sum^p_{i=1} \alpha_i =l,\ k)1,\ldots, q \Bigr \}$, $l=1,2,\ldots,
r$ and $F_A$ are sufficiently smooth functions of the given
arguments.

Let $G$ be a local transformation group, that acts on $M$ and is
the symmetry group of system (\ref{2.1}). Next, let the basis
operators of the Lie algebra $g$ of the group $G$ be of the form
\be
\label{2.2} X_a = \xi^i_a ({\bf x}, {\bf u})\partial_{x_i}
+\eta^a_j ({\bf x}, {\bf u})
\partial_{u^j}, \ a=1, \ldots, n,
\ee where $\xi^i_a, \eta^a_j$ are arbitrary smooth functions on
$M$, $\partial_{u^j} = \frac{\partial}{\partial u^j}$, $i=1,
\ldots, p, j\ =1, \ldots, q$. By definition, operators (\ref{2.2})
satisfy the commutation relations $$ [X_a, X_b]\equiv X_aX_b -
X_bX_a = C^c_{ab} X_c, \ a,b,c=1, \ldots, n, $$ where $C^c_{ab}$
are the structure constants, that determine uniquely the type of
the Lie algebra $g$.

We say that a solution ${\bf u} = {\bf f}({\bf x})$,\ (${\bf f} =
(f^1, \ldots, f^q)$) of system (\ref{2.1}), is $G$-invariant,
if the manifold ${\bf u} - {\bf f}({\bf x})=0$ is invariant
with respect to the action of the group $G$. This means that for
an arbitrary $g \in G$ the functions ${\bf f}$ and $g({\bf f})$
coincide in the intersection of the domains, where they are
defined. More precisely,  we can define $G$-invariant solution of
system (\ref{2.1}) as the solution ${\bf u} = {\bf f}({\bf x})$,
whose graph $\Gamma_{\bf f} = \{({\bf x}, {\bf f}({\bf x}))\}
\subset M$ is locally $G$-invariant subset of the set $M$.

If $G$ is the symmetry group of system (\ref{2.1}), then, under
some additional assumption of regularity of the action of the
group $G$, we can find all its $G$-invariant solutions by solving
the reduced system of differential equations $S/G$. Note that by
construction the system $S/G$ has fewer number of
in\-de\-pen\-dent variables, i.e., the dimension of the initial
system is reduced (by this very reason, the above procedure is
called the symmetry reduction method).

In the sequel, we will restrict our considerations to the case of
the projective action of the group $G$ in $M$. This means that all
the transformations $g$ from $G$ are of the form
$$
(\overline{\bf x}, \overline{\bf u}) = g(({\bf x} , {\bf u})) =
(\Psi_g({\bf x}), \Phi_g ({\bf x}, {\bf u})).
$$
In other words, the transformation law for the independent
variables ${\bf x}$ does not involve the dependent variables (for
the Lie algebra $g$ of the group $G$ this implies that in formulae
(\ref{1.2}) $\xi^i_a = \xi^i_a({\bf x}))$. This defines the
projective action of the group $G$ \ $\overline{\bf x} = g({\bf
x}) = \Psi_g ({\bf x})$ in an arbitrary subset $\Omega$ of the set
$X$.

In what follows, we will suppose that the action of the group $G$ in
$M$ and its projective action in $\Omega$ are regular and the
orbits of these actions have the same dimension $s$. This
dimension is called the rank of the group $G$ (or, alternatively,
the rank of the Lie algebra $g$). Note that the condition ${\rm
rank}\, G =s$ is equivalent to the requirement that the relation
\be \label{2.3} {\rm rank} \parallel \xi^i_a ({\bf x}_0) \parallel
= {\rm rank} \parallel \xi^i_a ({\bf x}_0), \eta^a_j ({\bf x}_0,
{\bf u}_0)\parallel =s \ee
holds in an arbitrary point $({\bf x}_0, {\bf u}_0) \in M$
\cite{m19}. And what is more, we will suppose that $s<p$ (the case
$s=p$ is trivial, and furthermore, $G$-invariant functions do not
exist under $s>p$).

If the above assumptions hold, then there are  $p-s$ fuctionally
independent invariants $y^1 = \omega^1({\bf x}), y^2 = \omega^2
({\bf x}), \ldots, y^{p-s} = \omega^{p-s} ({\bf x})$ (the first
set of invariants) of the group $G$ acting projectively in
$\Omega$, and what is more, each of them is the invariant of the
group $G$ acting in $M$. Furthermore, there are $q$ functionally
independent invariants $v^1 = g^1({\bf x},{\bf u}), \ v^2 =$ $ g^2
({\bf x}, {\bf u}), \ldots, v^q= g^q({\bf x}, {\bf u})$ of the
group $G$ acting in $M$ (the second set of invariants)
\cite{m19,m20}. Using the short-hand notation we represent the
full set of invariants of the group $G$ in the following way:
\be \label{2.4}
{\bf y} = {\bf w}({\bf x}), \quad {\bf v}={\bf g}({\bf x},
{\bf u}).
\ee

Owing to the validity of the relation
$$
{\rm rank}\, \biggl | \biggl | \frac{\partial g^j}{\partial u^i}
\biggr | \biggr | =q,\quad i,j=1, \ldots, q
$$
we can solve locally the second system of equations from
(\ref{2.4}) with respect to ${\bf u}$
 \be \label{2.5} {\bf u} = \tilde {\bf r} ({\bf x}, {\bf v}). \ee

Using the relation
$${\rm rank}\, \biggl |\biggl | \frac{\partial \omega^j}{\partial
x_i} \biggr | \biggr |=p-s,\quad j=1, \ldots, p-s,\quad i=1, \ldots,
p, $$
we choose $p-s$ independent variables $\tilde {\bf x} = ({\tilde
x}_1, \ldots, {\tilde x}_{p-s})$ so that
$${\rm rank}\, \biggl |\biggl | \frac{\partial \omega^j}{\partial
{\tilde x}_i}  \biggr |\biggr | =p-s, \ \ i, j=1, \ldots, p-s. $$
We call these variables principal. The remaining $s$ independent
variables ${\hat {\bf x} }= ({\hat x}_1, \ldots, {\hat x}_s)$ are
called parametrical (they enter all the subsequent for\-mu\-lae as
parameters).

Now we can solve the first system from (\ref{2.4}) with respect to
the principal variables
\be \label{2.6} {\tilde {\bf x}} = {\bf z}
({\hat {\bf x}},{\bf y}). \ee
Inserting (\ref{2.6}) into
(\ref{2.5}) we get the equality $$ {\bf u} = {\tilde {\bf r}}
({\hat {\bf x}}, {\bf z}, {\bf v}) $$ or \be \label{2.7} {\bf u} =
{\bf r} ({\hat {\bf x}}, {\bf y}, {\bf v}). \ee

Note that in (\ref{2.5})--(\ref{2.7}), ${\tilde {\bf r}}$ $=
({\tilde r}^1$, $\ldots$, ${\tilde r}^q)$,\ ${\bf r}$ $= (r^1$,
$\ldots$, $r^q)$,\ ${\bf z}$ $= (z^1$, $\ldots$, $z^{p-s})$. The
so constructed $G$-invariant function (\ref{2.7}) is called the
{\it ansatz}. Inserting ansatz (\ref{2.7}) into system (\ref{2.1})
yields the system of partial differential equa\-ti\-ons for the
functions ${\bf v}$ of the variables ${\bf y}$, that do not
involve explicitly the parametrical variables \cite{m19}. These
equations form the reduced (or factor) system $S/G$ having the
fewer number of independent variables $y^1, \ldots, y^{p-s}$, as
compared with the initial system (\ref{2.1}). Now, if we are given
a solution ${\bf v} = {\bf h} ({\bf y})$ of the reduced system,
then inserting it into (\ref{2.7}) yields a $G$-invariant solution
of system (\ref{2.1}).

Summing up, we formulate the algorithm of symmetry reduction and
construction of invariant solutions of systems of partial
differential equations, that admit non-trivial Lie symmetry.
\begin{enumerate}
\renewcommand{\labelenumi}{(\Roman{enumi})}
\item Using the infinitesimal Lie method we compute the maximal
sym\-me\-try group $G$ admitted by the equation under study.
\item We fix the symmetry degree $s$ of the invariant solutions to
be constructed and find the optimal system of subgroups of the group $G$
having the rank $s$. This is done with the use of the fact that
the subgroup classification problem reduces to classifying
inequivalent subalgebras of the rank $k$ of the Lie algebra $g$ of
the group $G$. This classification is performed within the action
of the inner automorphism group of the algebra $g$.
\item For each of the so obtained subgroups we construct the full
set of functionally-independent invariants, which yields the
invariant ansatz.
\item Inserting the above ansatz into the system of partial
differential equa\-ti\-ons under study reduces it to the one
having $n-s$ independent variables.
\item We investigate the reduced system and construct its exact solutions.
Each of them corresponds to the invariant solution of the initial
system.
\end{enumerate}

Symmetry properties of the overwhelming majority of physi\-ca\-lly
significant differential equations (including the Maxwell and
$SU(2)$ Yang-Mills equations) are well-known. The most important
symmetry groups are those isomorphic to the Euclid, Galilei,
Poincar\'e groups and their natural extensions (the Schr\"odinger
and conformal groups). This fact was a moti\-va\-ti\-on for
investigation of the subgroup structure of these fundamental
groups ini\-tia\-ted by the paper by J.Patera, P.Winternitz and
H.Zassenhaus \cite{m55}. They have suggested the general method
for classifying continuous subgroups of Lie groups and illustrated
its efficiency by re-deriving the known classification of
inequivalent subgroups of the Poincar\'e group $P(1,3)$.
Exploiting this method has enabled to get the full description of
continuous subgroups of a number of important symmetry groups
arising in theoretical and mathematical physics, including the
Euclid, Galilei, Poincar\'e, Schr\"odinger and conformal groups
(see, e.g., \cite{m30} and the references therein).

Thus to get the complete solution of the problem of symmetry
reduction within the framework of the above formulated algorithm
we need to be able to perform the remaining steps (III)--(V).
However, solving these problems for a system of partial
differential equations requires enormous amount of computations,
and what is more, these computations cannot be fully automatized
with the aid of symbolic computation rou\-tines. On the other
hand, it is possible to simplify drastically the computations, if
one notes that for the majority of physically important
realizations of the Euclid, Galilei, Poincar\'e groups and of
their extensions the corresponding invariant solutions admit
linear representation. It was this very idea that had enabled
construc\-ting  broad classes of invariant solutions of a
number of nonlinear spinor equations \cite{m31}--\cite{m33}.

In the sequel, we will concentrate on the case of the 15-parameter
conformal group $C(1,3)$, admitted both by the Maxwell and $SU(2)$
Yang-Mills equa\-ti\-ons. We emphasize that the same reasoning
applies directly to the case of the 11-parameter Schr\'odinger
group $Sch(1,3)$, which is the analogue of the conformal group in
the non-relativistic physics. The group $C(1,3)$ acts in the open
domain $M \subset R^{1,3} \times R^q$ of the four-dimensional
Minkowski space-time of the independent variables $x_0, {\bf x} =
(x_1, x_2, x_3)$ and of the $q$-dimensional space of dependent
variables ${\bf u} = {\bf u}(x_0, {\bf x}), \ {\bf u} = (u^1, u^2,
\ldots, u^q)$.

The Lie algebra $c(1,3)$ of the conformal group $C(1,3)$ is
spanned by the generators of the translation $P_\mu,\
(\mu=0,1,2,3)$, rotation $J_{ab},\ (a,b=1,2,3, \ a<b)$, Lorentz
rotation $J_{0a},\ (a=1,2,3)$, dilation $D$ and conformal
$K_\mu$,\ $ (\mu=0,1,2,3)$, transformations. The basis elements of
$c(1,3)$ satisfy the following commutation relations:
\begin{eqnarray}
&& [P_\mu, P_\nu]=0, \ \ [P_\mu, J_{\alpha \beta}] = g_{\mu
\alpha} P_\beta-g_{\mu \beta} P_\alpha, \nonumber \\ && [J_{\mu
\nu}, J_{\alpha \beta}] = g_{\mu \beta} J_{\nu \alpha} + g_{\nu
\alpha} J_{\mu \beta} -g_{\mu \alpha} J_{\nu \beta}-g_{\nu \beta}
J_{\mu \alpha}, \label{2.8} \\ && [P_\mu, D ] = P_\mu, \ \ [J_{\mu
\nu}, D] =0, \label{2.9} \\ && [K_\mu, J_{\alpha \beta}] = g_{\mu
\alpha} K_\beta -g_{\mu \beta} K_\alpha, \ [D, K_\mu] = K_\mu,
\nonumber \\ && [K_\mu, K_\nu] =0, \ [P_\mu, K_\nu] = 2 (g_{\mu
\nu} D-J_{\mu \nu}). \label{2.10}
\end{eqnarray}
Here $\mu, \nu, \alpha, \beta=0,1,2,3$ and $g_{\mu \nu}$ is the
metric tensor of the Minkowski space-time $R^{1,3}$, i.e., $$
g_{\mu \nu} = \cases{1, &$\mu=\nu=0$; \cr -1, & $\mu=\nu=1,2,3$;
\cr 0, & $\mu\not =\nu$. \cr} $$

The group $C(1,3)$ contains the following important subgroups:
\begin{enumerate}
\renewcommand{\labelenumi}{\arabic{enumi})}
\item The Poincar\'e group $P(1,3)$, whose Lie algebra $p(1,3)$
is spanned by the operators $P_\mu, J_{\mu \nu}$,\ $(\mu,
\nu=0,1,2,3)$ satisfying commutation relations (\ref{2.8});
\item the extended Poincar\'e group $\tilde P(1,3)$, whose
Lie algebra $\tilde p(1,3)$ is span\-ned by the operators $P_\mu,
J_{\mu \nu}, D,\ (\mu, \nu=0,1,2,3)$ satisfying
commu\-ta\-ti\-on relations (\ref{2.8}), (\ref{2.9}).
\end{enumerate}

Analysis of the symmetry groups of the equations of the
relativistic phy\-sics shows that for the majority of them the
generators of the Poincar\'e, extended Poincar\'e and conformal
groups can be represented in the following form (see, e.g.,
\cite{m19,m21,m33,m42}):
\begin{eqnarray} P_\mu &=&\partial_{x_\mu}, \nonumber \\ J_{\mu
\nu}&=& x^\mu \partial_{x_\nu} - x^\nu \partial_{x_\mu}-(S_{\mu
\nu} {\bf u} \cdot \partial_{\bf u}), \nonumber \\ D&=& x_\mu
\partial_{x_\mu} -k(E {\bf u}\cdot \partial_{\bf u}), \nonumber \\
K_0 &=& 2 x_0 D - (x_\nu x^\nu) \partial_{x_0} -2 x_a (S_{0a} {\bf
u} \cdot \partial_{\bf u}), \nonumber \\ K_1 &=&- 2 x_1 D - (x_\nu
x^\nu) \partial_{x_1}+2 x_0 (S_{01} {\bf u} \cdot \partial_{\bf
u}) \nonumber \\ && -2 x_2 (S_{12} {\bf u} \cdot \partial_{\bf
u}) -2 x_3 (S_{13} {\bf u} \cdot \partial_{\bf u}) , \label{2.11}
\\ K_2 &=&- 2 x_2 D - (x_\nu x^\nu) \partial_{x_2}+2 x_0 (S_{02}
{\bf u} \cdot \partial_{\bf u}) \nonumber \\ && +2 x_1 (S_{12}
{\bf u} \cdot \partial_{\bf u}) -2 x_3 (S_{23} {\bf u} \cdot
\partial_{\bf u}) , \nonumber  \\ K_3 &=&- 2 x_3 D - (x_\nu x^\nu)
\partial_{x_3}+2 x_0 (S_{03} {\bf u} \cdot \partial_{\bf u})
\nonumber \\ && +2 x_1 (S_{13} {\bf u} \cdot \partial_{\bf u})+2
x_2 (S_{23} {\bf u} \cdot \partial_{\bf u}) .\nonumber
\end{eqnarray}

In formulae (\ref{2.11}) $S_{\mu \nu}$ are constant $q\times q$
matrices, that realize a representation of the Lie algebra
$o(1,3)$ of the pseudo-orthogonal group $O(1,3)$ and satisfy the
commutation relations \be \label{2.12} [S_{\mu \nu}, S_{\alpha
\beta}] = g_{\mu \beta} S_{\nu \alpha} + g_{\nu \alpha} S_{\mu
\beta} -g_{\mu \alpha} S_{\nu \beta} -g_{\nu \beta} S_{\mu
\alpha}, \ee $\mu, \nu, \alpha, \beta =0,1,2,3;\ g_{\mu \nu}$ is
the metric tensor of the Minkowski space $R^{1,3};\ E$ is the unit
$q \times q$-matrix;\ ${\bf u} = (u^1, u^2, \ldots, u^q)^T$;\
$\partial_{\bf u} = ( \partial_{u^1}, \partial_{u^2}, \ldots,
\partial_{u^q})^T$; the symbol $(*\cdot *)$ stands for the scalar
product in the vector space $R^q$. We remind that the repeated
indices imply summation over the corresponding interval and
raising and lowering the indices is carried out with the help of
the metric $g_{\mu\nu}$. What is more, $k$ is some fixed real
number called the conformal degree of the group $c(1,3)$.

It follows from relations (\ref{2.11}) that the basis elements of
the Lie algebra $c(1,3)$ have the form (\ref{2.2}), where the
functions $\xi^i_a$ depend on ${\bf x} \in X= R^p$ only and the
functions $\eta^a_j$ are linear in ${\bf u}$. We will prove that
owing to these properties of the basis elements of $c(1,3)$ the
ansatzes invariant under subalgebras of the algebra (\ref{2.11})
admit linear representation.

Let a local transformation group $G$ act projectively in $M$,
and let $g = \langle X_1, \ldots, X_n \rangle$ be its Lie algebra
spanned by the infinitesimal operators of the form
\begin{equation} \label{2.13}
X_a = \xi^i_a({\bf x}) \partial_{x_i} +\rho^a_{jk} ({\bf x}) u^k
\partial_{u^j},
\end{equation}
where $a=1, \ldots, n,\ i=1, \ldots, p, \ j,k=1, \ldots, q$.

According to what was said above, the group $G$ has the two types
of invariants. The first set of invariants is formed by $p-s$
(where $s$ is the rank of the group $G$) functionally independent
invariants
\be \label{2.14} {\bf w} = {\bf w} ({\bf x}), \ {\bf w} =
(\omega^1, \ldots, \omega^{p-s}). \ee
The second set is formed by $q$ invariants
\be \label{2.15} {\bf h} = {\bf h} ({\bf x}, {\bf u}),
\ {\bf h} = (h^1, \ldots, h^{q}). \ee
And what is more, the functions ${\bf w}$ and ${\bf h}$ are
invariants of the group $G$ if and only if they are, respectively,
solutions of the following systems of partial differential
equations:
\begin{eqnarray}
&&\xi^i_a ({\bf x}) \frac{\partial \omega^b}{\partial x_i} =0,
 \label{2.16} \\ && \xi^i_a
({\bf x}) \frac{\partial h^l}{\partial x_i} +\rho^a_{jk} ({\bf x})
u^k \frac{\partial h^l}{\partial u^j}=0.
\label{2.17}
\end{eqnarray}
In (\ref{2.16}), (\ref{2.17}) the indices take the following
values,\ $a=1,\ldots, n$,\ $b=1, \ldots, p-s$,\ $i=1,\ldots, p$,\
$j,k,l =1,\ldots, q$.

Generically, a $G$-invariant ansatz has the form (\ref{2.6}),
where ${\bf v} \equiv {\bf h}$. However, provided the
infinitesimal operators of the group $G$ are of the form
(\ref{2.13}), $G$-invariant ansatz for the vector field ${\bf u}$
can be represented in the linear form \cite{m33} \be \label{2.18}
{\bf u} = \Lambda ({\bf x}) {\bf h} ({\bf w}), \ee where $\Lambda
({\bf x})$ is some $q\times q$ matrix non-singular in $\Omega
\subset M$,\ ${\bf u} = (u^1, \ldots, u^q)^T$, ${\bf h} = (h^1,
\ldots, h^q)^T$.

The matrix $\Lambda ({\bf x})$ from (\ref{2.18}) is obtained by
integrating the system of partial differential equations to be
derived below.

\bl \label{l1} Let a $G$-invariant ansatz be of the form
(\ref{2.18}). Then there is $q \times q$-matrix $H({\bf x}) =
\Lambda^{-1} ({\bf x})$ non-singular in $\Omega$ satisfying the
matrix partial differential equation \be \label{2.19} \xi^i_a
({\bf x}) \frac{\partial H({\bf x})}{\partial x_i} + H({\bf x})
\Gamma_a ({\bf x}) =0, \ee where $\Gamma_a({\bf x})$ is the
$q\times q$ matrices, whose $(i,j)$th entry reads as
$\rho^a_{ij}({\bf x})$,\ $i,j = 1,\ldots, q$. \el {\it Proof.}
Provided a $G$-invariant ansatz is of the form (\ref{2.18}), the
relation $${\bf h} = H({\bf x}) {\bf u}$$ with $H({\bf x}) =
\Lambda^{-1} ({\bf x})$ holds. So, the second set of invariants
(\ref{2.15}) of the group $G$ consists of the functions, which are
linear in $u^j$ and, consequently, can be represented in the form
$$h^b = h_{bl}({\bf x}) u^l, \ \ b,l=1, \ldots, q.$$

The function $h^b$ is the invariant of the group $G$, if and only
if, it satisfies equation (\ref{2.17}) $$\xi^i_a ({\bf x})
\frac{\partial h_{bl}({\bf x})}{\partial x_i} u^l
+\rho^a_{jl}({\bf x})u^l h_{bj}({\bf x}) =0.$$

Splitting this relation by $u^l$ yields that the system of
partial differential equations
\be \label{2.20} \xi^i_a ({\bf x}) \frac{\partial h_{bl}({\bf
x})}{\partial x_i} + h_{bj}({\bf x}) \rho^a_{jl}({\bf x}) =0,
\ee
holds for all the values of $b, l$. The indices in (\ref{2.20})
take the following values,\ $a=1, \ldots, n,\ i=1, \ldots, p,\
b,j,l=1, \ldots, q$.

It is readily seen that the second term of the left-hand side of
equation (\ref{2.20}) is the $(b,l)$th entry of the matrix $H({\bf
x}) \Gamma_a ({\bf x}),\ (a=1, \ldots, n)$. Hence it follows that
the matrix $H({\bf x})$ satisfies equation (\ref{2.19}). The lemma
is proved.

Below, we list the forms of the matrices $\Gamma_a$ for the basis
operators of the algebra $c(1,3)$
\begin{itemize}
\item matrices $\Gamma_a,\ a=1,2,3,4$ corresponding to the operators
$P_\mu,\ (\mu =0,1,2,3)$ are zero $q\times q$ matrices;
\item matrices $\Gamma_a,\ a=1,\ldots, 6$ corresponding to the
operators $J_{\mu \nu},\ (\mu, \nu=0,1,2,3)$ are equal to $-S_{\mu
\nu}$, where $S_{\mu \nu}$ are constant $q\times q$ matrices
realizing a representation of the algebra $o(1,3)$ and satisfying
commutation relations (\ref{2.12});
\item the matrix $\Gamma_1$ corresponding to the dilation operator
$D$ reads as $-kE$, where $k$ is the conformal degree of the
algebra $c(1,3)$ and $E$ is the unit $q\times q$ matrix;
\item matrices $\Gamma_a,\ a=1,2,3,4$ corresponding to the operators
$K_{\mu},\ (\mu=0,1,2,3)$ are given by the following formulae:
\begin{eqnarray*}
\Gamma_1&=& -2x_0 k E - 2 x_1 S_{01} - x_2 S_{02} - x_3 S_{03},
\\ \Gamma_2&=&2x_1 k E +2 x_0 S_{01}-2 x_2 S_{12} -2 x_3 S_{13}, \\
\Gamma_3 &=& 2x_2 k E +2 x_0 S_{02}+2 x_1 S_{12} -2 x_3 S_{23}, \\
\Gamma_4&=& 2x_3 k E +2 x_0 S_{03} + 2 x_1 S_{13} +2 x_2 S_{23}.
\end{eqnarray*}
\end{itemize}

With the explicit forms of the matrices $\Gamma_a$ in hand we can
determine the structure of the matrices $H = \Lambda^{-1}$ for
ansatz (\ref{2.18}) invariant under a subalgebra $g$ of the
conformal algebra $c(1,3)$.

If $g\in p(1,3) = \langle P_\mu, J_{\mu \nu} | $ $\mu, \nu=0,1,2,3
\rangle$, then the corresponding matrices $\Gamma_a$ are linear
combinations of the matrices $S_{\mu \nu}$. Hence it follows that
the matrix $H$ can be looked in the form
\be \label{2.21}
H = \tilde H = \mathop \Pi \limits_{\mu < \nu}\exp(\theta_{\mu\nu}
S_{\mu\nu}),
\ee
where $\theta_{\mu\nu} = \theta_{\mu\nu} (x_0, {\bf x})$ are
arbitrary smooth functions defined in $\tilde \Omega \subset
R^{1,3}$.

Next, if $g$ is a subalgebra of the conformal algebra $c(1,3)$
with a non-zero projection on the vector space spanned by the
operators $D, K_0, K_1, K_2, K_3$, then the corresponding
matrices $\Gamma_a$ are linear combinations of the matrices $E$ and
$S_{\mu \nu}$. That is why, the matrix $H$ should be looked for in
the more general form
\be \label{2.22} H = \exp (\theta E) \tilde H, \ee
where $\theta = \theta(x_0, {\bf x})$ is an arbitrary smooth
function defined in $\tilde \Omega$ and $\tilde H$ is the matrix
given in (\ref{2.21}).

\subsection{Subalgebras of the conformal algebra $c(1,3)$ of the
rank 3}

Now we turn to the problem of constructing conformally-invariant
ansatzes that reduce systems of partial differential equations
invariant under the group $C(1,3)$ to systems of ordinary
differential equations.

As a second step of the algorithm of symmetry reduction formulated
above, we have to describe the optimal system of subalgebras of
the algebra $c(1,3)$ of the rank $s=3$. Indeed, the initial system
has $p=4$ independent variables. It has to be reduced to a system
of differential equations in $4-s=1$ independent variables, so
that $s=3$.

Classification of inequivalent subalgebras of the algebras
$p(1,3)$, $\tilde p(1,3)$, $c(1,3)$ within actions of different
automorphism groups (including the groups $P(1,3)$, $\tilde
P(1,3)$ and $C(1,3)$) is already available (see, e.g.,
\cite{m30}). Since we will concentrate in the sequel on
conformally-invariant systems, it is natural to restrict our
considerations to the classification of subalgebras of $c(1,3)$
that are inequivalent within the action of the conformal group
$C(1,3)$.

In order to get the full lists of the subalgebras in question we
have to check that relation (\ref{2.3}) with $s=3$ holds for each
element of the lists of inequivalent subalgebras of the algebras
$p(1,3), \tilde p(1,3), c(1,3)$ given in \cite{m30}. Evidently, we
can restrict our considerations to subalgebras having the
dimension not less than 3.

Let $c(1,3)$ be the conformal algebra having the basis operators
(\ref{2.11}) and $c^{(1)}(1,3)$ be the conformal algebra spanned
by the operators
\begin{eqnarray} \label{2.23}
&& P^{(1)}_\mu = \partial_{x_\mu},\quad  J^{(1)}_{\mu \nu} = x^\mu
\partial_{x_\nu} -x^\nu \partial_{x_\mu},\quad   D^{(1)} = x_\mu
\partial_{x_\mu},\nonumber \\ &&  K^{(1)} = 2 x^\mu D^{(1)} -
(x_\nu x^\nu) \partial_{x_\mu},
\end{eqnarray}
where $\mu,\nu = 0,1,2,3$.

Note that the conformal group $C(1,3)$ generated by the
infinitesimal operators (\ref{2.23}) acts in the space of
independent variables $R^{1,3}$ only. That is why, the basis
operators of the algebra $c^{(1)}(1,3)$ act in the space of
dependent variables $R^q$ as zero operators.

\bl \label{l2} Let $L$ be a subalgebra of the algebra $c(1,3)$ of
the rank $s$ and let $s^{(1)}$ be the rank of the projection of
$L$ on $c^{(1)}(1,3)$. Then from the equality $s=s^{(1)}$ it
follows that ${\rm dim}\, L =s$. \el
{\it Proof.}\  Suppose that the reverse assertion holds, namely,
that $L\ne s$. As $L\ge s$, hence it follows that $L>s$. Choose
the basis elements $X_1, \ldots, X_m$ of the algebra $L$ so that
\begin{itemize}
\item the rank of the matrix $M$, whose entries are projections of
the operators $X_1, \ldots, X_m$ on $c^{(1)} (1,3)$, is equal to
$s$, and
\item the linear space spanned by the operators $X_1, \ldots, X_m$
contains $L \cap \langle P_0, P_1, P_2, P_3 \rangle$.
\end{itemize}

We denote as $S_0$ the point $(x^0_0, {\bf x}^0)$ $\in \tilde
\Omega$ in which the rank of the matrix $M$ equals to $s$. Let the
vector fields $X_i$ be equal to $X^0_1, \ldots, X^0_s$,
$X^0_{s+1}, \ldots X^0_m$ in $S_0$. Then there are constants
$\alpha_1, \ldots, \alpha_s$, such that the vector field $\alpha_1
X^0_1 + \ldots + \alpha_s X^0_s + X^0_{s+1}$ restricted to the
space of dependent variables $U = R^q$ is a non-zero operator.
Indeed, if this operator vanishes identically on $R^q$ for any
choice of $\alpha_1, \ldots, \alpha_s$, then the vector fields
$X^0_1,\ldots,X^0_s, X^0_{s+1}$ belong to the vector space
$\langle P_0, P_1, P_2, P_3 \rangle$ and this fact contradicts to
the assumptions that ${\rm dim}\, L>s,\ {\rm rank}\, L=s$.
Consequently, the matrix formed by the coefficients of the vector
fields $X^1, \ldots, X_s, \alpha_1X_1 + \ldots + \alpha_s X_s +
X_{s+1}$, has a non-zero minor of the order $s+1$ in some point
$(x^0_0, {\bf x}^0, {\bf u}^0)$ (the first four coordinates are
same as those of the point $S_0$). This contradicts to the
assumption that ${\rm dim}\, L > s$. Hence we conclude that ${\rm
dim}\, L = s$. The lemma is proved.

It follows from the above lemma that the validity of the relation
(\ref{2.3}) with $s=3$ should be ascertained only for the
three-dimensional subalgebras of the algebras $p(1,3), \tilde
p(1,3), c(1,3)$ given in \cite{m30}. And what is more, we can
restrict our considerations to checking the first condition from
(\ref{2.3}).

Consider the subalgebras of the algebra $p(1,3)$, whose basis
operators are of the form (\ref{2.11}). Among the
three-dimensional subalgebras of the algebra $p(1,3)$ listed in
\cite{m30} there are only five subalgebras $\langle G_1$, $P_0 +
P_3$, $P_1 \rangle$, $\langle J_{12}$, $P_1$, $P_2 \rangle$,
$\langle J_{03}, P_0, P_3 \rangle, \langle J_{12}, J_{13}, J_{23}
\rangle, $ $\langle J_{01}, J_{02}, J_{12} \rangle$, that do not
respect the first condition (\ref{2.3}). These subalgebras give
rise to the so called partially invariant solutions (see, e.g.,
\cite{m19}). Partially invariant solutions cannot be handled in a
generic way, they should always be considered within the context
of a specific system of partial differential equation to be
reduced. We exclude the partially invariant solutions from the
further considerations. The remaining inequivalent subalgebras are
listed in the assertion below.

\bt \label{t1} The list of subalgebras of the algebra $p(1,3)$ of
the rank 3, defined within the action of the inner automorphism
group of the algebra $c(1,3)$, is exhausted by the following
subalgebras:
\begin{eqnarray*}
L_1&=& \langle P_0, P_1, P_2 \rangle ; \ \ \ L_2 = \langle P_1,
P_2, P_3 \rangle ;\\ L_3&=& \langle M, P_1, P_2 \rangle ; \ \ \
L_4 = \langle J_{03}+\alpha J_{12}, P_1, P_2 \rangle; \\ L_5&=&
\langle J_{03}, M, P_1 \rangle ; \ \ \ L_6 = \langle J_{03}+P_1,
P_0, P_3 \rangle ;\\ L_7&=& \langle J_{03}+P_1, M, P_2 \rangle ; \
\ \ L_8 = \langle J_{12}+\alpha J_{03}, P_0, P_3 \rangle ;\\
L_9&=& \langle J_{12}+P_0, P_1, P_2 \rangle ; \ \ \ L^j_{10} =
\langle J_{12}+(-1)^j P_3, P_1, P_2 \rangle ;\\ L^j_{11}&=&
\langle J_{12}+(-1)^j 2 T, P_1, P_2 \rangle ; \ \ \ L_{12} =
\langle G_1,M, P_2+\alpha P_1 \rangle ;\\ L^j_{13}&=& \langle
G_{1}+(-1)^j P_2, M , P_1 \rangle ; \ \ \ L_{14} = \langle G_1+2
T, M , P_2 \rangle ;\\ L_{15}&=& \langle G_{1}+2T, M,  P_1+\alpha
P_2 \rangle ; \ \ \ L_{16} = \langle J_{12}, J_{03},M \rangle ;\\
L^j_{17}&=& \langle G^j_{1}, G^j_2, M \rangle ; \ \ \ L_{18} =
\langle J_{03}, G_1 , P_2 \rangle ;\\ L_{19}&=& \langle G_{1},
J_{03}, M \rangle ; \ \ \ L_{20} = \langle G_1, J_{03}+ P_2, M
\rangle ;\\ L_{21}&=& \langle G_{1}, J_{03}+P_1+\alpha P_2 , M
\rangle ; \ \ \ L_{22} = \langle G_1, G_2, J_{03}+ \alpha J_{12}
\rangle ,
\end{eqnarray*}
where $\alpha \in {\bf R}; \ M = P_0+P_3, \ T =
\frac{\textstyle{1}}{\textstyle{2}} (P_0-P_3), \ G_a = J_{0a}
-J_{a3},\ (a=1,2)$; $G^j_1 = G_1+(-1)^j P_2, $ $G^j_2 = G_2
-(-1)^j P_1 +\alpha P_2$; j=1,2. \et

In the same way, we handle the three-dimensional subalgebras of
the algebras $\tilde p(1,3)$ and $c(1,3)$. We have skipped from
the list of subalgebras of the algebra $\tilde p(1,3)$ those
conjugate to subalgebras of $p(1,3)$. Furthermore, we have skipped
from the list of subalgebras of the conformal algebra those
conjugate to subalgebras of the algebra $\tilde p(1,3)$. The
results obtained are presented in the two assertions below.

\bt \label{t2} The list of subalgebras of the algebra $\tilde
p(1,3)$ of the rank 3, defined within the action of the inner
automorphism group of the algebra $c(1,3)$, is exhausted by the
subalgebras given in Assertion \ref{t1} and by the following
subalgebras:
\begin{eqnarray*} F_1&=& \langle D, P_0, P_3 \rangle;
\ \ F_2 = \langle J_{12} +\alpha D, P_0, P_3 \rangle; \\ F_3&=&
\langle J_{12}, D, P_0 \rangle; \ \ F_4 = \langle J_{12}, D, P_3
\rangle; \\ F_5&=& \langle J_{03}+\alpha D,  P_0, P_3  \rangle; \
\ F_6 = \langle J_{03}+\alpha D, P_1, P_2 \rangle; \\ F_7&=&
\langle J_{03}+\alpha D, M,  P_1 \rangle \ (a \not =0) ; \\ F_8&
=& \langle J_{03}+ D+(-1)^j 2 T , P_1, P_2 \rangle; \\ F_9&=&
\langle J_{03}+ D+(-1)^j 2 T,  M,  P_1 \rangle  ; \ \ F_{10} =
\langle J_{03}, D , P_1 \rangle; \\ F_{11}&=& \langle J_{03}, D, M
\rangle  ; \ \ F_{12} = \langle J_{12}+ \alpha J_{03}+\beta  D ,
P_0, P_3 \rangle \ (\alpha \not =0); \\ F_{13}&=& \langle J_{12}
+\alpha J_{03} +\beta D, P_1, P_2 \rangle \ (\alpha \not =0); \\
F_{14}&=& \langle J_{12} +\alpha( J_{03}+ D+2T), P_1, P_2 \rangle
\ (\alpha \not =0); \\ F_{15}&=& \langle J_{12} +\alpha J_{03}, D,
M \rangle \ (\alpha \not =0); \\ F_{16}&=& \langle J_{03} +\alpha
D, J_{12}+\beta D,M \rangle \ (0 \le |\alpha | \le 1, \ \beta \ge
0, |\alpha |+|\beta |\not =0); \\ F_{17}&=& \langle J_{03} +
D+(-1)^j 2T, J_{12}+2\alpha T,M \rangle \ (\alpha \in R); \\
F_{18}&=& \langle J_{03} + D, J_{12}+(-1)^j 2T ,M \rangle; \ \
F_{19} =\langle J_{03}, J_{12}, D \rangle ; \\ F_{20}&=& \langle
G_1, J_{03} +\alpha  D, P_2 \rangle\ (0< |\alpha| \le 1); \\
F_{21}&=& \langle J_{03} + D, G_1 +(-1)^j P_2,M \rangle;\\
F_{22}&=& \langle J_{03} -D +(-1)^j M, G_1, P_2  \rangle;\\
F_{23}&=& \langle J_{03} + 2 D, G_1 +(-1)^j 2 T ,M \rangle;\\
F_{24}&= &\langle J_{03} + 2 D, G_1 +(-1)^j 2 T , P_2\rangle.
\end{eqnarray*}
Here $M = P_0 + P_3, \ G_1 = J_{01} - J_{13}$,\ $T =
\frac{1}{2}(P_0 - P_3)$, the parameters $\alpha, \beta$ are
positive (if otherwise is not indicated);\ $j=1,2$. \et

\bt \label{t3} The list of subalgebras of the algebra $c(1,3)$ of
the rank 3, defined within the action of the inner automorphism
group of the algebra $c(1,3)$, is exhausted by the subalgebras of
the algebras $p(1,3), \tilde p(1,3)$ given in Assertions \ref{t1}
and \ref{t2} and by the following subalgebras:
\begin{eqnarray*}
C_1 &=& \langle S +T +J_{12}, G_1+P_2, M \rangle ; \\ C_2& =&
\langle S +T+J_{12} +G_1+P_2, G_2-P_1, M \rangle ; \\ C_3&=&
\langle J_{12}, S+T, M \rangle; \ \ C_4 = \langle S+T, Z, M
\rangle ; \\ C_5 &=& \langle S+T+\alpha J_{12}, Z, M \rangle \
(\alpha \not =0); \\ C_6 &=& \langle S+T+ J_{12}+\alpha Z,
G_1+P_2, M \rangle \ (\alpha \not =0); \\ C_7 &=& \langle S+T+
J_{12}, Z, G_1+P_2 \rangle ; \\ C_8 &=& \langle S+T+ \beta Z,
J_{12}+\alpha Z,  M \rangle \ \ (\alpha , \beta \in R, |\alpha|
+|\beta| \not =0); \\ C_9&=& \langle J_{12}, S+T, Z \rangle ; \ \
C_{10} = \langle D -J_{03}, S, T \rangle ; \\ C_{11} &=& \langle
P_2 + K_2 +\sqrt{3}(P_1+K_1)+K_0-P_0, \\ && -D
+J_{02}-\sqrt{3}J_{01}, P_0+K_0-2(K_2-P_2) \rangle ; \\ C_{12}&=&
\langle P_0 +K_0 \rangle \oplus \langle J_{12}, K_3 -P_3 \rangle ;
\\ C_{13} &=& \langle 2 J_{12} +K_3- P_3, 2 J_{13} -K_2 +P_2, 2
J_{23}+K_1-P_1 \rangle ; \\ C_{14}&=& \langle P_1 +K_1 +2 J_{03},
P_2 +K_2+K_0-P_0, 2 J_{12} +K_3 -P_3 \rangle,
\end{eqnarray*}
where $M = P_0 + P_3,\ G_{0a} = J_{0a} - J_{a3},\ (a=1,2),\ Z =
J_{03} + D, S = \frac{\textstyle{1}}{\textstyle{2}}(K_0 + K_3)$,\
$ T = \frac{\textstyle{1}}{\textstyle{2}} (P_0-P_3)$. \et

\noindent {\bf Remark.} While classifying subalgebras of the
extended Poincar\'e algebra $\tilde p(1,3)$, the discrete
equivalence transformations $\Phi_1, \Phi_2, \Phi_3$, that leave
the algebra $\tilde p(1,3)$ invariant, were exploited in
\cite{m30}. The result of the action of these groups on the
operators of the algebra $\tilde p(1,3)$ is given in Table 2.1.
That is why, we have completed the list of subalgebras of the
algebras $p(1,3), \tilde p(1,3)$ obtained in \cite{m30} by the
subalgebras obtainable by acting on these subalgebras with the
discrete transformation groups $\Phi_1, \Phi_2, \Phi_3$.
\vspace{2mm}

\begin{center}
Table 2.1. \vskip 4mm
\begin{tabular}{|c|c|c|c|} \hline
Operators & \multicolumn{3}{|c|}{Action on $\tilde p(1,3)$ }
\\ \cline{2-4} & $\Phi_1$ &$ \Phi_2 $ & $ \Phi_3 $  \\ \hline
$P_0$ &$-P_0$ & $P_0$ & $-P_0$ \\ \hline $P_1$ &$-P_1$ & $-P_1$ &
$P_1$
\\ \hline $P_a\ \ (a=2,3)$ &$-P_a$ & $P_a$ & $-P_a$ \\  \hline
$J_{03}$ &$J_{03}$ & $J_{03}$ & $J_{03}$ \\ \hline $J_{12}$
&$J_{12}$ & $-J_{12}$ & $-J_{12}$ \\ \hline $G_1$ &$G_1$ & $-G_1$
& $-G_1$ \\ \hline $G_2$ &$G_2$ & $G_2$ & $G_2$ \\ \hline $M$
&$-M$ & $M$ & $-M$ \\ \hline $T$ &$-T$ & $T$ & $-T$ \\ \hline $D$
&$D$ & $D$ & $D$ \\ \hline
\end{tabular}
\end{center}

\subsection{Construction of conformally-invariant ansatzes}

Now we turn to constructing $C(1,3)$-invariant ansatzes that
reduce confor\-ma\-lly-invariant systems of partial differential
equations to systems of ordinary differential equations. To this
end, we use the lists of subalgebras of the algebra $c(1,3)$ given
in Assertions \ref{t1}--\ref{t3}. Note that all the subsequent
computations are performed under supposition that the basis
operators of $c(1,3)$ are of the form (\ref{2.11}).

As shown in Subsection 2.1, the ansatzes in question can be looked
for in the linear form (\ref{2.18}), matrices $H = \Lambda^{-1}$
being searched for in the form (\ref{2.22}). According to Lemma
\ref{l1}, the matrix $H$ has to satisfy equations (\ref{2.19}),
whose coefficients are defined uniquely by the choice of a
subalgebra of the conformal algebra of the rank 3. So that, the
problem of complete description of conformally-invariant ansatzes
reduces to solving system of partial differential equations
(\ref{2.16}), (\ref{2.19}) for each of the subalgebras of the
conformal algebra, which requires the huge amount of computations.
The calculations simplify essentially, if we take into account the
general structure of the subalgebras listed in Assertions
\ref{t1}--\ref{t3}.

For the further convenience, we will use the following basis of
the algebra $o(1,3)$:\ $S_{03}$, $S_{12}$, $H_a$, $\tilde H_a,\
(a=1,2)$, where $H_a = S_{0a} - S_{a3},\ \tilde H_a = S_{0a} +
S_{a3},\ (a=1,2)$. It is not difficult to check that these
matrices satisfy the commutation relations
\begin{eqnarray} \label{2.23a}
&& [S_{03}, S_{12}] = [H_1, H_2] = [\tilde H_1, \tilde H_2] =0,
\nonumber \\ && [H_a, S_{03}] = H_a, [\tilde H_a, S_{03}] =
-\tilde H_a,\ (a=1,2), \nonumber \\ && [H_1, S_{12}] = -H_2, [H_2,
S_{12}] = H_1, \\ && [\tilde H_1, S_{12}] = -\tilde H_2, \ [\tilde
H_2, S_{12}] = \tilde H_1, \nonumber \\ && [H_1, \tilde H_1] =
[H_2, \tilde H_2] = -2 S_{03}, \nonumber \\ && [\tilde H_2, H_1] =
[H_2, \tilde H_1] = 2 S_{12}. \nonumber
\end{eqnarray}
In particular, relations (\ref{2.23a}) imply that the matrices
$H_1, H_2, S_{12}, S_{03}$ and $\tilde H_1, \tilde H_2, S_{12},
S_{03}$ realize two matrix representations of the Euclid algebra
$\tilde e(2)$ (here the matrix $S_{03}$ is identified with the
dilation generator and the matrices $H_1, H_2$ and $\tilde H_1,
\tilde H_2$ are identified with the translation generators).
Furthermore, as $E$ is the unit matrix, it commutes with all the
basis elements of $o(1,3)$, namely, \be \label{2.24} [E, S_{12}] =
[E, S_{03}] = [E, H_a] = [E, \tilde H_{a}] =0, \ee where $a=1,2$.

Analyzing the structure of the basis elements of the subalgebras
of the conformal algebra given in Assertions \ref{t1}--\ref{t3} we
see that the corresponding matrices $\Gamma_a$ are most
conveniently represented in terms of the matrices $S_{03}$,
$S_{12}$, $H_a$, $\tilde H_a,\ (a=1,2)$. Hence we conclude that
the matrix $H = H(x_0, {\bf x}) = \Lambda^{-1} (x_0, {\bf x})$ can
be looked for in the form
\begin{eqnarray} \label{2.25}
H& =& \exp\{(-\ln \theta) E \} \exp (\theta_0 S_{03})
\exp(-\theta_3 S_{12}) \exp(-2 \theta_1 H_1)\nonumber \\
&& \times\exp(-2 \theta_2 H_2)  \exp(-2 \theta_4 \tilde
H_1)\exp(-2 \theta_5 \tilde H_2),
\end{eqnarray}
where $\theta = \theta(x_0, {\bf x}), \ \theta_0 = \theta_0 (x_0,
{\bf x}), \theta_m=\theta_m (x_0,{\bf x}), \ (m=1,2,\ldots, 5)$
are arbitrary smooth functions defined in an open domain $\tilde
\Omega \subset R^{1,3}$ of the Minkowski space of the independent
variables $x_0, {\bf x} = (x_1, x_2, x_3)$.

Let $L = \langle X_a| a=1,2,3 \rangle$ be a subalgebra of the
algebra $c(1,3)$ of the rank 3. By assumption, the basis operators
of $L$ can be written in the following form:
\be \label{2.26} X_a = \xi^\mu_a (x_0, {\bf x}) \partial_{x_\mu} +
(\tilde \Gamma_a {\bf u} \cdot \partial_{\bf u}),  \ a=1,2,3, \ee
and what is more,
\be \label{2.27} \tilde \Gamma_a = f^a E + f^a_0 S_{03} + f^a_1
H_1 + f^a_2 H_2 + f^a_3 S_{12} + f^a_4 \tilde H_1 + f^a_5 \tilde
H_2, \ (a=1,2,3), \ee
where $f^a = f^a(x_0, {\bf x}), \ f^a_0 = f^a_0(x_0, {\bf x}),
f^a_m =f^a_m(x_0, {\bf x}),\ (m=1, \ldots, 5)$ are some fixed
smooth functions. In particular, if the operator $X_a$ is a linear
combination of the translation generators, then $\Gamma_a =0$, and
therefore, $f^a = f^a_0 = f^a_m=0$ in (\ref{2.27}).

Owing to Lemma \ref{l1}, in order to construct ansatz (\ref{2.18})
invariant under the subalgebra $L$, we have to solve systems
(\ref{2.16}), (\ref{2.19}), which in the case under consideration
read as
\begin{eqnarray}
&& \xi^\mu_a \frac{\partial \omega}{\partial x_\mu}=0,
\label{2.28} \\ && \xi^\mu_a \frac{\partial H}{\partial x_\mu} + H
\tilde \Gamma_a=0, \label{2.29}
\end{eqnarray}
where\ $a=1,2,3,\ \mu=0,1,2,3$. The functions $\xi^\mu_a$ $=
\xi^\mu_a(x_0, {\bf x})$ and the variable matrices $\tilde
\Gamma_a$ $= \tilde \Gamma (x_0, {\bf x})$ are the coefficients of
the basis operators of the subalgebra $L$ (note that $\tilde
\Gamma_a$ is of the form (\ref{2.27})). Matrix function
$H$ (\ref{2.25}) and the scalar function $\omega = \omega(x_0,
{\bf x})$ are to be determined while integrating (\ref{2.28}),
(\ref{2.29}).

Next, we will prove a technical assertion to be used in the sequel
for simplifying the form of system (\ref{2.29}).

\bl \label{l3} Let $H$ be of the form (\ref{2.25}). Then the
following identity holds true:
\begin{eqnarray*}
\xi^\mu_a \frac{\partial H}{\partial x_\mu} &=& H \Bigl \{
-\theta^{-1} \xi^\mu_a \frac{\partial \theta}{\partial x_\mu} E +
\xi^\mu_a \frac{\partial \theta_0}{\partial x_\mu} [(1+8 \theta_1
\theta_4 +8 \theta_2 \theta_5) S_{03} \\ && +8(\theta_1
\theta_5-\theta_2 \theta_4) S_{12}+2 \theta_1 H_1+2 \theta_2 H_2
-2(\theta_4 +4\theta_1 \theta^2_4+8 \theta_2 \theta_4 \theta_5 \\
&&- 4 \theta_1 \theta^2_5) \tilde H_1-2(\theta_5+4 \theta_2
\theta^2_5 +8 \theta_1 \theta_4 \theta_5-4 \theta_2 \theta^2_4)
\tilde H_2] \\ && -\xi^\mu_a \frac{\partial \theta_3}{\partial
x_\mu}[8 (\theta_2 \theta_4 -\theta_1 \theta_5) S_{03}+(1+8
\theta_1 \theta_4 +8 \theta_2 \theta_5) S_{12} \\ && + 2 \theta_2
H_1 -2 \theta_1 H_2+2(\theta_5 +4 \theta_2 \theta^2_5-4 \theta_2
\theta^2_4 +8 \theta_1 \theta_4 \theta_5) \tilde H_1 \\ &&
-2(\theta_4 +4 \theta_1 \theta^2_4 -4 \theta_1 \theta^2_5+8
\theta_2 \theta_4 \theta_5) \tilde H_2]\\ && -2 \xi^\mu_a
\frac{\partial \theta_1}{\partial x_\mu} [4 \theta_4 S_{03} +4
\theta_5 S_{12} +H_1+4(\theta^2_5 -\theta^2_4) \tilde H_1 -8
\theta_4 \theta_5 \tilde H_2] \\ && -2 \xi^\mu_a \frac{\partial
\theta_2}{\partial x_\mu} [4 \theta_5 S_{03} -4 \theta_4 S_{12}
+H_2-8 \theta_4 \theta_5 \tilde H_1 +4(\theta^2_4 -\theta^2_5)
\tilde H_2] \\ && -2 \xi^\mu_a \frac{\partial \theta_4}{\partial
x_\mu} \tilde H_1 -2 \xi^\mu_a \frac{\partial \theta_5}{\partial
x_\mu} \tilde H_2]\Bigr \},
\end{eqnarray*}
where $a=1,2,3, \ \mu=0,1,2,3$. \el {\it Proof.}\ Acting by the
linear differential operator $\xi^\mu_a \partial_{x_\mu}$ on
matrix $H$\ (\ref{2.25}) yields the equality, whose right-hand
side can be decomposed into the sum of seven terms having the same
structure
\be \label{2.31}
\xi^\mu_a \frac{\partial H}{\partial x_\mu} = \sum^7_{i=1} D_i.
\ee
As each of the terms $D_i$ is handled in the same way, we give the
calculation details for one of them, say, for
\be \label{2.32} D_4 = \exp\{ (-\ln \theta)E \}
\mathop \Pi \limits_{i=1}^3 \Lambda_i (-2 \xi^\mu_a \frac{\partial
\theta_1}{\partial x_\mu} H_1) \mathop \Pi \limits_{j=4}^6
\Lambda_j.
\ee

Note that in (\ref{2.32}) we use the following designations:
\begin{eqnarray} \label{2.33}
&&\Lambda_1=\exp(\theta_0 S_{03}),\quad \Lambda_2 = \exp(-\theta_3
S_{12}), \nonumber \\ &&\Lambda_3=\exp(-2 \theta_1 H_1), \quad
\Lambda_4 = \exp(-2 \theta_3 H_2), \\ &&\Lambda_5=\exp(-2 \theta_4
\tilde H_1),\quad \Lambda_6 = \exp(-2 \theta_5 \tilde H_2).
\nonumber
\end{eqnarray}
Having multiplied the right-hand side of (\ref{2.32}) by the
matrix $H H^{-1}$ on the left, we arrive at the equality
\be
\label{2.34} D_4 = H\Bigl (-2 \xi^\mu_a \frac{\partial
\theta_1}{\partial x_\mu} \Bigr ) \Lambda^{-1}_6 \Lambda^{-1}_5
\Lambda^{-1}_4 H_1 \Lambda_4 \Lambda_5 \Lambda_6, \ee where the
matrices $\Lambda_4, \Lambda_5, \Lambda_6$ are given in
(\ref{2.33}).

To simplify the right-hand side of (\ref{2.34}) we exploit the
Campbell-Haus\-dorff formula
\begin{eqnarray*}
&&  \exp(\tau A) B \exp(-\tau A) = \sum^\infty_{n=0}
\frac{\tau^n}{n!} \{ A ,  B\}^n, \\ && \{ A,B\}^n = [A, \{ A,
B\}^{n-1}], \ \ \{ A, B\}^0=B,
\end{eqnarray*}
that holds for arbitrary square matrices $A, B$.

With account of commutation relations (\ref{2.23a}), (\ref{2.24})
we get
$$\Lambda^{-1}_4 H_1 \Lambda_1 = \exp(2 \theta_2 H_2) H
\exp(-2 \theta_2 H_2) = H_1, $$
whence
\begin{eqnarray*}
&& \Lambda^{-1}_5 \Lambda^{-1}_4 H \Lambda_4 \Lambda_5 =
\Lambda^{-1}_5 H_1 \Lambda_5 = \exp(2 \theta_4 \tilde H_1) H_1
\exp(-2 \theta_4 \tilde H_1) \\ && \quad = H_1 +4 \theta_4 S_{03}
-4 \theta^2_4 \tilde H_1.
\end{eqnarray*}
Consequently,
\begin{eqnarray*}
&& \Lambda^{-1}_6 \Lambda^{-1}_5 \Lambda^{-1}_4 H_1 \Lambda_4
\Lambda_5 \Lambda_6 =\exp (2 \theta_5 \tilde H_2) (H_1 +4 \theta_4
S_{03} -4 \theta^2_2 \tilde H_1) \\ && \quad \times  \exp(-2
\theta_5 \tilde H_2) =H_1 +4 \theta_4 S_{03} +4 \theta_5 S_{12}
+4(\theta^2_5 -\theta^2_4) \tilde H_1 -8 \theta_4 \theta_5 \tilde
H_2.
\end{eqnarray*}
Finally, we have $$D_4 = H(-2 \xi^\mu_a \frac{\partial
\theta_2}{\partial x_\mu}) [H_1 +4 \theta_4 S_{03}+4 \theta_5
S_{12} +4(\theta^2_5 -\theta^2_4) \tilde H_1 -8 \theta_4 \theta_5
\tilde H_2].$$

The same reasonings, when applied to the remaining terms of the
right-hand side of the equality (\ref{2.31}), complete the proof
of the lemma. \vspace{2mm}

\begin{tver} \label{tt1}
System (\ref{2.29}) is equivalent to the system of partial
differential equations for the functions $\theta, \theta_0,
\theta_m $,\ $ (m=1,2,\ldots, 5)$
\begin{eqnarray} \label{2.35}
\xi^\mu_a \frac{\partial \theta}{\partial x_\mu} &=& f^a \theta,
\nonumber\\ \xi^\mu_a \frac{\partial \theta_0}{\partial x_\mu}
&=& 4(\theta_4 f^a_1 +\theta_5 f^a_2) -f^a_0, \nonumber\\
\xi^\mu_a \frac{\partial \theta_1}{\partial x_\mu} &=& 4(\theta_1
\theta_4+\theta_2 \theta_5 )f^a_1 \nonumber\\ && +4(\theta_1
\theta_5-\theta_2 \theta_4) f^a_2-\theta_1f^a_0-\theta_2 f^a_3
+\frac{1}{2} f^a_1, \nonumber \\  \xi^\mu_a \frac{\partial
\theta_2}{\partial x_\mu} &=& 4(\theta_2 \theta_4-\theta_1 \theta_5
)f^a_1 \nonumber\\ && +4(\theta_2 \theta_5+\theta_1 \theta_4)
f^a_2-\theta_2 f^a_0+\theta_1 f^a_3 +\frac{1}{2} f^a_2, \nonumber
\\ \xi^\mu_a \frac{\partial \theta_3}{\partial x_\mu} &=&
4(\theta_4 f^a_2 -\theta_5 f^a_1)+f^a_3, \\ \xi^\mu_a
\frac{\partial \theta_4}{\partial x_\mu} &=& \theta_4 f^a_0
-2(\theta^2_4-\theta^2_5) f^a_1 -4 \theta_4 \theta_5
f^a_2-\theta_5 f^a_3 +\frac{1}{2}f^a_4, \nonumber \\ \xi^\mu_a
\frac{\partial \theta_5}{\partial x_\mu} &=& \theta_5 f^a_0 -4
\theta_4 \theta_5 f^a_1+2(\theta^2_4-\theta^2_5) f^a_2 +\theta_4
f^a_3 +\frac{1}{2}f^a_5. \nonumber
\end{eqnarray}
In (\ref{2.35}) $\mu =0,1,2,3; \ a=1,2,3$. The coefficients of
linear differential operators $\xi^\mu_a \partial_{x_\mu}$ and the
functions $f^a, f^a_0$, $ f^a_m,\ (m=1,2, \ldots, 5)$ are defined
by the coefficients of the basis operators of the subalgebra $L$
of the algebra $c(1,3)$ of the rank 3.
\end{tver}
{\it Proof.} Inserting the expression for $\xi^\mu_a
\frac{\partial H}{\partial x_\mu}$ given in Lemma \ref{l3} into
the left-hand side of (\ref{2.29}) and multiplying the obtained
equation by the inverse of the non-singular matrix $H$ we arrive
at the system of matrix equations, whose left-hand sides are the
linear combinations of the linearly independent matrices $E,
S_{01}, S_{12}, H_a, \tilde H_a,\ (a=1,2)$. Splitting the system
obtained by these matrices, and taking into account the forms of
the matrices $\tilde \Gamma_a$, and performing some simplifications
yield system of equations (\ref{2.35}). The assertion is proved.

Summarizing we conclude that the problem of constructing
conformally-invariant ansat\-zes reduces to finding the
fundamental solution of the system of linear partial differential
equations (\ref{2.28}) and particular solutions of first-order
system of nonlinear partial differential equations (\ref{2.35}).

The next subsections are devoted to constructing the ansatzes
invariant under the subalgebras of the Poincar\'e, extended
Poincar\'e and conformal algebras given in Assertions
\ref{t1}--\ref{t3}. The solution procedure is based on the above
derived identities and, essentially, on Assertion \ref{tt1}.


\subsubsection{$P(1,3)$-invariant ansatzes}

Subalgebras listed in Assertion \ref{t1} give rise to $P(1,3)$-
(Poincar\'e-) invariant ansatzes. Analysis of the structure of
these subalgebras shows that we can put $\theta=1, \theta_4
=\theta_5 =0$ in formula (\ref{2.25}) for the matrix $H$. What is
more, the form of the basis elements of these subalgebras imply
that in formulae (\ref{2.27}), (\ref{2.34})\ $f^a = f^a_4 = f^a_5
=0$, for all the values of $a=1,2,3$. Owing to these facts, system
(\ref{2.35}) for the matrix $H$ takes the form of twelve
first-order partial differential equations for the functions
$\theta_0, \theta_1, \theta_2, \theta_3$
\begin{eqnarray} \label{2.36}
&& \xi^\mu_a \frac{\partial \theta_0}{\partial x_\mu} = -f^a_0,
\quad \xi^\mu_a \frac{\partial \theta_3}{\partial x_\mu} =
f^a_3, \nonumber \\ && \xi^\mu_a \frac{\partial \theta_1}{\partial
x_\mu} = -\theta_1 f^a_0-\theta_2 f^a_3 +\frac{1}{2}f^a_1, \\ &&
\xi^\mu_a \frac{\partial \theta_2}{\partial x_\mu} = -\theta_2
f^a_0+\theta_1 f^a_3 +\frac{1}{2}f^a_2, \nonumber
\end{eqnarray}
where $\mu=0,1,2,3; a=1,2,3$.

We integrate system (\ref{2.28}) for the case of the subalgebra
$L_{22}$ $= \langle G_1$, $G_2$, $J_{03}$ $+ \alpha J_{12} \rangle
$,\ $(\alpha \in R)$ (all other cases are handled in a similar way).

System (\ref{2.28}) for finding the function $\omega = \omega(x_0,
{\bf x})$ reads as
\begin{eqnarray} \label{2.37}
&& G^{(1)}_1 \omega = [(x_0-x_3) \partial_{x_1}
+x_1(\partial_{x_0} +\partial_{x_3})] \omega =0, \nonumber \\ &&
G^{(1)}_2 \omega = [(x_0-x_3) \partial_{x_2} +x_2(\partial_{x_0}
+\partial_{x_3})] \omega =0, \\ && (J^{(1)}_{03} +\alpha
J^{(1)}_{12}) \omega = [x_0 \partial_{x_3} +x_3
\partial_{x_0}+\alpha(x_2 \partial_{x_1}-x_1 \partial_{x_2})]
\omega =0, \ \ \alpha \in R. \nonumber
\end{eqnarray}
On making the change of variables
\begin{eqnarray*}
y_0&=& (x_0+x_3) (x_0-x_3), \quad y_1 = \sqrt{x^2_1+x^2_2},
\\ y_2&=& \arctan \frac{x_2}{x_1}, \quad y_3 = x_0-x_3,
\end{eqnarray*}
reduces system (\ref{2.37}) to the form
\begin{eqnarray*}
&& y_1 \frac{\partial \omega}{\partial y_1} +2 y^2_1
\frac{\partial \omega}{\partial y_0} -\tan y_2 \frac{\partial
\omega}{\partial y_2} =0, \\ && y_1 \frac{\partial
\omega}{\partial y_1} +2 y^2_1 \frac{\partial \omega}{\partial
y_0} -(\tan y_2 )^{-1} \frac{\partial \omega}{\partial y_2} =0, \\
&& y_3 \frac{\partial \omega}{\partial y_3} +\alpha \frac{\partial
\omega}{\partial y_2}=0.
\end{eqnarray*}
The fundamental solution of the above system reads as $\omega =
y_0 - y^2_1.$ Returning back to the initial variables, we get the
fundamental solution of system (\ref{2.37}),\ $\omega = x_\mu
x^\mu$ $= x_0^2 - x_1^2 - x_2^2 - x_3^2$.

Next, taking into account the forms of the basis elements of the
subalgebra $L_{22}$, we get the expressions for the functions
$f^a_\mu,\ (\mu=0,1,2,3; \ a=1,2,3)$
\begin{eqnarray*}
G_1&:& f^1_0=f^1_2=f^1_3=0,\quad f^1_1 =-1; \\ G_2&:&
f^2_0=f^2_1=f^2_3=0,\quad f^2_2 =-1; \\ J_{03} + \alpha J_{12} &:&
f^3_0 =-1,\quad f^3_1 = f^3_2 =0,\quad f^3_3=-\alpha,\quad (\alpha \in R).
\end{eqnarray*}
So that, system (\ref{2.36}) takes the form
\begin{eqnarray} \label{2.38}
&& G^{(1)}_1 \theta_0 = G^{(1)}_1 \theta_2 = G^{(1)}_1 \theta_3
=0,\quad G^{(1)}_1 \theta_1 = -\frac{1}{2}, \nonumber \\ &&
G^{(1)}_2 \theta_0 = G^{(1)}_2 \theta_1 = G^{(1)}_2 \theta_3 =0, \
\ G^{(1)}_2 \theta_2 = -\frac{1}{2}, \nonumber \\ && (J^{(1)}_{03}
+\alpha J^{(1)}_{12}) \theta_0 =1,\quad (J^{(1)}_{03} + \alpha
J^{(1)}_{12}) \theta_3 =-\alpha, \\ && (J^{(1)}_{03} + \alpha
J^{(1)}_{12}) \theta_1 = \theta_1 + \alpha \theta_2, \quad
(J^{(1)}_{03} + \alpha J^{(1)}_{12}) \theta_2 = \theta_2 -
\alpha \theta_1.\nonumber
\end{eqnarray}
As we have already mentioned, to construct the matrix $H$ it
suffices to find particular solutions of system (\ref{2.38}). The
system for determination of the function $\theta_0$ reads as
\begin{eqnarray} \label{2.39}
&& (x_0-x_3) \frac{\partial \theta_0}{\partial x_a} + x_a \Bigl (
\frac{\partial \theta_0}{\partial x_0} + \frac{\partial
\theta_0}{\partial x_3} \Bigr) =0,\ (a=1,2), \\ && x_0
\frac{\partial \theta_0}{\partial x_3} + x_3\frac{\partial
\theta_0}{\partial x_0} + \alpha  \Bigl (x_2 \frac{\partial
\theta_0}{\partial x_1} - x_1\frac{\partial \theta_0}{\partial
x_2} \Bigr) = 1.\nonumber
\end{eqnarray}
We look for its particular solution of the form $\theta_0 =
f(x_0-x_3)$. By the direct check we become convinced of the fact
that this function satisfies the first two equations of system
(\ref{2.39}), and furthermore, the third one reduces to the
ordinary differential equation
$$
-\xi \frac{ d f}{d \xi} =1, \quad \xi = x_0 - x_3,
$$
whose solution reads as $f=-\ln|\xi|$.

Thus we can choose $\theta_0 = -\ln|x_0-x_3|$. The first two
equations for the function $\theta_3$ coincide with those from
(\ref{2.39}) and the third equation $$x_0 \frac{\partial
\theta_3}{\partial x_3} +x_3 \frac{\partial \theta_3}{\partial
x_0}+\alpha\Bigl ( x_2 \frac{\partial \theta_3}{\partial x_1} -x_1
\frac{\partial \theta_3}{\partial x_2} \Bigr ) = -\alpha$$ differs
from the third equation from system (\ref{2.39}) by the constant
$-\alpha$ in the right-hand side. Owing to these remarks, we
easily get the final form of the particular solution of
(\ref{2.39}) $$\theta_3 = \alpha \ln|x_0-x_3|.$$

According to (\ref{2.38}) the system for finding the functions
$\theta_1, \theta_2$ has the form
\begin{eqnarray}\label{2.40}
&& (x_0 -x_3) \frac{\partial \theta_1}{\partial x_1} +x_1 \Bigl
(\frac{\partial \theta_1}{\partial x_0} +\frac{\partial
\theta_1}{\partial x_3} \Bigr ) =-\frac{1}{2}, \nonumber \\ &&
(x_0 -x_3) \frac{\partial \theta_2}{\partial x_1} +x_1 \Bigl
(\frac{\partial \theta_2}{\partial x_0} +\frac{\partial
\theta_2}{\partial x_3} \Bigr ) =0, \nonumber \\ && (x_0 -x_3)
\frac{\partial \theta_1}{\partial x_2} +x_2 \Bigl (\frac{\partial
\theta_1}{\partial x_0} +\frac{\partial \theta_2}{\partial x_3}
\Bigr ) =0,  \\ && (x_0 -x_3) \frac{\partial \theta_2}{\partial
x_2} +x_2 \Bigl (\frac{\partial \theta_2}{\partial x_0}
+\frac{\partial \theta_2}{\partial x_3} \Bigr ) =-\frac{1}{2},
\nonumber \\ && x_0\frac{\partial \theta_1}{\partial x_3} +x_3
\frac{\partial \theta_1}{\partial x_0} -\alpha \Bigl (x_1
\frac{\partial \theta_1}{\partial x_2}-x_2\frac{\partial
\theta_1}{\partial x_1} \Bigr ) =\theta_1 +\alpha \theta_2,
\nonumber \\ && x_0\frac{\partial \theta_2}{\partial x_3} +x_3
\frac{\partial \theta_2}{\partial x_0} -\alpha \Bigl (x_1
\frac{\partial \theta_2}{\partial x_2}-x_2\frac{\partial
\theta_2}{\partial x_1} \Bigr ) =\theta_2-\alpha \theta_1.
\nonumber
\end{eqnarray}

We seek for its solutions of the form
\be \label{2.41} \theta_1 = g(\xi, x_1),\quad \theta_2 = h(\xi,
x_2),\quad \xi= x_0 - x_3.
\ee
Inserting functions (\ref{2.41}) into system (\ref{2.40})
reduces it to the form
\begin{eqnarray*}
&& \xi \frac{\partial g}{\partial x_1} = -\frac{1}{2}, \quad
\xi \frac{\partial h}{\partial x_2} = -\frac{1}{2},\\ && -\xi
\frac{\partial g}{\partial \xi} +\alpha x_2 \frac{\partial
g}{\partial x_1} = g +\alpha h, \\ && -\xi \frac{\partial
h}{\partial \xi} -\alpha x_1 \frac{\partial h}{\partial x_2} =h
-\alpha g.
\end{eqnarray*}
By the direct check we verify that the functions $g = -x_1 (2
\xi)^{-1}, \ h = -x_2 (2 \xi)^{-1}$ satisfy this system. So that
we can choose
$$ \theta_1 = -\frac{1}{2} x_1 (x_0-x_3)^{-1}, \quad
\theta_2 = -\frac{1}{2} x_2(x_0-x_3)^{-1}.
$$

Performing the same calculations for the remaining subalgebras
listed in Assertion \ref{t1}, we arrive at the following
statement.

\bt \label{t4} Each subalgebra $L_j,\ (j=1,2, \ldots, 22)$ from
the list  given in Assertion \ref{t1} yields invariant ansatz
(\ref{2.18}) with
$$ \Lambda^{-1} = H = \exp(\theta_0 S_{03}) \exp(-\theta_3 S_{12})
\exp(-2\theta_1 H_1) \exp(-2 \theta_2 H_2). $$
And what is more, the functions $\theta_\mu = \theta_\mu (x_0,
{\bf x})$, $(\mu =0,1,2,3),\ \omega = \omega(x_0, {\bf x})$ are
given by one of the corresponding formulae below:
\begin{eqnarray*}
L_1&:& \theta_\mu =0,\ (\mu=0,1,2,3), \ \omega = x_3; \\ L_2&:&
\theta_\mu =0 ,\ (\mu=0,1,2,3), \ \omega = x_0; \\ L_3&:&
\theta_\mu =0,\ (\mu=0,1,2,3), \ \omega = \xi; \\ L_4&:& \theta_0
=-\ln |\xi|, \ \ \theta_1=\theta_2=0, \ \ \theta_3 = \alpha \ln
|\xi|, \ \ \omega = \xi \cdot \eta ; \\ L_5&:& \theta_0 =-\ln
|\xi|, \ \ \theta_1=\theta_2=\theta_3 =0, \ \ \omega = x_2; \\
L_6&:& \theta_0 =x_1, \ \ \theta_1=\theta_2=\theta_3 =0, \ \
\omega = x_2; \\ L_7&:& \theta_0 =x_1, \ \
\theta_1=\theta_2=\theta_3 =0, \ \ \omega = x_1+\ln |\xi|; \\
L_8&:& \theta_0 =\alpha \arctan x_1 x^{-1}_2, \ \
\theta_1=\theta_2=0, \\ && \theta_3 =-\arctan x_1 x^{-1}_2, \ \
\omega = x^2_1+x^2_2; \\ L_9&:& \theta_0 = \theta_1=\theta_2=0, \
\ \theta_3 =-x_0, \ \ \omega = x_3; \\ L_{10}&:& \theta_0 =
\theta_1=\theta_2=0, \ \ \theta_3 =-(-1)^i x_3, \ \ \omega = x_0;
\\ L_{11}&:& \theta_0 = \theta_1=\theta_3=0, \ \ \theta_2
=-\frac{(-1)^i}{2}\xi , \ \ \omega = \eta; \\ L_{12}&:& \theta_0 =
\theta_2=\theta_3=0, \ \ \theta_1 =-\frac{1}{2}(x_1-\alpha
x_2)\xi^{-1} , \ \ \omega = \xi; \\ L_{13}&:& \theta_0 =
\theta_2=\theta_3=0, \ \ \theta_1 =-\frac{(-1)^i}{2}x_2 , \ \
\omega = \xi; \\ L_{14}&:& \theta_0 = \theta_2=\theta_3=0, \ \
\theta_1 =-\frac{1}{4}\xi , \ \ \omega = \xi^2-4 x_1; \\ L_{15}&:&
\theta_0 = \theta_2=\theta_3=0, \ \ \theta_1 =-\frac{1}{4}\xi , \
\ \omega = \alpha \xi^2-4(\alpha  x_1-x_2); \\ L_{16}&:& \theta_0
= -\ln|\xi|, \ \theta_1 = \theta_2 =0, \ \ \theta_3 = -\arctan x_1
x^{-1}_2, \ \ \omega =x^2_1 +x^2_2; \\ L_{17}&:& \theta_0 =
\theta_3=0, \ \ \theta_1 =-\frac{1}{2}[(-1)^i x_2 +(\alpha +\xi)
x_1] [1+(\alpha+\xi) \xi]^{-1}, \\ && \theta_2 = \frac{1}{2}
[(-1)^i x_1 -x_2 \xi][1+(\alpha +\xi) \xi]^{-1},   \ \ \omega =
\xi; \\ L_{18}&:& \theta_0 = -\ln|\xi|, \ \theta_1 = -\frac{1}{2}
x_1 \xi^{-1}, \ \ \theta_2 = \theta_3 = 0, \ \ \omega =\xi\eta
-x^2_1; \\ L_{19}&:& \theta_0 = -\ln|\xi|, \ \theta_1 =
-\frac{1}{2} x_1 \xi^{-1}, \ \ \theta_2 = \theta_3 = 0, \ \ \omega
=x_2; \\ L_{20}&:& \theta_0 = -\ln|\xi|, \ \theta_1 = -\frac{1}{2}
x_1 \xi^{-1}, \ \ \theta_2 = \theta_3 = 0, \ \ \omega =\ln
|\xi|+x_2; \\ L_{21}&:& \theta_0 = -\ln|\xi|, \ \theta_1 =
-\frac{1}{2} (x_1+\ln | \xi|) \xi^{-1}, \\ && \theta_2 = \theta_3
= 0, \ \ \omega =\alpha \ln |\xi| +x_2; \\ L_{22}&:& \theta_0 =
-\ln|\xi|, \ \theta_1 = -\frac{1}{2} x_1 \xi^{-1}, \ \ \theta_2 =
-\frac{1}{2} x_2\xi^{-1}, \theta_3 = \alpha \ln |\xi|,  \\ &&
\omega =x_\mu x^\mu \ (\mu=0,1,2,3).
\end{eqnarray*}
Here $i=1,2; \ \alpha \in R;\ \xi = x_0-x_3,\ \eta = x_0 + x_3$.
\et

\subsubsection{$\tilde P(1,3)$-invariant ansatzes}

Generically, the list of $\tilde P(1,3)$-invariant ansatzes is
exhausted by $P(1,3)$-invariant ansatzes given in Assertion
\ref{t4} and by ansatzes invariant with respect to the subalgebras
$F_j,\ (j=1,2, \ldots, 24)$ listed in Assertion \ref{t2}. By this
reason, to construct all inequivalent $\tilde P(1,3)$-invariant
ansatzes, it suffices to consider the cases of the subalgebras $F_j,\
(j=1,2, \ldots 24)$ only.

A preliminary analysis of these algebras shows that for the
algebras $F_j$ with $j$ taking the values $2,3,4,12,13, \ldots,
19$ we can choose $\theta_1 = \theta_2 = \theta_4 = \theta_5 =0$
in (\ref{2.25}) and, in addition, we can put $f^a_1 = f^a_2 =
f^a_4 = f^a_5=0$,\ $(a=1,2,3)$ in (\ref{2.35}). As a consequence,
system (\ref{2.35}) for the subalgebras in question reads as
$$
\xi^\mu_a \frac{\partial \theta}{\partial x_\mu} = f^a \theta,
\quad \xi^\mu_a \frac{\partial \theta_0}{\partial x_\mu} = -f^a_0,
\quad \xi^\mu_a \frac{\partial \theta_3}{\partial x_\mu} = f^a_3,
$$
where $\mu=0,1,2,3;\ a=1,2,3$.

For the remaining subalgebras from the list given in Assertion
\ref{t2} the following equalities hold, $\theta_b =0, f^a_b=0$,\
$(b=2,3,4,5; \ a=1,2,3)$, and system (\ref{2.35}) takes the form
$$
\xi^\mu_a \frac{\partial \theta}{\partial x_\mu} = f^a \theta,
\quad \xi^\mu_a \frac{\partial \theta_0}{\partial x_\mu} = -f^a_0,
\quad \xi^\mu_a \frac{\partial \theta_1}{\partial x_\mu} =
-\theta_1 f^a_0+\frac{1}{2}f^a_1,
$$
where $\mu=0,1,2,3; \ a=1,2,3$.

Summing up, we conclude that the problem of construction of
$\tilde P(1,3)$-invariant ansatzes reduces to finding solutions of
linear systems of first-order partial differential equations, that
are integrated by rather standard methods of the general theory of
partial differential equations.

We omit the cumbersome intermediate calculations, which are very
much the same as those performed in the previous subsection, and
give the final result.

\bt \label{t5} Each subalgebra $F_j,\ (j=1,2, \ldots, 24)$ from
the list given in Assertion \ref{t2} yields invariant ansatz
(\ref{2.18}) with $$\Lambda^{-1} = H = \exp\{(-\ln \theta) E
\}\exp (\theta_0 S_{03}) \exp(-\theta_3 S_{12}) \exp(-2\theta_1
H_1), \ \ \theta_1\theta_3 =0.$$ And what is more, the functions
$\theta = \theta(x_0, {\bf x})$, $\theta_0 = \theta_0(x_0, {\bf
x})$, $ \theta_1 = \theta_1(x_1, {\bf x})$, $\theta_3 =
\theta_3(x_3, {\bf x})$, $\omega = \omega (x_0, {\bf x})$ are
given by one of the corresponding formulae below
\begin{eqnarray*}
F_1&:& \theta = |x_1|^{-k}, \ \ \theta_0 = \theta_1 = \theta_3 =0,
\ \ \omega = x_2 x^{-1}_1; \\ F_2&:& \theta =
(x^2_1+x^2_2)^{-\frac{k}{2}}, \ \ \theta_0 = \theta_1 = 0, \ \
\theta_3 = \arctan x_2 x^{-1}_1, \\ &&  \omega= \ln (x^2_1 +x^2_2)
+2 \alpha \arctan x_2 x^{-1}_1, \ \alpha >0; \\ F_3&:& \theta =
|x_3|^{-k},  \theta_0 = \theta_1 = 0,  \theta_3 = \arctan x_2
x^{-1}_1,  \omega = (x^2_1+x^2_2) x^{-2}_3; \\ F_4&:& \theta =
|x_0|^{-k},  \theta_0 = \theta_1 = 0,  \theta_3 = \arctan x_2
x^{-1}_1,  \omega = (x^2_1+x^2_2) x^{-2}_0; \\ F_5&:& \theta =
|x_1|^{-k}, \ \ \theta_0 = \alpha^{-1} \ln |x_1|, \ \ \theta_1
=\theta_3 = 0, \ \ \omega = x_2 x^{-1}_1, \ \ \alpha>0; \\ F_6&:&
\theta = |\xi \eta |^{-\frac{k}{2}}, \ \ \theta_0 =
\frac{\textstyle{1}}{\textstyle{2}}\ln |\eta \xi^{-1}|, \ \
\theta_1 =\theta_3 = 0, \\ && \omega = (1-\alpha) \ln |\eta|
+(1+\alpha) \ln|\xi|, \ \ \alpha>0; \\ F_7&:& \theta = |x_2|^{-k},
\theta_0 = \alpha^{-1} \ln |x_2|, \theta_1 =\theta_3 = 0,  \omega
= |\xi|^{\alpha} |x_2|^{1-\alpha}, \alpha>0; \\ F_8&:& \theta = |
\eta |^{-\frac{k}{2}}, \ \ \theta_0 =
\frac{\textstyle{1}}{\textstyle{2}}\ln |\eta |, \ \ \theta_1
=\theta_3 = 0, \\ && \omega = \xi -(-1)^j \ln|\eta|, \ \ j=1,2; \\
F_9&:& \theta = |x_2|^{-k}, \ \ \theta_0 = \ln |x_2|, \ \ \theta_1
=\theta_3 = 0, \\ && \omega = \xi- 2(-1)^j \ln |x_2|,   \ \ j=1,2;
\\ F_{10}&:& \theta = |x_2|^{-k}, \ \ \theta_0 = \ln |\eta
x^{-1}_2|, \ \ \theta_1 =\theta_3 = 0, \ \ \omega = \xi \eta
x^{-2}_2; \\ F_{11}&:& \theta = |x_2|^{-1}, \ \ \theta_0 = -\ln
|\xi x^{-1}_2|, \ \ \theta_1 =\theta_3 = 0, \ \ \omega =  x_2
x^{-1}_1; \\ F_{12}&:& \theta = (x^2_1+x^2_2)^{-\frac{k}{2}}, \ \
\theta_0 = -\alpha \arctan x_2  x^{-1}_1, \ \ \theta_1 = 0, \\ &&
\theta_3 =\arctan x_2 x^{-1}_1, \ \ \omega = \ln (x^2_1 +x^2_2)+
2\beta \arctan x_2 x^{-1}_1 , \\ && \alpha \not =0, \ \ \beta >0;
\\ F_{13}&:& \theta = |\xi \eta |^{-\frac{k}{2}}, \ \ \theta_0 =
-\frac{\textstyle{1}}{\textstyle{2}}\ln |\eta \xi^{-1}|, \ \
\theta_1 = 0, \\ && \theta_3
=-\frac{\textstyle{1}}{\textstyle{2\alpha}} \ln |\eta \xi^{-1}|, \
\ \omega = (\alpha -\beta) \ln |\eta| +(\alpha+\beta) \ln |\xi|,
\\ &&  \alpha \not =0, \ \ \beta >0;\\ F_{14}&:& \theta = | \eta
|^{-\frac{k}{2}}, \ \ \theta_0 =
\frac{\textstyle{1}}{\textstyle{2}}\ln |\eta |, \ \ \theta_1 = 0,
\ \ \theta_3 =-\frac{\textstyle{1}}{\textstyle{2}} \ln |\eta |, \
\ \omega = \xi - \ln |\eta|;\\ F_{15}&:& \theta =
(x^2_1+x^2_2)^{-\frac{k}{2}}, \ \ \theta_0 = -\alpha \arctan x_2
x^{-1}_1, \ \ \theta_1 = 0, \\ && \theta_3 =\arctan x_2 x^{-1}_1,
\ \ \omega = \ln (x^2_1 +x^2_2)\xi^{-2} + 2\alpha \arctan x_2
x^{-1}_1, \ \alpha \not =0; \\ F_{16}&:& \theta =
(x^2_1+x^2_2)^{-\frac{k}{2}}, \ \ \theta_0 =
\frac{\textstyle{1}}{\textstyle{2}} \ln (x^2_1 +x^2_2) \xi^{-2}, \
\ \theta_1 =0, \\ && \theta_3 =\arctan x_2 x^{-1}_1, \ \ \omega =
\ln (x^2_1 +x^2_2)^{1-\alpha} \xi^{2\alpha } + 2\beta \arctan x_2
x^{-1}_1, \\ &&  0 \leq |\alpha| \leq 1, \beta \geq  0,
|\alpha|+|\beta|  \not =0; \\ F_{17}&:& \theta =
(x^2_1+x^2_2)^{-\frac{k}{2}}, \ \ \theta_0 =
\frac{\textstyle{1}}{\textstyle{2}} \ln (x^2_1 +x^2_2), \ \
\theta_1 =0, \ \ \theta_3 =\arctan x_2 x^{-1}_1, \\ &&  \omega =
\xi -(-1)^j \ln (x^2_1 +x^2_2) + 2\alpha \arctan x_2 x^{-1}_1, \ \
\alpha \in R, \ \ j=1,2; \\ F_{18}&:& \theta =
(x^2_1+x^2_2)^{-\frac{k}{2}}, \ \ \theta_0 =
\frac{\textstyle{1}}{\textstyle{2}} \ln (x^2_1 +x^2_2), \ \
\theta_1 =0, \\ && \theta_3 =\arctan x_2 x^{-1}_1, \ \ \omega =
\xi +2(-1)^j \arctan x_2 x^{-1}_1, \ \ j=1,2; \\ F_{19}&:& \theta
= (x^2_1+x^2_2)^{-\frac{k}{2}}, \ \ \theta_0 =
-\frac{\textstyle{1}}{\textstyle{2}}\ln |\xi \eta^{-1} |, \ \
\theta_1 = 0, \\ && \theta_3 =\arctan x_2 x^{-1}_1, \ \ \omega =
(x^2_1+x^2_2)(\xi \eta)^{-1};\\ F_{20}&:& \theta = |\xi \eta
-x^2_1|^{-\frac{k}{2}}, \ \ \theta_0 =
\frac{\textstyle{1}}{\textstyle{2\alpha }} \ln |\xi \eta -x^2_1|,
\ \ \theta_1 =-\frac{\textstyle{1}}{\textstyle{2}} x_1 \xi^{-1},
\\ && \theta_3 =0, \ \ \omega = |\xi|^{2 \alpha} |\xi \eta
-x^2_1|^{1-\alpha}, \ 0 \leq |\alpha| \leq 1; \\ F_{21}&:& \theta
= |x_1 -(-1)^j \xi x_2|^{-k}, \ \ \theta_0 =  \ln |x_1 -(-1)^j \xi
x_2|, \ \ \theta_1 =-\frac{\textstyle{(-1)^j}}{\textstyle{2}} x_2,
\\ && \theta_3 =0, \ \ \omega = \xi, \ \ j=1,2; \\ F_{22}&:&
\theta = |\xi |^{-\frac{k}{2}}, \ \ \theta_0 =
-\frac{\textstyle{1}}{\textstyle{2 }} \ln |\xi |, \ \ \theta_1
=-\frac{\textstyle{1}}{\textstyle{2}} x_1 \xi^{-1}, \\ && \theta_3
=0, \ \ \omega = \eta - x^2_1 \xi^{-1}+(-1)^j \ln |\xi|,  \ \
j=1,2; \\ F_{23}&:& \theta = |x_2|^{-k}, \ \ \theta_0 =
\frac{\textstyle{1}}{\textstyle{2}}\ln | x_2|, \ \ \theta_1
=-\frac{\textstyle{(-1)^j}}{\textstyle{4}} \xi^{-1}, \\ &&
\theta_3 =0, \ \ \omega = (\xi^2-4(-1)^j x_1) x^{-1}_2, \ \ j=1,2;
\\ F_{24}&:& \theta = |\xi^2 -4 (-1)^j  x_1|^{-k}, \ \ \theta_0 =
\frac{\textstyle{1}}{\textstyle{2}}\ln |\xi^2 -4 (-1)^j | x_1|, \
\ \theta_1 =-\frac{\textstyle{(-1)^j}}{\textstyle{4}} \xi, \\ &&
\theta_3 =0,  \omega = (\eta -(-1)^j x_1
\xi+\frac{\textstyle{1}}{\textstyle{6}}\xi^3)^2 (\xi^2 -4(-1)^j
x_1)^{-3},  j=1,2.
\end{eqnarray*}
Here $k$ is an arbitrarily fixed constant (the conformal degree of
the algebra $c(1,3)$),\ $\xi = x_0 - x_3,\ \eta = x_0 + x_3$. \et

\subsubsection{$C(1,3)$-invariant ansatzes}

To obtain the full description of conformally-invariant ansatzes
it suffices to consider the subalgebras $C_j,\ (j=1,2, \ldots,
14)$ listed in Assertion \ref{t3}.

The preliminary analysis of these subalgebras shows that we can
put $\theta_4 = \theta_5 = f^a_4 = f^a_5 = 0$,\ $(a=1,2,3)$ for the
subalgebras $C_j,\ (j=1,2, \ldots, 10)$. As a result, system
(\ref{2.35}) corresponding to these subalgebras takes the
following form:
\begin{eqnarray*}
&& \xi^\mu_a \frac{\partial \theta}{\partial x_\mu} = f^a \theta,
\quad \xi^\mu_a \frac{\partial \theta_0}{\partial x_\mu} = -f^a_0,
\quad \xi^\mu_a \frac{\partial \theta_3}{\partial x_\mu} = f^a_3,
\\ && \xi^\mu_a \frac{\partial \theta_1}{\partial x_\mu} =
-\theta_1 f^a_0-\theta_2 f^a_3 + \frac{\textstyle{1}}
{\textstyle{2}} f^a_1, \quad \xi^\mu_a
\frac{\partial \theta_2}{\partial x_\mu} = -\theta_2
f^a_0 + \theta_1 f^a_3 + \frac{\textstyle{1}}{\textstyle{2}}f^a_2,
\end{eqnarray*}
where $a=1,2,3$.

Thus the problem of constructing ansatzes invariant under the
subalgebras $C_j,\ (j=1,2, \ldots, 10)$ is again reduced to
solving linear first-order partial differential equations.
However, for the remaining subalgebras $C_j,\ (j=11,12,13,14)$
system (\ref{2.35}) is not linear. It has been solved for the case
of the spinor field in \cite{m33}. The obtained expressions for
the functions are so cumbersome that they prove to be useless
within the context of symmetry reduction of the
conformally-invariant nonlinear Dirac equation. By this reason, we
do not give here the ansatzes corresponding to the subalgebras
$C_j,\ (j=11,12,13,14)$.

\bt \label{t6} Each subalgebra $C_j,\ (j=1,2, \ldots,10)$ from the
list given in Assertion \ref{t3} yields invariant ansatz
(\ref{2.18}) with
\begin{eqnarray*}
&&\Lambda^{-1} = H = \exp\{(-\ln \theta) E \}\exp (\theta_0
S_{03}) \exp(-\theta_3 S_{12}) \exp(-2 \theta_1 H_1)\\ && \quad
\times \exp(-2 \theta_2 H_2).
\end{eqnarray*}
What is more, the functions $\theta = \theta(x_0, {\bf x}),\
\theta_\mu = \theta_\mu (x_0, {\bf x}),\ (\mu=0,1,2,3),\ \omega =
\omega (x_0, {\bf x})$ are given by one of the corresponding
formulae below.
\begin{eqnarray*}
C_1&:& \theta = (1+\xi^2) ^{-\frac{k}{2}}, \ \ \theta_0 =
-\frac{\textstyle{1}}{\textstyle{2}} \ln (1+\xi^2), \\ && \theta_1
= -\frac{\textstyle{1}}{\textstyle{2}} (x_2+x_1 \xi)(1+\xi^2)
^{-1}, \ \ \theta_2 = \frac{\textstyle{1}}{\textstyle{2}} (x_1-
\xi x_2)(1+\xi^2) ^{-1}, \\ && \theta_3 = -\arctan \xi, \ \ \omega
= (x_1-x_2 \xi) (1+\xi^2) ^{-1}; \\ C_2&:& \theta = (1+\xi^2)
^{-\frac{k}{2}}, \ \ \theta_0 =
-\frac{\textstyle{1}}{\textstyle{2}} \ln (1+\xi^2), \\ && \theta_1
= -\frac{\textstyle{1}}{\textstyle{2}} (x_2+x_1 \xi)(1+\xi^2)
^{-1}, \ \ \theta_2 = \frac{\textstyle{1}}{\textstyle{2}} (x_1-
x_2\xi )(1+\xi^2) ^{-1}, \\ && \theta_3 = -\arctan \xi, \ \ \omega
= (x_2+x_1 \xi) (1+\xi^2) ^{-1}-\arctan \xi; \\ C_3&:& \theta =
(1+\xi^2) ^{-\frac{k}{2}}, \ \ \theta_0 =
-\frac{\textstyle{1}}{\textstyle{2}} \ln (1+\xi^2), \\ && \theta_1
= -\frac{\textstyle{1}}{\textstyle{2}} x_1 \xi(1+\xi^2) ^{-1}, \ \
\theta_2 = -\frac{\textstyle{1}}{\textstyle{2}} x_2 \xi (1+\xi^2)
^{-1}, \\ && \theta_3 = \arctan x_2 x^{-1}_1, \ \ \omega =
(1+\xi^2)(x^2_1 +x^2_2)^{-1} ; \\ C_4&:&\theta = |x_1|^{-k}, \ \ \
\theta_0 = \ln |x_1| -\ln(1+\xi^2), \\ && \theta_1 =
-\frac{\textstyle{1}}{\textstyle{2}} x_1 \xi(1+\xi^2) ^{-1}, \ \
\theta_2 =- \frac{\textstyle{1}}{\textstyle{2}} x_2 \xi (1+\xi^2)
^{-1}, \\ && \theta_3 = 0, \ \ \omega =  x_2 x^{-1}_1; \\ C_5&:&
\theta = ((x^2_1+x^2_2)(1+\xi^2)) ^{-\frac{k}{2}}, \ \ \theta_0 =
\frac{\textstyle{1}}{\textstyle{2}}\ln
(x^2_1+x^2_2)(1+\xi^2)^{-1}, \\ && \theta_1 =
-\frac{\textstyle{1}}{\textstyle{2}} x_1 \xi(1+\xi^2) ^{-1}, \ \
\theta_2 = -\frac{\textstyle{1}}{\textstyle{2}} x_2  \xi(1+\xi^2)
^{-1}, \\ && \theta_3 = \arctan x_2 x^{-1}_1, \ \ \omega = \arctan
x_2 x^{-1}_1+\alpha \arctan  \xi, \ \ \alpha \not =0; \\ C_6&:&
\theta = [(x_1-x_2\xi)^2(1+\xi^2)^{-1}] ^{-\frac{k}{2}}, \ \
\theta_0 = \frac{\textstyle{1}}{\textstyle{2}}\ln
[(x_1-x_2\xi)^2(1+\xi^2)^{-3}], \\ && \theta_1 =
-\frac{\textstyle{1}}{\textstyle{2}} (x_2+x_1\xi)(1+\xi^2)^{-1}, \
\ \theta_2 = \frac{\textstyle{1}}{\textstyle{2}}
(x_1-x_2\xi)(1+\xi^2)^{-1}, \\ && \theta_3 = -\arctan \xi , \ \
\omega = \alpha \arctan  \xi- \ln [(x_1-x_2\xi)(1+\xi^2)^{-1}] , \
\ \alpha \not =0; \\ C_7&:& \theta =
[(x_1-x_2\xi)^2(1+\xi^2)^{-1}] ^{-\frac{k}{2}}, \ \ \theta_0 =
\frac{\textstyle{1}}{\textstyle{2}}\ln
[(x_1-x_2\xi)^2(1+\xi^2)^{-3}], \\ && \theta_1 =
-\frac{\textstyle{1}}{\textstyle{2}} (x_2+x_1\xi)(1+\xi^2)^{-1}, \
\ \theta_2 = \frac{\textstyle{1}}{\textstyle{2}}
(x_1-x_2\xi)(1+\xi^2)^{-1}, \\ && \theta_3 = -\arctan \xi , \\ &&
\omega = [\eta(1+\xi^2)^{2}-2 x_1 (x_2 +x_1
\xi)-\xi(x^2_1\xi^2-x^2_2)][x_1 -\xi x_2]^{-2} -\xi; \\ C_8&:&
\theta = (x^2_1+x^2_2)^{-\frac{k}{2}}, \ \ \theta_0 =
\frac{\textstyle{1}}{\textstyle{2}}\ln
[(x^2_1+x^2_2)(1+\xi^2)^{-2}], \\ && \theta_1 =
-\frac{\textstyle{1}}{\textstyle{2}} x_1 \xi (1+\xi^2)^{-1}, \ \
\theta_2 = -\frac{\textstyle{1}}{\textstyle{2}} x_2  \xi(1+\xi^2)
^{-1}, \\ && \theta_3 = \arctan x_2 x^{-1}_1, \\ && \omega = \ln
(x^2_1+x^2_2)(1+\xi^2)^{-1}+2 \alpha \arctan  x_2 x^{-1}_1-2\beta
\arctan \xi, \\ && \alpha, \beta \in R, \ \ |\alpha|+|\beta| \not
=0; \\ C_9&:& \theta = (x^2_1+x^2_2)^{-\frac{k}{2}}, \ \ \theta_0
= \frac{\textstyle{1}}{\textstyle{2}}\ln (x^2_1+x^2_2)-\ln
(1+\xi^2), \\ && \theta_1 = -\frac{\textstyle{1}}{\textstyle{2}}
x_1 \xi (1+\xi^2)^{-1}, \ \ \theta_2 =
-\frac{\textstyle{1}}{\textstyle{2}} x_2  \xi(1+\xi^2) ^{-1}, \\
&& \theta_3 = \arctan x_2 x^{-1}_1, \ \ \omega = \eta
(1+\xi^2)(x^2_1+x^2_2)^{-1} - \xi; \\ C_{10}&:& \theta =
(x^2_1+x^2_2)^{-\frac{k}{2}}, \ \ \theta_0 =
-\frac{\textstyle{1}}{\textstyle{2}}\ln (x^2_1+x^2_2), \\ &&
\theta_1 = -\frac{\textstyle{1}}{\textstyle{2}} x_1 \eta
(x^2_1+x^2_2)^{-1}, \ \ \theta_2 =
-\frac{\textstyle{1}}{\textstyle{2}} x_2  \eta (x^2_1+x^2_2)^{-1},
\\ && \theta_3 = 0, \ \ \omega = x_2 x^{-1}_1.
\end{eqnarray*}
Here $k$ is an arbitrarily fixed constant (the conformal degree of
the algebra $c(1,3)$),\ $\xi=x_0-x_3, \eta = x_0 +x_3$. \et

\section{Exact solutions of the Yang-Mills equations}
\setcounter{equation}{0}
\setcounter{tver}{0}
\setcounter{lema}{0}

In this section we apply the above described technique in order to
perform in-depth analysis of the problems of symmetry reduction
and construction of exact invariant solutions of the $SU(2)$
Yang-Mills equations in the (1+3) dimensional Minkowski space of
independent variables. Since the general method to be used relies
heavily upon symmetry properties of the equations under study, we
will review briefly the group-theoretical properties of the
$SU(2)$ Yang-Mills equations.

\subsection{Symmetry properties of the Yang-Mills equa\-ti\-ons}

The classical Yang-Mills equations of $SU(2)$ gauge theory in the
Minkowski space-time $R^{1,3}$ form the system of twelve nonlinear
second-order partial differential equations of the form
\begin{eqnarray} \label{3.1} && \partial_\nu \partial^\nu{\bf
A}_\mu -\partial^\mu \partial_\nu {\bf A}_\nu +e[(\partial_\nu
{\bf A}_\nu) \times {\bf A}_\mu -2(\partial_\nu {\bf A}_\mu)
\times {\bf A}_\nu  \\ && \quad + (\partial^\mu {\bf A}_\nu)
\times {\bf A}^\nu] + e^2 {\bf A}_\nu \times ({\bf A}^\nu \times
{\bf A}_\mu)=0. \nonumber \end{eqnarray}
Hereafter in this section, the indices $\mu, \nu, \alpha, \beta,
\gamma, \delta, \sigma$ take the values $0, 1, 2, 3$; $\partial_\mu
= \partial_{x_\mu} = \frac{\partial}{\partial x_\mu}$; rising and
lowering the indices is performed with the use of the metric
tensor $g_{\mu \nu}$ of the Minkowski space and the summation
convention over the repeated indices is used. Furthermore, ${\bf
A}_\mu = {\bf A}_\mu (x_0, {\bf x})= (A^1_\mu (x_0, {\bf x}),
A^2_\mu (x_0, {\bf x}), A^3_\mu (x_0, {\bf x}))^{\bf T}$ is the
vector-potential of the Yang-Mills field (for brevity it is called
in the sequel the Yang-Mills field) and $e$ is the gauge coupling
constant.

The maximal symmetry group admitted by equations (\ref{3.1}) is the
group $C(1,3) \otimes SU(2)$ \cite{m17}, where $C(1,3)$ is the
$15$-parameter conformal group generated by the following vector
fields:
\begin{eqnarray} \label{3.2} P_\mu &=& \partial_{x_\mu}, \nonumber
\\ J_{\mu \nu}&=& x^\mu \partial_{x_\nu} -x^\nu \partial_{x_\mu}
+A^{a\mu} \partial_{A^a_\nu} -\partial A^{a\nu}
\partial_{A^a_\mu}, \nonumber \\ D&=& x_\mu \partial_{x_\mu}
-A^a_\mu \partial_{A^a_\mu}, \\ K_\mu&=& 2 x^\mu D -(x_\nu x^\nu)
\partial_{x_\mu}+2 A^{a\mu} x_\nu \partial_{A^a_\nu} - 2 A^a_\nu
x^\nu \partial_{A^a_\mu} \nonumber \end{eqnarray}
and $SU(2)$ is the infinite-parameter unitary gauge transformation
group having the generator
\be \label{3.3} Q = \left(\varepsilon_{abc} A^b_\mu \omega ^c (x_0,
{\bf x}) + e^{-1} \partial_{x_\mu} \omega^a (x_0, {\bf x})\right)
\partial_{A^a_\mu}. \ee
In formulae (\ref{3.2}), (\ref{3.3}), $\partial_{A^a_\mu} =
\frac{\partial}{\partial A^a_\mu}$,\ $\omega^c (x_0, {\bf x})$
stand for arbitrary real functions,\ $a,b,c =1,2,3$ and
$\varepsilon_{abc}$ is the anti-symmetric third-order tensor with
$\varepsilon_{123} =1$.

It is not difficult to check that vector fields (\ref{3.2}) can be
rewritten in the form (\ref{2.11}) if we put
\begin{eqnarray} \label{3.4}
S_{01} & = &\left(\matrix{0&-I&0&0\cr -I&0&0&0\cr 0&0&0&0\cr
0&0&0&0\cr}\right), \  S_{02} = \left(\matrix{0&0&-I&0\cr
0&0&0&0\cr -I&0&0&0\cr 0&0&0&0\cr}\right), \nonumber \\ [3mm]
 S_{03}& =& \left(\matrix{0&0&0&-I\cr
0&0&0&0\cr 0&0&0&0\cr -I&0&0&0\cr}\right), \ S_{12} =
\left(\matrix{0&0&0&0\cr 0&0&-I&0\cr 0&I&0&0\cr
0&0&0&0\cr}\right), \\ [3mm]
  S_{13} & =& \left(\matrix{0&0&0&0\cr 0&0&0&-I\cr
0&0&0&0\cr 0&I&0&0\cr}\right), \ \ S_{23} =
\left(\matrix{0&0&0&0\cr 0&0&0&0\cr 0&0&0&-I\cr
0&0&I&0\cr}\right),\nonumber
\end{eqnarray}
where $0$ and $I$ are the zero and unit $3 \times3$ matrices,
correspondingly. Next, we choose the matrix $E$ to be the
$12\times 12$ unit matrix and the conformal degree $k$ of the
algebra $c(1,3)$ to be equal to 1.

One of the important applications of the symmetry admitted by the
Yang-Mills equations is a possibility of getting new exact
solutions with the help of the solution generation formulae. This
method is based upon the fact that the symmetry group maps the set
of solutions of an equation admitting this group into itself. We
give the corresponding formulae without proof (see,
\cite{m20,m21,m33} for further details).

\bt \label{t31} Let
\begin{eqnarray*}
\overline{x}_{i}&=& f_{i}({\bf x}, {\bf u}, \tau),\quad i = 1,2,
\ldots, p, \\ \overline{u}_{j}& =& g_{j}({\bf x}, {\bf u}, \tau),\quad
j = 1,2, \ldots, q,
\end{eqnarray*}
where $\tau = (\tau_{1}, \tau_{2},\dots, \tau_{r})$, be the
$r$-parameter invariance group admitted by a system of partial
differential equations and $U_{j}({\bf x}),\ j = 1,2, \ldots, q$
be a particular solution of the latter. Then the $q$-component
function ${\bf u}({\bf x})= (u^1(x),\ldots, u^q(x))$, defined in
implicit way by the formulae
$$
U_{j}({\bf f}({\bf x}, {\bf u}, \tau)) = g_{j}({\bf x}, {\bf u},
\tau)
$$
with ${\bf f} = (f_1,\ldots, f_{p}),\ j = 1,2,\ldots, q$, is
also a solution of the system in question. \et

In order to be able to take advantage of Assertion \ref{t31}, we
need the formulae for the final transformations generated by the
basis operators (\ref{3.2}), (\ref{3.3}) of the Lie algebra of the
group $C(1,3) \otimes SU(2)$. We give these formulae following
\cite{m2,m21}.
\begin{enumerate}
\renewcommand{\labelenumi}{\arabic{enumi})}
\item the translation group (the generator is $ X = \tau_{\mu}
P_{\mu})$ $$ \overline{x}_{\mu} = x_{\mu} + \tau_{\mu},
\overline{A}_{\mu}^{d} = A_{\mu}^{d};$$
\item
the Lorentz group $O(1,3)$
\begin{enumerate}
\renewcommand{\labelenumii}{(\alph{enumii})}
\item the rotation group
(the generator is $ X = \tau J_{ab})$
\begin{eqnarray*}
\overline{x}_{0} &=& x_0,\quad \overline{x}_{c} = x_{c},\quad c
\not= a,\quad c\not=b,\\ \overline{x}_{a} &=& x_{a} \cos{\tau} +
x_{b}\sin{\tau},\\ \overline{x}_{b} &=& x_{b}\cos{\tau} - x_{a}
\sin{\tau},\\ \overline{A}_{0}^{d} &=& A_{0}^{d},\/
\overline{A}_{c}^{d}=A_{c}^{d},\quad c\not=a,\ c\not=b,\\
\overline{A}_{a}^{d}& =& A_{a}^{d}\cos{\tau} +
A_{b}^{d}\sin{\tau},\\ \overline{A}_{b}^{d}& =&
A_{b}^{d}\cos{\tau} - A_{a}^{d}\sin{\tau};
\end{eqnarray*}
\item the Lorentz transformations (the generator is $ X = \tau
J_{0a}$)
\begin{eqnarray*}
\overline{x}_{0} &=& x_{0} \cosh{\tau} + x_{a}\sinh{\tau},\\
\overline{x}_{a} &=& x_{a}\cosh{\tau} + x_{0}\sinh{\tau},\\
\overline{A}_{0}^{d} &=& A_{0}^{d}\cosh{\tau} +
A_{a}^{d}\sinh{\tau},\\ \overline{A}_{a}^{d} &=&
A_{a}^{d}\cosh{\tau} + A_{0}^{d}\sinh{\tau},\\ \overline{x}_b &=
&x_b, \ \ \overline{A}_{b}^{d} = A_{b}^{d},  \ \ b \not =a;
\end{eqnarray*}
\end{enumerate}
\item the scale transformation group (the generator is $X = \tau D$)
$$ \overline{x}_{\mu} = x_{\mu} \, e^{\tau},\ \ \
\overline{A}_{\mu}^{d} = A_{\mu}^{d} e^{-\tau}.$$
\item the group of conformal transformations (the generation is $X
= \tau_{\mu} K^{\mu}$)
\begin{eqnarray*}
\overline{x}_{\mu} &=& ( x_{\mu} -
\tau_{\mu}x_{\nu}x^{\nu})\sigma^{-1}(x_0, {\bf x}),\\
\overline{A}_{\mu}^{d} &=& [g_{\mu\nu}\sigma(x_0, {\bf x}) + 2
(x_{\mu}\tau_{\nu} -x_{\nu}\tau_{\mu} \\ &&+
2\tau_{\alpha}x^{\alpha}\tau_{\mu}x_{\nu} -
x_{\alpha}x^{\alpha}\tau_{\mu}\tau_{\nu} -
\tau_{\alpha}\tau^{\alpha}x_{\mu}x_{\nu}]A^{d\nu}.
\end{eqnarray*}
\item the gauge transformation group (the generator is $X = Q$)
\begin{eqnarray*}
\overline{x}_{\mu} &=& x_{\mu},\\ \overline{A}_{\mu}^{d}& =&
A_{\mu}^{d}\cos{\omega} +
\varepsilon_{dbc}A_{\mu}^{b}n^{c}\sin{\omega} + 2
n^{d}n^{b}A_{\mu}^{b}\sin^{2}\frac{\textstyle{\omega}}{\textstyle{2}}
\\ && + e^{-1}\lbrack
\frac{\textstyle{1}}{\textstyle{2}}n^{d}\partial_{x_\mu}\omega +
\frac{\textstyle{1}}{\textstyle{2}}(\partial_{x_\mu}n^{d})\sin
\omega + \varepsilon_{dbc}(\partial_{x_\mu}n^{b})n^{c} \rbrack.
\end{eqnarray*}
\end{enumerate}
In the above formulae $\sigma(x_0, {\bf x}) = 1 -
\tau_{\alpha}\tau^{\alpha} + (\tau_{\alpha}\tau^{\alpha})
(x_{\beta}x^{\beta}), n^{a} = n^{a}(x_0, {\bf x})$ are the
components of the unit vector given by the relations
$\omega^{a}(x_0, {\bf x}) = \omega(x_0, {\bf x}) n^{a}(x_0, {\bf
x})$ with $a,b,c,d= 1,2,3$.

Using Assertion \ref{t31}, it is not difficult to derive the
formulae for generating solutions of the Yang-Mills equations by
the above enumerated transformation groups. We give these
following \cite{m33}.
\begin{enumerate}
\renewcommand{\labelenumi}{\arabic{enumi})}
\item the translation group
\[
A_{\mu}^{a}(x) = u_{\mu}^{a}(x + \tau);
\]
\item  the Lorentz group
\begin{eqnarray*}
A_{\mu}^{d}(x) &=& a_{\mu} u_{0}^{d}(ax,bx,cx,dx) +
b_{\mu}u_{1}^{d}(ax,bx,cx,dx)\\ && + c_{\mu}u_{2}^{d}(ax,bx,cx,dx)
+ d_{\mu}u_{3}^{d}(ax,bx,cx,dx);
\end{eqnarray*}
\item the scale transformation group
\[
A_{\mu}^{d}(x) = e^{\tau}u_{\mu}^{d}(xe^{\tau});
\]
\item the group of conformal transformations
\begin{eqnarray*}
A_{\mu}^{d}(x) &=& \lbrack{g_{\mu \nu}\sigma^{-1}(x) +
2\sigma^{-2}(x)(x_{\mu}\tau_{\nu} - x_{\nu}\tau_{\mu} +
2\tau_{\alpha}x^{\alpha}\tau_{\mu}x_{\nu}} \\ && -
x_{\alpha}x^{\alpha}\tau_{\mu}\tau_{\nu} - \tau_{\alpha}
\tau^{\alpha}x_{\mu}x_{\nu})\rbrack
u^{d\nu}((x-\tau(x_{\alpha}x^{\alpha}))\sigma^{-1}(x)).
\end{eqnarray*}
\item the gauge transformation group
\begin{eqnarray*}
A_{\mu}^{d}(x) &=& u_{\mu}^{d}\cos \omega +
\varepsilon_{dbc}u_{\mu}^{b} n^{c}\sin \omega + 2n^{d}
n^{b}u_{\mu}^{b}\sin^2\frac{\textstyle{\omega}}{\textstyle{2}} \\
&& + e^{-1} \lbrack \frac{1}{2}n^{d}\partial_{x_\mu}\omega +
\frac{1}{2}(\partial_{x_\mu} n^d)\sin \omega +
\varepsilon_{dbc}(\partial_{x_\mu}n^{b})n^{c}\rbrack.
\end{eqnarray*}
\end{enumerate}
Here $u_{\mu}^{d}(x)$is an arbitrary particular solution of the
Yang-Mills equations;\ $x = (x_0, {\bf x})$;\ $\tau,\tau_{\mu}$
are arbitrary parameters;\ $a_{\mu},b_{\mu}, c_{\mu},d_ {\mu}$ are
arbitrary constants satisfying the relations
\begin{eqnarray}
a_{\mu}a^{\mu} &=& - b_{\mu}b^{\mu} = - c_{\mu}c^{\mu} =
-d_{\mu}d^{\mu} = 1,\nonumber \\ a_{\mu}b^{\mu}&=& a_{\mu}c^{\mu}
= a_{\mu}d^{\mu} = b_{\mu}c^{\mu} = b_{\mu}d^{\mu} =
c_{\mu}d^{\mu} = 0. \label{comm}
\end{eqnarray}
In addition, we use the following notations:
\begin{eqnarray*}
&& x + \tau = \lbrace x_{\mu} + \tau_{\mu}, \mu = 0,1,2,3 \rbrace,
\\ && ax = a_{\mu}x^{\mu},\quad bx = b_{\mu}x^{\mu},\quad cx =
c_{\mu}x^{\mu},\quad dx = d_{\mu}x^{\mu}.
\end{eqnarray*}

Thus, using the solution generation formulae enables extending a
single solution of the Yang-Mills equations to a multi-parameter
family of exact solutions.

Let us also discuss briefly the discrete symmetries of equations
(\ref{3.1}). It is straightforward to check that the Yang-Mills
equations admit the following groups of discrete transformations:
\begin{eqnarray*}
\Psi_1 &:& \overline{x}_\mu = -x_\mu, \ \  \overline{\bf A}_\mu =
-{\bf A}_\mu ;\\ \Psi_2 &:&  \overline{x}_0 = -x_0, \ \
\overline{x}_1=-x_1, \overline{x}_2= x_2, \  \overline{x}_3=x_3,\\
&&  \overline{\bf A}_0= {\bf A}_0, \  \overline{\bf A}_1= -{\bf
A}_1, \ \overline{\bf A}_2= {\bf A}_2, \  \overline{\bf A}_3= {\bf
A}_3 ;\\ \Psi_3 &:&   \overline{x}_0 = -x_0, \ \
\overline{x}_1=x_1, \overline{x}_2= -x_2, \ \overline{x}_3=-x_3,\\
&&  \overline{\bf A}_0= -{\bf A}_0, \ \overline{\bf A}_1= {\bf
A}_1, \ \overline{\bf A}_2= {\bf A}_2, \ \overline{\bf A}_3= -{\bf
A}_3 .
\end{eqnarray*}
Action of these transformation groups on the basis elements
(\ref{3.2}) of the symmetry algebra admitted by equations
(\ref{3.1}) is described in Table 3.1, where $G_m = J_{0m} -J_{m3}$,
$(m=1,2)$, $M = P_0 + P_3$, $T = \frac{\textstyle{1}}
{\textstyle{2}} (P_0-P_3)$.

While classifying the subalgebras of the algebras $p(1,3)$ and
$\tilde p(1,3)$ of the rank 3 we have exploited the discrete
symmetries $\Phi_a$ given in Table 2.1. Comparing Tables 2.1 and
3.1 we see that the actions of the discrete symmetries $\Phi_a$
and $\Psi_a$ on the operators $P_\mu, J_{\mu \nu}, D$ give
identical results, namely,
$$
\Phi_a P_\mu = \Psi_a P_\mu,\quad \Phi_a J_{\mu\nu} = \Psi_a
J_{\mu\nu},\quad \Phi_a D = \Psi_a D
$$
for all $a=1,2,3$. This fact makes it possible to use the discrete
symmetries in order to simplify the forms of the basis operators of
subalgebras of the algebras $p(1,3)$, $\tilde p(1,3)$.
\vspace{2mm}

\centerline{{\bf Table 3.1.}\ Discrete symmetries of equations
(\ref{3.1})}

\begin{center}
\begin{tabular}{|c|c|c|c|} \hline
Generators&\multicolumn{3}{|c|}{ Action of $\Psi_a$ } \\
\cline{2-4} & $\Psi_1$&$\Psi_{2}$&$\Psi_{3}$\\ \hline
$P_{0}$&$-P_{0}$&$P_{0}$&$-P_{0}$\\ \hline
$P_{1}$&$-P_{1}$&$-P_{1}$&$P_{1}$\\ \hline $P_{k}\
(k=2,3)$&$-P_{k}$&$P_{k}$&$-P_{k}$\\ \hline
$J_{03}$&$J_{03}$&$J_{03}$&$J_{03}$\\ \hline
$J_{12}$&$J_{12}$&$-J_{12}$&$-J_{12}$\\ \hline
$G_{1}$&$G_{1}$&$-G_{1}$&$-G_{1}$\\ \hline
$G_{2}$&$G_{2}$&$G_{2}$&$G_{2}$\\ \hline $M$&$-M$&$M$&$-M$\\
\hline $T$&$-T$&$T$&$-T$\\ \hline $D$&$D$&$D$&$D$\\ \hline
$K_0$&$-K_0$&$K_0$&$-K_0$ \\ \hline $K_1$&$-K_1$&$-K_1$&$K_1$ \\
\hline $K_m \ (m=2,3)$&$-K_m$&$K_m$&$-K_m$\\ \hline
\end{tabular}
\end{center}
\vspace{2mm}

\subsection{Ansatzes for the Yang-Mills field}

Conformally-invariant ansatzes for the Yang-Mills field, that
reduce equations (\ref{3.1}) to systems of ordinary differential
equations, can be represented in the linear form
\be \label{3.5}
{\bf A}_\mu (x_0, {\bf x}) = \Lambda_{\mu\nu} {\bf B}_\nu
(\omega),
\ee
where $\Lambda_{\mu\nu} = \Lambda_{\mu\nu}(x_0, {\bf x})$ are some
fixed non-singular $3\times 3$ matrices and ${\bf B}_\nu (\omega)
= (B^1_\nu (\omega)$, $ B^2_\nu (\omega), B^3_\nu (\omega))^T$ are
new unknown vector functions of the new independent variable
$\omega = \omega(x_0, {\bf x})$. In the sequel, we will denote the
$12\times 12$ matrix having the matrix entries $\Lambda_{\mu\nu}$
as $\Lambda$.

Due to the space limitations, we restrict our considerations to
the ansatzes invariant under the subalgebras of the Poincar\'e
algebra. Let us note that the case of the extended Poincar\'e
algebra is handled in \cite{m53}.

The structure of the matrix $\Lambda$ for the case of arbitrary
vector field is described in Assertion \ref{t4}. Adapting the
formula for $\Lambda$ to the case in hand, we have
$$
\Lambda = \exp(2 \theta_1 H_1) \exp(2 \theta_2 H_2) \exp
(-\theta_0 S_{03}) \exp(\theta_3 S_{12}),
$$
where $\theta_\mu = \theta_\mu(x_0, {\bf x})$ are some
real-valued functions, $H_1 = S_{01}-S_{13}, H_2 = S_{02} -S_{23}$
and $S_{\mu \nu}$ are matrices (\ref{3.4}), that realize the
matrix representation of the Lie algebra $o(1,3)$ of the Lorentz
group $O(1,3)$.

Computing the exponents with the help of the Campbell-Hausdorff
formula yields
$$
\Lambda = \left ( \matrix{ [\cosh \theta_0 +\Phi] & -2 [\Psi_1] &
2 [\Psi_2] & [\sinh \theta_0 -\Phi] \cr [-2 \theta_1
e^{-\theta_0}] & [\cos \theta_3] & [-\sin \theta_3] & [2 \theta_1
e^{-\theta_0}] \cr [-2 \theta_2 e^{-\theta_0}] & [\sin  \theta_3]
& [\cos \theta_3] & [2 \theta_2 e^{-\theta_0}] \cr [\sinh \theta_0
+\Phi] & -2 [\Psi_1] & 2 [\Psi_2] & [\cosh \theta_0 +\Phi] \cr}
\right ),
$$
where $\Phi= 2 (\theta^2_1 +\theta^2_2) e^{-\theta_0}, \ \Psi_1 =
\theta_1 \cos \theta_3 +\theta_2 \sin \theta_3, \ \Psi_2 =
\theta_1 \sin \theta_3 -\theta_2 \cos \theta_3$ and the symbol
$[f]$ stands for $f I$, $I$ being the unit $3\times 3$
matrix.

Inserting the obtained expression for the matrix $\Lambda$
into (\ref{3.5}) yields the final form of the Poincar\'e-invariant
ansatz for the Yang-Mills field
\begin{eqnarray} \label{3.6}
{\bf A}_0& = & \cosh \theta_0 {\bf B}_0 +\sinh \theta_0 {\bf B}_3
-2(\theta_1 \cos \theta_3 + \theta_2 \sin \theta_3) {\bf
B}_1\nonumber \\ & & +2(\theta_1 \sin \theta_3 -\theta_2 \cos
\theta_3) {\bf B}_2 +2 (\theta^{2}_{1} +\theta^{2}_{2})
e^{\textstyle{-\theta_0}}({\bf B}_0 -{\bf B}_3), \nonumber \\ {\bf
A}_1& = & \cos \theta_3 {\bf B}_1 -\sin \theta_3 {\bf B}_2
-2\theta_1 e^{\textstyle{-\theta_0}}({\bf B}_0 -{\bf B}_3), \\
{\bf A}_2& = & \sin \theta_3 {\bf B}_1 +\cos \theta_3 {\bf B}_2
-2\theta_2 e^{\textstyle{-\theta_0}}({\bf B}_0 -{\bf B}_3),
\nonumber \\ {\bf A}_3& = & \sinh \theta_0 {\bf B}_0 +\cosh
\theta_0 {\bf B}_3 -2(\theta_1 \cos \theta_3 + \theta_2 \sin
\theta_3) {\bf B}_1 \nonumber \\ & & +2(\theta_1 \sin \theta_3
-\theta_2 \cos \theta_3) {\bf B}_2 +2 (\theta^{2}_{1}
+\theta^{2}_{2}) e^{\textstyle{-\theta_0}}({\bf B}_0 -{\bf B}_3),
\nonumber
\end{eqnarray}
where ${\bf B}_\mu= {\bf B}_\mu(\omega)$ and the forms of the
functions $\theta_\mu, \omega$ are given in Assertion \ref{t4}.

Inserting (\ref{3.6}) into (\ref{3.1}) yields a system of ordinary
differential equations for the functions ${\bf B}_\mu (\omega)$.
If we will succeed in constructing its general or particular
solution, then substituting it into (\ref{3.6}) gives an exact
solution of the Yang-Mills equations (\ref{3.1}). However, the so
constructed solution will have an unpleasant feature of being
asymmetric in the variables $x_\mu$, while equations (\ref{3.1})
are symmetric in these.

To get exact solutions, that are symmetric in all the variables,
we exploit the formulae for generating solutions by Lorentz
transformations (see, the previous subsection) and thus come to
the following general form of the Pojncar\'e-invariant ansatz:
\be\label{3.7}
{\bf A}_\mu (x) =a_{\mu \nu}(x) {\bf
B}^\nu (\omega), \ee
where
\begin{eqnarray}
a_{\mu \nu}(x) &=&  (a_{\mu}a_{\nu} -
d_{\mu}d_{\nu})\cosh\theta_{0} + (d_{\mu}a_{\nu} -
d_{\nu}a_{\mu})\sinh\theta_0 \nonumber\\ && + 2(a_{\mu} +
d_{\mu})[( \theta_{1}\cos\theta_{3} +
\theta_{2}\sin\theta_{3})b_{\nu} + ( \theta_{2}\cos\theta_{3}
-\theta_{1}\sin\theta_{3})c_{\nu}\nonumber \\ && +(\theta_{1}^{2}
+ \theta_{2}^{2})e^{-\theta_{0}}(a_{\nu} + d_{\nu})] +
(b_{\mu}c_{\nu} - b_{\nu}c_{\mu})\sin\theta_{3} \label{3.8} \\ &&
-(c_{\mu}c_{\nu} + b_{\mu}b_{\nu})\cos\theta_{3} -2
e^{-\theta_{0}}(\theta_{1}b_{\mu} + \theta_{2}c_{\mu})(a_{\nu} +
d_{\nu}).\nonumber
\end{eqnarray}
Here $\mu, \nu$ $=$ $0, 1, 2, 3$;\ $x=(x_0, {\bf x})$ and $a_\mu,
b_\mu, c_\mu, d_\mu$ are arbitrary parameters that satisfy
relations (\ref{comm}). Thus, we have represented
Poincar\'e-invariant ansatzes (\ref{3.6}) in the explicitly
covariant form.

Before giving the corresponding forms of the functions
$\theta_\mu, \omega$ for the above ansatz, we remind that using
the discrete symmetries $\Phi_a,\ (a=1,2,3)$ enables simplifying
the forms of the subalgebras of the algebra  $p(1,3)$. Namely, at
the expense of these symmetries we can put $j=2$ in the
subalgebras $L^j_i,\ (i=10, 11, 13, 17)$. Consequently, for the
corresponding ansatzes we have $(-1)^j = 1$. With this remark the
forms of the functions $\theta_\mu, \omega$, defining ansatzes
(\ref{3.7}), (\ref{3.8}) invariant with respect to subalgebras
from Assertion \ref{t1}, read as
\begin{eqnarray}
L_{1} &:& \theta_{\mu} = 0,\quad \omega = dx;\nonumber\\
L_{2} &:& \theta_{\mu} = 0,\quad \omega = ax;\nonumber\\
L_{3} &:& \theta_{\mu} = 0,\quad \omega = kx;\nonumber\\
L_{4} &:& \theta_{0} = -\ln|kx|,\quad \theta_{1} = \theta_{2} = 0,
\quad \theta_{3} = \alpha \ln|kx|,\quad
\omega = (ax)^{2} - (dx)^{2};\nonumber \\
L_{5} &:& \theta_{0} = -\ln|kx|,\quad \theta_{1} = \theta_{2} =
\theta_{3} = 0,\quad \omega = cx;\nonumber \\
L_{6} &:& \theta_{0} = - bx,\quad \theta_{1} = \theta_{2} =
\theta_{3} = 0,\ \omega = cx;\nonumber \\
L_{7} &:& \theta_{0} = - bx,\quad \theta_{1} = \theta_{2} =
\theta_{3} = 0,\quad \omega = bx - \ln|kx|;\nonumber \\
L_{8} &:& \theta_{0} = \alpha \arctan(bx(cx)^{-1}),\quad
\theta_{1} = \theta_{2} =0,\nonumber \\
&& \theta_{3} = - \arctan(bx(cx)^{-1}),\quad
\omega = (bx)^{2} + (cx)^{2};\nonumber \\
L_{9} &:& \theta_{0} = \theta_{1} = \theta_{2} = 0,\quad
\theta_{3} = - ax,\quad \omega = dx;\nonumber \\
L_{10} &:& \theta_{0} = \theta_{1} = \theta_{2} = 0,\quad
\theta_{3} =  dx,\quad \omega = ax;\nonumber \\
L_{11} &:& \theta_{0} = \theta_{1} = \theta_{3} = 0,\quad
\theta_{2} = -\frac{1}{2} kx ,\quad \omega = ax - dx;\nonumber \\
L_{12} &:& \theta_{0} = 0,\quad \theta_{1} =
\frac{1}{2}(bx - \alpha cx ) (kx)^{-1},\quad
\theta_{2} = \theta_{3} = 0,\quad \omega = kx;\nonumber \\
L_{13} &:& \theta_{0} = \theta_{2} = \theta_{3} = 0,
\quad \theta_{1} =  \frac{1}{2}cx,\quad \omega = kx;\nonumber \\
L_{14} &:& \theta_{0} = \theta_{2} = \theta_{3} = 0,\quad
\theta_{1} = - \frac{1}{4} kx,\quad \omega = 4 bx + (kx)^{2};
\nonumber \\
L_{15} &:& \theta_{0} = \theta_{2} = \theta_{3} = 0,
\quad \theta_{1} = - \frac{1}{4} kx,\quad \omega = 4(\alpha bx -
cx) + \alpha (kx)^{2};\nonumber \\
L_{16} &:& \theta_{0} = - \ln|kx|,\quad \theta_{1} =
\theta_{2} = 0,\quad \theta_{3} = - \arctan (bx(cx)^{-1}),
\label{3.9}\\
&& \omega = (bx)^{2} + (cx)^{2};\nonumber \\
L_{17} &:& \theta_{0} = \theta_{3} = 0,\quad
\theta_{1} =  \frac{1}{2}(cx + (\alpha + kx)bx)
(1 + kx (\alpha + kx))^{-1},\nonumber\\
&& \theta_{2} = -\frac{1}{2}(bx - cx \cdot kx)(1 + kx
(\alpha + kx))^{-1},\quad \omega = kx;\nonumber \\
L_{18} &:& \theta_{0} = - \ln|kx|,\quad \theta_{1} =
\frac{1}{2}bx(kx)^{-1},\nonumber \\
&& \theta_{2} = \theta_{3} = 0,\quad \omega = (ax)^{2} -
(bx)^{2} - (dx)^{2};\nonumber \\
L_{19} &:& \theta_{0} = - \ln|kx|,\quad \theta_{1} =
\frac{1}{2}bx (kx)^{-1},\quad \theta_{2} = \theta_{3} = 0,
\quad \omega = cx ;\nonumber \\
L_{20} &:& \theta_{0} = - \ln |kx|,\quad \theta_{1} =
\frac{1}{2}bx (kx)^{-1},\nonumber \\
&&\theta_{2} = \theta_{3} = 0,\quad \omega = \ln |kx|
- cx;\nonumber \\
L_{21} &:& \theta_{0} = - \ln |kx|,\quad
\theta_{1} = \frac{1}{2} (bx - \ln |kx|) (kx)^{-1},\nonumber \\
&& \theta_{2} = \theta_{3} = 0,\quad \omega = \alpha
\ln |kx|- cx;\nonumber \\
L_{22} &:& \theta_{0} = -\ln |kx|,\quad
\theta_{1} = - \frac{1}{2} bx (kx)^{-1},\quad
\theta_{2} = - \frac{1}{2} cx (kx)^{-1},\nonumber \\
&&\theta_{3} = \alpha\ln|kx|,\quad \omega = (ax)^{2} -
(bx)^{2} - (cx)^{2} - (dx)^{2}.\nonumber
\end{eqnarray}
As earlier, we use the short-hand notations for the scalar product
in the Minkowski space:
$$
ax = a_{\mu}x^{\mu}, \ bx =
b_{\mu}x^{\mu},\ cx = c_{\mu}x^{\mu}, \ dx = d_{\mu}x^{\mu},
$$
and what is more, $kx = ax + dx$.


\subsection{Symmetry reduction of the Yang-Mills equations}

Ansatzes (\ref{3.7})--(\ref{3.9}) are given in the explicitly
covariant form. This fact enables us to perform symmetry reduction
of equations (\ref{3.1}) in a unified way. First of all, we give without
derivation three important identities for the tensor $a_{\mu \nu}$
(see, e.g., \cite{m45})
\begin{eqnarray}
&& a^{\gamma}_{\mu} a_{\gamma \nu} =
g_{\mu \nu},\label{3.10} \\
& & a^{\gamma}_{\mu} \frac{\textstyle{\partial a_{\gamma
\nu}}}{\textstyle{\partial x_\delta}} = -(a_\mu d_\nu -a_\nu
d_\mu) \frac{\textstyle{\partial \theta_0}}{\textstyle{\partial
x_\delta}} +(b_\mu c_\nu -c_\mu b_\nu) \frac{\textstyle{\partial
\theta_3}}{\textstyle{\partial x_\delta}} \nonumber \\ & &\quad +2
e^{-\theta_0} [(k_\mu b_\nu-k_\nu b_\mu) \cos \theta_3-
 (k_\mu c_\nu -k_\nu c_\mu) \sin \theta_3] \frac{\textstyle{\partial
\theta_1}}{\textstyle{\partial x_\delta}}\label{3.11}  \\ & &\quad +2
e^{-\theta_0}[(k_\mu b_\nu-k_\nu b_\mu) \sin \theta_3+(k_\mu
c_\nu-k_\nu c_\mu) \cos \theta_3]\frac{\textstyle{\partial
\theta_2}}{\textstyle{\partial x_\delta}}, \nonumber\\
& & a^{\gamma}_{\mu} \square a_{\gamma \nu} = (a_\mu a_\nu -d_\mu
d_\nu) \frac{\textstyle{\partial \theta_0}}{\textstyle{\partial
x_\gamma}}\frac{\textstyle{\partial \theta_0}}{\textstyle{\partial
x^\gamma}} -(a_\mu d_\nu -a_\nu d_\mu) \square \theta_0 \nonumber
\\ & & \quad + 2 e^{-\theta_0} k_\mu b_\nu[(\square \theta_1) \cos
\theta_3 + (\square \theta_2) \sin \theta_3 - 2
\frac{\textstyle{\partial \theta_1}}{\textstyle{\partial
x_\gamma}} \frac{\textstyle{\partial \theta_3}}
{\textstyle{\partial x^\gamma}}\sin \theta_3 \nonumber
\\ & &\quad + 2 \frac{\textstyle{\partial \theta_2}}{\textstyle{\partial
x_\gamma}}\frac{\textstyle{\partial \theta_3}}{\textstyle{\partial
x^\gamma}}\cos \theta_3]+ 2 e^{-\theta_0} k_\mu c_\nu[(\square
\theta_2) \cos \theta_3 -(\square \theta_1) \sin \theta_3
\nonumber  \\ & &\quad -2 \frac{\textstyle{\partial
\theta_1}}{\textstyle{\partial x_\gamma}}
\frac{\textstyle{\partial \theta_3}}{\textstyle{\partial
x^\gamma}}\cos \theta_3-2 \frac{\textstyle{\partial
\theta_2}}{\textstyle{\partial x_\gamma}}\frac{\textstyle{\partial
\theta_3}}{\textstyle{\partial x^\gamma}}\sin \theta_3]  + 4
e^{-2\theta_0} k_\mu k_\nu \label{3.12} \\ && \quad \times (
\frac{\textstyle{\partial \theta_1}} {\textstyle{\partial
x_\gamma}} \frac{\textstyle{\partial
\theta_1}}{\textstyle{\partial x^\gamma}}+
\frac{\textstyle{\partial \theta_2}}{\textstyle{\partial
x_\gamma}}\frac{\textstyle{\partial \theta_2}}{\textstyle{\partial
x^\gamma}})+ (b_\mu b_\nu + c_\mu c_\nu)\frac{\textstyle{\partial
\theta_3}}{\textstyle{\partial x_\gamma}}\frac{\textstyle{\partial
\theta_3}}{\textstyle{\partial x^\gamma}}\nonumber
\\ && \quad + (b_\mu c_\nu -c_\mu b_\nu) \square \theta_3.\nonumber
\end{eqnarray}

Hereafter, we denote the derivatives of the functions in one
variable $\omega$ by the dots over the symbols of the functions,
for example,
$$
\frac{\textstyle{d f}}{\textstyle{d \omega}} =\dot
f,\quad \frac{\textstyle{d^2 f}}{\textstyle{d \omega^2}}=\ddot f.
$$.
\begin{tver} \label{tt31}
Let ansatz (\ref{3.7}) reduce system (\ref{3.1}) to a system of
second-order ordinary differential equations. Then the reduced
system is necessarily of the form
\begin{eqnarray} \label{3.13}
& &  k_{\mu\gamma} \ddot {\bf B}^{\gamma} + l_{\mu\gamma}\dot {\bf
B}^{\gamma} + m_{\mu\gamma}{\bf B}^{\gamma} + e
g_{\mu\nu\gamma}\dot {\bf  B}^{\nu}\times{\bf B}^{\gamma}  \\ & &
\quad + e h_{\mu\nu\gamma} {\bf B}^{\nu}\times{\bf B}^{\gamma} +
e^{2}{\bf B}_{\gamma}\times ({\bf B}^{\gamma}\times{\bf B}_{\mu})
= 0, \nonumber
\end{eqnarray}
its coefficients being given by the relations
\begin{eqnarray} \label{3.14}
& & k_{\mu\gamma} = g_{\mu\gamma} F_{1} - G_{\mu}G_{\gamma},\quad
 l_{\mu\gamma} = g_{\mu\gamma} F_{2} + 2 S_{\mu\gamma} -
 G_{\mu}H_{\gamma} - G_{\mu}{\dot G}_\gamma,\nonumber \\
& & m_{\mu\gamma} = R_{\mu\gamma} - G_{\mu}{\dot H}_{\gamma},\quad
g_{\mu\nu\gamma} = g_{\mu\gamma} G_{\nu} + g_{\nu\gamma} G_{\mu} -
2 g_{\mu\nu} G_{\gamma},\\ & & h_{\mu\nu\gamma} =
\frac{1}{2}(g_{\mu\gamma} H_{\nu} - g_{\mu\nu} H_{\gamma}) -
T_{\mu\nu\gamma}, \nonumber
\end{eqnarray}
where $F_{1}, F_{2}, G_{\mu}, H_\mu, S_{\mu \nu}, R_{\mu \nu},
T_{\mu\nu\gamma}$ are functions of $\omega$ presented below,
\begin{eqnarray} \label{3.15}
& & F_1 = \frac{\textstyle{\partial\omega}}{\textstyle{\partial
x_\mu}} \frac{\textstyle{\partial \omega}}{\textstyle{\partial
x^\mu}}, \quad F_2 = \square \omega, \quad G_\mu = a_{\gamma \mu}
\frac{\textstyle{\partial
\omega}}{\textstyle{\partial x_\gamma}}, \nonumber \\ & & H_\mu
=\frac{\textstyle{\partial a_{\gamma \mu}}}{\textstyle{\partial
x_\gamma}}, \quad S_{\mu \nu}= a^{\gamma}_{\mu}
\frac{\textstyle{\partial a_{\gamma \nu}}}{\textstyle{\partial
x_\delta}}\frac{\textstyle{\partial \omega}}{\textstyle{\partial
x^\delta}}, \quad R_{\mu \nu} = a^{\gamma}_{\mu}\square
a_{\gamma \nu}, \\ & & T_{\mu \nu \gamma} =
a^{\delta}_{\mu} \frac{\textstyle{\partial a_{\delta
\nu}}}{\textstyle{\partial x_\sigma}} a_{\sigma \gamma}
+a^{\delta}_{\nu} \frac{\textstyle{\partial a_{\delta
\gamma}}}{\textstyle{\partial x_\sigma}} a_{\sigma \mu}
+a^{\delta}_{\gamma} \frac{\textstyle{\partial a_{\delta
\mu}}}{\textstyle{\partial x_\sigma}} a_{\sigma \nu}. \nonumber
\end{eqnarray}
\end{tver}
{\bf Proof.} Inserting ansatz (\ref{3.7}) into equation
(\ref{3.1}) and performing some simplifications yield the
following identities:
\begin{eqnarray}
& & \square {\bf A}_\mu -\partial^\mu
(\partial_\nu {\bf A}_\nu) =\Bigl (\square
a_{\mu \gamma} -\frac{\textstyle{ \partial^2 a_{\nu
\gamma}}}{\textstyle{\partial x^\mu
\partial x_\nu}} \Bigr ){\bf B}^\gamma \nonumber \\ & & \quad +(2
\frac{\textstyle{\partial a_{\mu \gamma}}}{\textstyle{ \partial
x_\nu}} \frac{\textstyle{\partial \omega}}{\textstyle{\partial
x^\nu}} + a_{\mu \gamma} \square \omega -
\frac{\textstyle{\partial a_{\nu
\gamma}}}{\textstyle{\partial x_\nu}}
\frac{\textstyle{\partial \omega}}{\textstyle{\partial x^\mu}} -
 \frac{\textstyle{\partial a_{\nu \gamma}}}{\textstyle{\partial x^\mu}}
\frac{\textstyle{\partial \omega}}{\textstyle{\partial x_\nu}}
 \label{3.16} \\ & & \quad - a_{\nu \gamma} \frac{\textstyle{\partial^2
\omega}}{\textstyle {\partial x_\nu \partial x^\mu}}) \dot  {\bf
B}^\gamma +\Bigl (a_{\mu \gamma} \frac{\textstyle{\partial
\omega}}{\textstyle{\partial x_\nu}} \frac{\textstyle{\partial
\omega}}{\textstyle{\partial x^\nu}}- a_{\nu \gamma}
\frac{\textstyle{\partial \omega}}{\textstyle{\partial x_\nu}}
\frac{\textstyle{\partial \omega}}{\textstyle{\partial
x^\mu}}\Bigr )\ddot{\bf B}^\gamma;\nonumber \\
& & (\partial_\nu {\bf A}_\nu) \times {\bf A}_\mu - 2 (\partial_\nu
{\bf A}_\mu) \times {\bf A}_\nu +(\partial^\mu {\bf A}_\nu) \times
{\bf A}^\nu \nonumber \\ & & \quad = (a_{\mu \gamma}
\frac{\textstyle{\partial a_{\nu \alpha}}}{\textstyle{\partial
x_\nu}} -2 \frac{\textstyle{\partial a_{\mu
\alpha}}}{\textstyle{\partial x_\nu}} a_{\nu \gamma}
+\frac{\textstyle{\partial a_{\nu \alpha}}}{\textstyle{\partial
x^\mu}} a^{\nu}_{\gamma}) {\bf B}^\alpha \times {\bf B}^\gamma
\label{3.17} \\ & & \quad
+ (a_{\mu \gamma} a_{\nu \alpha} \frac{\textstyle{\partial
\omega}}{\textstyle{\partial x_\nu}} - 2 a_{\mu \alpha} a_{\nu \gamma}
\frac{\textstyle{\partial \omega}}{\textstyle{\partial x_\nu}}+
a_{\nu \alpha} a^{\nu}_{\gamma} \frac{\textstyle{\partial
\omega}}{\textstyle{\partial x^\mu}}) \dot{\bf B}^\alpha \times
{\bf B}^\gamma; \nonumber \\
&& {\bf A}_\nu \times ({\bf A}^\nu \times {\bf A}_\mu) =
a_{\nu \beta} a^{\nu}_{\alpha} a_{\mu \gamma} {\bf B}^\beta\times
({\bf B}^\alpha \times {\bf B}^\gamma).  \label{3.18}
\end{eqnarray}
Here $\alpha, \beta = 0, 1, 2, 3$.

Convoluting the left- and right-hand sides of the obtained
expressions with\ $a^{\mu}_{\delta}$ and taking into account
(\ref{3.10}) yield
\begin{eqnarray*}
& & a^{\mu}_{\delta}{\bf A}_\nu \times({\bf A}^\nu \times {\bf
A}_\mu) = a^{\mu}_{\delta} a_{\nu \beta} a^{\nu}_{\alpha} a_{\mu
\gamma} {\bf B}^\beta \times ({\bf B}^\alpha \times {\bf
B}^\gamma) \\ & & \quad =g_{\beta \alpha} g_{\delta \gamma} {\bf
B}^\beta \times ({\bf B}^\alpha \times {\bf B}^\gamma) = {\bf
B}_\alpha \times ({\bf B}^\alpha \times {\bf B}_\delta).
\end{eqnarray*}
Consequently, convoluting (\ref{3.16}) and (\ref{3.17}) with
$a^{\mu}_{\delta}$ we get the equalities such that their
right-hand sides are linear combinations of\ ${\bf B}^\gamma, \dot
{\bf B}^\gamma, \ddot {\bf B}^{\gamma}$, $ {\bf B}^\alpha \times
{\bf B}^\gamma, \dot{\bf  B}^\alpha \times {\bf B}^\gamma$. And
furthermore, the coefficients of these combinations are the
functions of $\omega$ only. Consider first equality (\ref{3.16}).
The coefficients of ${\bf B}^\gamma, \dot{\bf B}^\gamma, \ddot{\bf
B}^{\gamma}$ read as
\begin{eqnarray}
{\bf B}^\gamma &:& a^{\mu}_{\delta} (\partial_\nu \partial^\nu)
a_{\mu \gamma} -a^{\mu}_{\delta} \frac{\textstyle{\partial^2
a_{\nu \gamma}}}{\textstyle{ \partial x^\mu \partial x_\nu}} =
F_{\delta \gamma}(\omega); \label{3.19}\\
\dot{\bf  B}^\gamma &:& 2 a^{\mu}_{\delta}
\frac{\textstyle{\partial a_{\mu \gamma}}}{\textstyle{\partial
x_\nu}} \frac{\textstyle{\partial \omega}}{\textstyle{\partial
x^\nu}} +g_{\delta \gamma} (\partial_\nu
\partial^\nu )\omega -a^{\mu}_{\delta} \frac{\textstyle{\partial
a_{\nu
\gamma}}}{\textstyle{\partial x_\nu}}\frac{\textstyle{\partial
\omega}}{\textstyle{\partial x^\mu}}\nonumber \\ & & \quad
-a^{\mu}_{\delta} \frac{\textstyle{\partial a_{\nu
\gamma}}}{\textstyle{\partial x^\mu}} \frac{\textstyle{\partial
\omega}}{\textstyle{\partial x_\nu}}-a^{\mu}_{\delta} a_{\nu
\gamma} \frac{\textstyle{\partial^2 \omega}}{\textstyle{\partial
x_\nu
\partial x^\mu}} = G_{\delta \gamma} (\omega); \label{3.20}
\\
\ddot{\bf B}^\gamma &:& g_{\delta \gamma}
\frac{\textstyle{\partial \omega}}{\textstyle{\partial x_\nu}}
\frac{\textstyle{\partial \omega}}{\textstyle{\partial
x^\nu}}-a^{\mu}_{\delta} a_{\nu \gamma} \frac{\textstyle{\partial
\omega}}{\textstyle{\partial x_\nu}} \frac{\textstyle{\partial
\omega}}{\textstyle{\partial x^\mu}} = H_{\delta \gamma} (\omega).
\label{3.21}
\end{eqnarray}

Turn now to coefficient (\ref{3.21}). Convoluting the function
$ H_{\delta \gamma}(\omega)$ with the metric tensor $g^{\delta
\gamma}=g_{\delta \gamma}$ yields that
\begin{eqnarray*}
& & g^{\delta \gamma}  H_{\delta \gamma} (\omega)  =
4\frac{\textstyle{\partial \omega}}{\textstyle{\partial x_\nu}}
\frac{\textstyle{\partial \omega}}{\textstyle{\partial x^\nu}}
-a_{\mu \delta}a^{\delta}_{\nu} \frac{\textstyle{\partial
\omega}}{\textstyle{\partial x_\nu}} \frac{\textstyle{\partial
\omega}}{\textstyle{\partial x_\mu}} =4
\frac{\textstyle{\partial \omega}}{\textstyle{\partial x_\nu}}
\frac{\textstyle{\partial \omega}}{\textstyle{\partial
x^\nu}}-g_{\mu \nu} \frac{\textstyle{\partial
\omega}}{\textstyle{\partial x_\nu}} \frac{\textstyle{\partial
\omega}}{\textstyle{\partial x_\mu}}\\
& & \quad = 4 \frac{\textstyle{\partial
\omega}}{\textstyle{\partial x_\nu}} \frac{\textstyle{\partial
\omega}}{\textstyle{\partial x^\nu}}-\frac{\textstyle{\partial
\omega}}{\textstyle{\partial x_\nu}} \frac{\textstyle{\partial
\omega}}{\textstyle{\partial x^\nu}} =3 \frac{\textstyle{\partial
\omega}}{\textstyle{\partial x_\nu}} \frac{\textstyle{\partial
\omega}}{\textstyle{\partial x^\nu}}.
\end{eqnarray*}
Hence we get that\ $\frac{\textstyle{\partial \omega}}
{\textstyle{\partial x_\nu}} \frac{\textstyle{\partial
\omega}}{\textstyle{\partial x^\nu}}$\ is the function of
$\omega$ only
\be \label{3.22}
\frac{\textstyle{\partial \omega}}{\textstyle{\partial x_\nu}}
\frac{\textstyle{\partial \omega}}{\textstyle{\partial x^\nu}}
=F_1(\omega).
\ee
Therefore,
$$a^{\mu}_{\delta} a_{\nu \gamma} \frac{\textstyle{\partial
\omega}}{\textstyle{\partial x_\nu}} \frac{\textstyle{\partial
\omega}}{\textstyle{\partial x^\mu}} = a_{\mu \delta}
\frac{\textstyle{\partial \omega}}{\textstyle{\partial x_\mu}}
a_{\nu \gamma} \frac{\textstyle{\partial
\omega}}{\textstyle{\partial x_\nu}} = {\tilde H}_{\delta \gamma}
(\omega), $$
whence
\be \label{3.23}
a_{\mu \delta} \frac{\textstyle{\partial \omega}}
{\textstyle{\partial x_\mu}} = G_\delta(\omega).
\ee
In view of (\ref{3.22}), (\ref{3.23}) we get the equality
$$
H_{\delta \gamma}(\omega) = g_{\delta \gamma} F_1
-G_\delta G_\gamma.
$$
Thus the function $H_{\delta \gamma}(\omega)$ coincides with
$k_{\mu \gamma}$ from (\ref{3.14}).

Convoluting (\ref{3.20}) with the metric tensor $g^{\delta
\gamma}$ gives
\begin{eqnarray} \label{3.24}
&& g^{\delta \gamma} G_{\delta \gamma}(\omega) = 2a^{\mu}_{\delta}
\frac{\textstyle{\partial a^{\delta}_{\mu}}}{\textstyle{\partial
x_\nu}} \frac{\textstyle{\partial \omega}}{\textstyle{\partial
x^\nu}} +4 (\partial_\nu
\partial^\nu )\omega \nonumber
\\ & & \quad - a_{\mu \delta }\frac{\textstyle{\partial
\omega}}{\textstyle{\partial x_\mu}}
\frac{\textstyle{\partial a^{\delta}_{\nu}}}{\textstyle{\partial
x_\nu}}-a_{\mu \delta } \frac{\textstyle{\partial
a^{\delta}_{\nu}}}{\textstyle{\partial
x_\nu}}\frac{\textstyle{\partial \omega}}{\textstyle{\partial
x_\mu}}-a_{\mu \delta} a^{\delta}_{\nu}
\frac{\textstyle{\partial^2 \omega}}{\textstyle {\partial x_\mu
\partial x_\nu}}.
\end{eqnarray}
Now using (\ref{3.10}) we ensure that the relation
\be \label{3.25}
 a^{\mu}_{\delta} \frac{\textstyle{\partial
 a^{\delta}_{\mu}}}{\textstyle{\partial x_\nu}} = \frac{1}{2}
\frac{\textstyle{\partial}}{\textstyle{\partial
x_\nu}}(a^{\mu}_{\delta} a^{\delta}_{\mu}) = \frac{1}{2}
\frac{\textstyle{\partial}}{\textstyle{\partial
x_\nu}}(g^{\beta}_{\beta})=0,
\ee
as well as the relation
$$\frac{\textstyle{\partial}}{\textstyle{\partial x_\nu}}\Bigl
[a_{\mu \delta} a^{\delta}_{\nu} \frac{\textstyle{\partial  \omega
}}{\textstyle{\partial x_\mu}} \Bigr ] = \frac{\textstyle{\partial
a_{\mu \delta}}}{\textstyle{\partial x_\nu}} a^{\delta}_{\nu}
\frac{\textstyle{\partial \omega }}{\textstyle{\partial x_\mu}}
+a_{\mu \delta} \frac{\textstyle{\partial a^{\delta}_{\nu}
}}{\textstyle{\partial x_\nu}} \frac{\textstyle{\partial \omega
}}{\textstyle{\partial x_\mu}} +a_{\mu \delta} a^{\delta}_{\nu}
\frac{\textstyle{\partial^2 \omega }}{\textstyle{\partial
x_\mu\partial x_\nu}}
$$
hold true. Owing to the fact that
$$
\frac{\textstyle{\partial
a_{\mu \delta}}}{\textstyle{\partial x_\nu}} a^{\delta}_{\nu}
\frac{\textstyle{\partial \omega}}{\textstyle{\partial x_\mu}} =
\frac{\textstyle{\partial a^{\delta}_{\mu}}}{\textstyle{\partial
x_\nu}} a_{\nu \delta} \frac{\textstyle{\partial
\omega}}{\textstyle {\partial x_\mu}},
$$
we make sure that the relation
$$
\frac{\textstyle{\partial a_{\mu \delta}}}{\textstyle{\partial
x_\nu}} a^{\delta}_{\nu} \frac{\textstyle{\partial
\omega}}{\textstyle{\partial x_\mu}} =a_{\mu
\delta}\frac{\textstyle{\partial
a^{\delta}_{\nu}}}{\textstyle{\partial x_\mu}}
\frac{\textstyle{\partial \omega}}{\textstyle{\partial x_\nu}}
$$
is valid, whence
\begin{eqnarray} \label{3.26}
& & a_{\mu \delta}\frac{\textstyle{\partial
\omega}}{\textstyle{\partial x_\mu}} \frac{\textstyle{\partial
a^{\delta}_{\nu}}}{\textstyle{\partial x_\nu}} +
a_{\mu\delta}\frac{\textstyle{\partial
a^{\delta}_{\nu}}}{\textstyle{\partial x_\nu}}
\frac{\textstyle{\partial \omega}}{\textstyle {\partial x_\nu}}=
\frac{\textstyle{\partial}}{\textstyle{\partial x_\nu}}\Bigl
[a_{\mu \delta} a^{\delta}_{\nu} \frac{\textstyle{\partial
\omega}}{\textstyle{\partial x_\mu}}\Bigr ]\\ & & \quad - a_{\mu \delta}
a^{\delta}_{\nu} \frac{\textstyle{\partial^2
\omega}}{\textstyle{\partial x_\mu \partial x_\nu}} =
\frac{\textstyle{\partial}}{\textstyle{\partial x_\nu}}\Bigl
[g_{\mu \nu} \frac{\textstyle{\partial
\omega}}{\textstyle{\partial x_\mu}}\Bigr ]-g_{\mu \nu}
\frac{\textstyle{\partial^2 \omega}}{\textstyle{\partial x_\mu
\partial x_\nu}}=0. \nonumber
\end{eqnarray}
With account of (\ref{3.24}), (\ref{3.25}) and (\ref{3.26}) we
get that
$$
g^{\delta \gamma} G_{\delta \gamma}(\omega) = 4 (\square
\omega -g_{\mu \nu} \frac{\textstyle{\partial^2
\omega}}{\textstyle{\partial x_\mu \partial x_\nu}} = 3
\square \omega.
$$
Consequently, the relation
\be \label{3.27}
\square \omega = F_2(\omega)
\ee
holds.

Next, making sure that the equalities
\begin{eqnarray*}
& & a_{\mu \delta}\frac{\textstyle{\partial
a_{\nu\gamma}}}{\textstyle{\partial
x_\mu}}\frac{\textstyle{\partial \omega}}{\textstyle{\partial
x_\nu}} +a_{\mu\delta}a_{\nu \gamma} \frac{\textstyle{\partial^2
\omega}}{\textstyle{\partial x_\nu \partial x_\mu}}= a_{\mu
\delta} \frac{\textstyle{\partial }}{\textstyle {\partial
x_\mu}}(a_{\nu \gamma} \frac{\textstyle{\partial
\omega}}{\textstyle{\partial x_\nu}} ) \\ & & \quad = a_{\mu \delta}
\frac{\textstyle{\partial }}{\textstyle{\partial x_\nu}}
G_\gamma(\omega)= a_{\mu \delta} \dot{G}_{\gamma}(\omega)
\frac{\textstyle{\partial \omega}}{\textstyle{\partial x^\mu}}
=\dot{G}_\gamma(\omega) G_{\delta} (\omega).
\end{eqnarray*}
hold and taking into account (\ref{3.20}) yield
\be \label{3.28}
2a^{\mu}_{\delta} \frac{\textstyle{\partial a_{\mu
\gamma}}}{\textstyle{\partial x_\nu}} \frac{\textstyle{\partial
\omega}}{\textstyle{\partial x^\nu}} -G_\delta
\frac{\textstyle{\partial a_{\nu \gamma}}}{\textstyle{\partial
x^\nu}} =\tilde G_{\delta \gamma}(\omega).
\ee
Now, convoluting (\ref{3.28}) with $g_{\mu \nu}$, we have
$$g_{\mu \nu} \tilde
G_{\delta \gamma}(\omega) = g_{\mu \nu} G_{\delta} (\omega) \Bigl
(2 \frac{\textstyle{\partial a_{\mu \nu}}}{\textstyle{\partial
x_\mu}} -g_{\mu \nu} \frac{\textstyle{\partial a_{\nu
\gamma}}}{\textstyle{\partial \textstyle{\partial x_\nu}}} \Bigr
)$$
or, equivalently,
\begin{equation}
\label{3.29}
\frac{\textstyle{\partial a_{\mu \mu}}}{ \textstyle{\partial
x_\mu}} = H_\nu(\omega),\quad
a^{\mu}_{\delta} \frac{\textstyle{\partial a_{\mu
\gamma}}}{\textstyle{\partial x_\nu}} \frac{\textstyle{\partial
\omega }}{\textstyle{\partial x^\nu}} = S_{\delta \gamma}(\omega).
\end{equation}

With account of (\ref{3.23}), (\ref{3.27}), (\ref{3.29}), we get
that the coefficient of $\dot{\bf B}^\gamma$ in the reduced system
(\ref{3.13}) coincides with $l_{\mu \gamma}$ (\ref{3.14}).

Finally, from the relation
$$ a_{\mu \delta} \frac{\textstyle{\partial^2
a_{\mu \gamma}}}{\textstyle{\partial x_\mu \partial x_\nu}}=
a_{\mu \delta} \frac{\textstyle{\partial }}{\textstyle{\partial
x_\nu}} \Bigl( \frac{\textstyle{\partial a_{\nu
\gamma}}}{\textstyle{\partial x_\nu}}\Bigr) =a_{\mu \delta}
\frac{\textstyle{\partial }}{\textstyle{\partial x_\mu}} (H_\gamma
(\omega)) =G_\delta (\omega) {\dot H}_\gamma( \omega), $$
with account of (\ref{3.19}) it follows that
$$
a^{\mu}_{\delta} \square a_{\mu \gamma} = R_{\delta \gamma}
(\omega).
$$
Consequently, the function in the right-hand side of (\ref{3.19})
coincides with $m_{\mu \gamma}$ from (\ref{3.14}).

Analysis of (\ref{3.17}) is carried out in the same way (we do not
present here the corresponding calculations). The assertion is
proved.

Thanks to the above assertion, the problem of symmetry reduction
of the Yang-Mills equations by the subalgebras of the algebra
$p(1,3)$ reduces to routine substitution of the corresponding
expressions for $a_{\mu \nu}, \omega$ into (\ref{3.15}). We give
below the final forms of the coefficients (\ref{3.14}) of the
reduced system of ordinary differential equations (\ref{3.13}) for
each of the subalgebras of the algebra $p(1,3)$:
\begin{eqnarray} \label{3.31}
L_{1} &:& k_{\mu\gamma} = - g_{\mu\gamma} -
d_{\mu}d_{\gamma},\quad l_{\mu\gamma} = m_{\mu\gamma} =
0,\nonumber \\ & &  g_{\mu\nu\gamma} = g_{\mu\gamma}d_{\nu} +
g_{\nu\gamma}d_{\mu} - 2g_{\mu\nu}d_{\gamma},\quad
h_{\mu\nu\gamma} = 0 ;\nonumber \\ L_{2} &:& k_{\mu\gamma} =
g_{\mu\gamma} - a_{\mu}a_{\gamma},\quad l_{\mu\gamma} =
m_{\mu\gamma} = 0,\nonumber \\ & & g_{\mu\nu\gamma} =
g_{\mu\gamma}a_{\nu} + g_{\nu\gamma}a_{\mu} - 2
g_{\mu\nu}a_{\gamma}, \ \ h_{\mu\nu\gamma} = 0 ;\nonumber \\ L_{3}
&:& k_{\mu\gamma} = k_{\mu}k_{\gamma},\quad l_{\mu\gamma} =
m_{\mu\gamma} = 0,\nonumber \\ & &  g_{\mu\nu\gamma} =
g_{\mu\gamma}k_{\nu} + g_{\nu\gamma}k_{\mu} -
2g_{\mu\nu}k_{\gamma},\quad h_{\mu\nu\gamma} = 0 ;\nonumber \\
L_{4} &:& k_{\mu\gamma} = 4g_{\mu\gamma}\omega  -
a_{\mu}a_{\gamma}(\omega  + 1)^{2} - d_\mu d_\gamma
(\omega-1)^2\nonumber \\ & &\quad -(a_{\mu}d_{\gamma} +
a_{\gamma}d_\mu)(\omega ^{2} - 1),\nonumber \\ & & l_{\mu\gamma} =
4(g_{\mu\gamma} +\alpha (b_{\mu}c_{\gamma} - c_\mu b_{\gamma})) -
2k_\mu (a_{\gamma}-d_\gamma +k_\gamma \omega), \nonumber \\ & &
m_{\mu \gamma} = 0, \nonumber \\ & & g_{\mu\nu\gamma} = \epsilon
(g_{\mu\gamma}(a_{\nu}-d_\nu +k_\nu \omega) +g_{\nu \gamma} (a_\mu
-d_\mu +k_\mu \omega) \nonumber \\ & &\quad -2 g_{\mu \nu}(a_\gamma
-d_\gamma +k_\gamma \omega)), \nonumber \\ & & h_{\mu\nu\gamma} =
\frac{\epsilon}{2}[g_{\mu\gamma}k_{\nu} -
g_{\mu\nu}k_{\gamma}]+\alpha \epsilon [(b_\mu c_\nu-c_\mu b_\nu)
k_\gamma \nonumber \\ & &\quad +(b_\nu c_\gamma -c_\nu b_\gamma) k_\mu
+(b_{\gamma} c_\mu -c_\gamma b_\mu) k_\nu];\nonumber \\ L_{5} &:&
k_{\mu\gamma} = - g_{\mu\gamma} - c_{\mu}c_{\gamma},\quad
l_{\mu\gamma} = -\epsilon c_{\mu}k_{\gamma},\quad m_{\mu\gamma} =
0,\nonumber \\ & & g_{\mu\nu\gamma} = g_{\mu\gamma}c_{\nu} +
g_{\nu\gamma}c_{\mu} - 2g_{\mu\nu}c_{\gamma},\nonumber \\ & &
h_{\mu\nu\gamma} =\frac{\epsilon}{2} (g_{\mu\gamma}k_{\nu} -
g_{\mu\nu}k_{\gamma}) ;\nonumber \\ L_{6} &:& k_{\mu\gamma} = -
g_{\mu\gamma} - c_{\mu}c_{\gamma},\quad l_{\mu\gamma} =
0,\nonumber \\ & & m_{\mu\gamma} =  -(a_{\mu}a_{\gamma} -
d_{\mu}d_{\gamma}),\quad  g_{\mu\nu\gamma} = g_{\mu\gamma}c_{\nu}
+ g_{\nu\gamma}c_{\mu} - 2g_{\mu\nu}c_{\gamma},\nonumber  \\ & &
h_{\mu\nu\gamma} =-\lbrack(a_{\mu}d_{\nu} - a_{\nu}d_{\mu})
b_{\gamma} + (a_{\nu}d_{\gamma} - a_{\gamma}d_{\nu})b_{\mu}
\nonumber \\ & &\quad +(a_{\gamma}d_{\mu}
-a_{\mu}d_{\gamma})b_{\nu}\rbrack ;\nonumber  \\ L_{7} &:&
k_{\mu\gamma} = - g_{\mu\gamma} - (b_\mu -\epsilon k_\mu
e^\omega)(b_{\gamma} -\epsilon k_\gamma e^\omega),\nonumber \\ & &
l_{\mu\gamma} = -2(a_{\mu}d_{\gamma} - a_{\gamma}d_{\mu})+\epsilon
e^\omega (b_\mu -\epsilon k_\mu e^\omega) k_\gamma,\nonumber \\ &
& m_{\mu\gamma} = -(a_\mu a_{\gamma} -
d_{\mu}d_{\gamma}),\nonumber
\\ & & g_{\mu\nu\gamma} = g_{\mu\gamma}(b_{\nu}-\epsilon k_\nu
e^\omega) +g_{\nu \gamma} (b_\mu -\epsilon k_\mu e^\omega)
\nonumber \\
& & \quad -2 g_{\mu \nu}(b_\gamma -\epsilon k_\gamma e^\omega), \quad
h_{\mu\nu\gamma} = -\lbrack(a_{\mu}d_{\nu}
 -a_{\nu}d_{\mu})b_{\gamma} \nonumber \\
& & \quad + (a_{\nu}d_{\gamma} - a_{\gamma}d_{\nu})b_{\mu} +
(a_{\gamma}d_{\mu} - a_{\mu}d_{\gamma})b_{\nu}\rbrack ;\nonumber
\\ L_{8} &:& k_{\mu\gamma} = - 4\omega (g_{\mu\gamma} +
c_{\mu}c_{\gamma}),\quad l_{\mu\gamma} = - 4(g_{\mu\gamma} +
c_{\mu}c_{\gamma}),\nonumber \\ & &  m_{\mu\gamma} =
-\frac{1}{\omega} (\alpha^2(a_{\mu}a_{\gamma}-d_\mu d_\gamma)
+b_\mu b_\gamma), \nonumber \\ & & g_{\mu\nu\gamma} = 2
\sqrt{\omega }( g_{\mu\gamma}c_{\nu} + g_{\nu\gamma}c_{\mu} - 2
g_{\mu\nu}c_{\gamma}),\nonumber \\ & &  h_{\mu\nu\gamma} =
\frac{\textstyle{1}}{\textstyle{2 \sqrt{\omega}}}
(g_{\mu\gamma}c_{\nu} -
g_{\mu\nu}c_{\gamma})+\frac{\textstyle{\alpha}}
{\textstyle{\sqrt{\omega}}} ((a_\mu d_\nu -a_\nu d_\mu) b_\gamma
\nonumber \\ & & \quad +(a_\nu d_\gamma -d_\nu a_\gamma) b_\mu
+(a_\gamma d_\mu -a_\mu d_\gamma) b_\nu);\nonumber \\ L_{9} &:&
k_{\mu\gamma} = - g_{\mu\gamma} -d_{\mu}d_{\gamma},\quad
l_{\mu\gamma} = 0,\nonumber \\ & & m_{\mu\gamma} =
b_{\mu}b_{\gamma} + c_{\mu}c_{\gamma}, \nonumber
\\ & &  g_{\mu\nu\gamma} = g_{\mu\gamma}d_{\nu} +
g_{\nu\gamma}d_{\mu}-2 g_{\mu \nu} d_\gamma,\nonumber \\ & &
h_{\mu\nu\gamma} = a_{\gamma}(b_{\mu}c_{\nu} - c_{\mu}b_{\nu}) +
a_{\mu} (b_{\nu}c_{\gamma} - c_{\nu}b_{\gamma}) \nonumber \\ & &
\quad + a_{\nu}(b_{\gamma}c_{\mu} -c_{\gamma}b_{\mu}) ;\nonumber \\
L_{10} &:& k_{\mu\gamma} = g_{\mu\gamma} -a_{\mu}a_{\gamma}, \quad
l_{\mu\gamma} = 0,\nonumber \\ & & m_{\mu\gamma} = -
(b_{\mu}b_{\gamma} + c_{\mu}c_{\gamma}), \nonumber \\ & &
g_{\mu\nu\gamma} = g_{\mu\gamma}a_{\nu} + g_{\nu\gamma}a_{\mu} - 2
g_{\mu\nu}a_{\gamma},\nonumber \\ & &  h_{\mu\nu\gamma} = -
\lbrack d_{\gamma}(b_{\mu}c_{\nu} - c_{\mu}b_{\nu})
+d_{\mu}(b_{\nu}c_{\gamma} - c_{\nu}b_{\gamma}) \nonumber \\ & &
\quad + d_{\nu}(b_{\gamma}c_{\mu} -c_{\gamma}b_{\mu})\rbrack ; \\ L_{11}
&:& k_{\mu\gamma} = - (a_\mu -d_{\mu})(a_\gamma -d_\gamma),\quad
l_{\mu\gamma} = 2(b_{\mu}c_{\gamma} -c_{\mu}b_{\gamma}),\nonumber
\\ & & m_{\mu\gamma} = 0, \nonumber \\ & & g_{\mu\nu\gamma} =
g_{\mu\gamma}(a_{\nu} -d_\nu)+ g_{\nu\gamma}(a_{\mu}-d_\mu) -
2g_{\mu\nu}(a_{\gamma}-d_\gamma),\nonumber \\ & & h_{\mu\nu\gamma}
= \frac{1}{2}\lbrack(k_{\gamma}(b_\mu c_\nu-c_\mu b_\nu) +k_\mu
(b_\nu c_\gamma -c_\nu b_\gamma) \nonumber \\ & & \quad
+k_\nu(b_{\gamma}c_{\mu} - c_{\gamma}b_{\mu})\rbrack; \nonumber \\
L_{12} &:& k_{\mu\gamma} = - k_{\mu}k_{\gamma},\quad l_{\mu\gamma}
= - \omega^{-1 }k_{\mu}k_{\gamma},\quad m_{\mu\gamma}=
-\alpha^2\omega^{-2} k_\mu k_\gamma,\nonumber \\ & &
g_{\mu\nu\gamma} = g_{\mu\gamma}k_{\nu} + g_{\nu\gamma}k_{\mu} -
2g_{\mu\nu}k_{\gamma},\nonumber \\ & &  h_{\mu\nu\gamma} =
\frac{1}{2}\omega^{-1} (g_{\mu\gamma}k_{\nu} -
g_{\mu\nu}k_{\gamma})+\alpha\omega^{-1}((k_\mu b_\nu-k_\nu b_\mu)
c_\gamma \nonumber \\ & &\quad +(k_\nu b_\gamma -k_\gamma b_\nu) c_\mu
+(k_\gamma b_\mu -k_\mu b_\gamma) c_\nu) ;\nonumber \\ L_{13} &:&
k_{\mu\gamma} = - k_{\mu}k_{\gamma},\quad l_{\mu\gamma} = 0,\quad
m_{\mu\gamma} = - k_{\mu}k_{\gamma},\nonumber \\ & &
g_{\mu\nu\gamma} = g_{\mu\gamma}k_{\nu} + g_{\nu\gamma}k_{\mu} -
2g_{\mu\nu}k_{\gamma},\nonumber \\ & &  h_{\mu\nu\gamma} =-
((k_{\mu}b_\nu -k_\nu b_\mu) c_\gamma + (k_\nu b_{\gamma}-k_\gamma
b_\nu) c_\mu \nonumber \\ & &\quad +(k_\gamma b_\mu -k_\mu b_\gamma)
c_\nu);\nonumber \\ L_{14} &:& k_{\mu\gamma} = - 16(g_{\mu \gamma}
+b_{\mu}b_{\gamma}),\quad l_{\mu \gamma} =m_{\mu\gamma} =
h_{\mu\nu \gamma} = 0,\nonumber \\ & & g_{\mu\nu\gamma} =
4(g_{\mu\gamma}b_{\nu} + g_{\nu\gamma}b_{\mu} -
2g_{\mu\nu}b_{\gamma}), \nonumber \\ L_{15} &:& k_{\mu\gamma} = -
16[(1+\alpha^2)g_{\mu\gamma} + (c_{\mu}-\alpha b_{\mu})(c_\gamma
-\alpha b_\gamma)], \nonumber \\ & &  l_{\mu\gamma} =
m_{\mu\gamma} =h_{\mu\nu\gamma} = 0,\nonumber \\ & &
g_{\mu\nu\gamma} = -4[g_{\mu\gamma}(c_\nu -\alpha b_{\nu}) +
g_{\nu\gamma}(c_\mu -\alpha b_{\mu})\nonumber \\ & & \quad
-2g_{\mu\nu}(c_\gamma -\alpha b_{\gamma})];\nonumber \\ L_{16}
&:& k_{\mu\gamma} = - 4\omega(g_{\mu\gamma}+c_{\mu}c_{\gamma}),\quad
 l_{\mu\gamma} = -4(g_{\mu \gamma} +c_\mu c_\gamma) -2 \epsilon
 k_\gamma
c_\mu \sqrt{\omega}, \nonumber \\ & & m_{\mu\gamma} =-\omega^{-1}
b_\mu b_\gamma,\quad
 g_{\mu \nu\gamma}= 2 \sqrt{\omega}(g_{\mu \gamma} c_\nu +g_{\nu
 \gamma}
c_\mu -2 g_{\mu \nu} c_\gamma), \nonumber \\ & & h_{\mu\nu\gamma}
= \frac{1}{2}[\epsilon(g_{\mu \gamma} k_\nu -g_{\mu \nu} k_\gamma)
+{1\over \sqrt{\omega}}(g_{\mu \gamma} c_\nu -g_{\mu \nu}
c_\gamma)];\nonumber \\ L_{17} &:& k_{\mu\gamma} = - k_\mu
k_\gamma,\ \ l_{\mu\gamma} = - \frac{\textstyle{2 \omega
+\alpha}}{\textstyle{\omega(\omega+\alpha)+1}} k_\mu k_\gamma, \nonumber
\\ & & m_{\mu\gamma} = - 4k_{\mu}k_{\gamma}(1+\omega(\alpha
+\omega))^{-2}, \nonumber \\
& &  g_{\mu\nu\gamma} = g_{\mu\gamma}k_{\nu} +
g_{\nu\gamma}k_{\mu} - 2g_{\mu\nu}k_{\gamma},\nonumber \\ & &
h_{\mu\nu\gamma} = \frac{1}{2}(\alpha +2
\omega)(g_{\mu\gamma}k_{\nu} - g_{\mu\nu}k_{\gamma})(1+\omega(\alpha
+\omega ))^{-1}\nonumber \\ & &\quad-2(1+\omega(\omega+\alpha))^{-1}((k_\mu
b_\nu-k_\nu b_\mu)c_\gamma \nonumber \\ & &\quad +(k_\nu b_\gamma
-k_\gamma b_\nu) c_\mu +(k_\gamma b_\mu -k_\mu
b_\gamma)c_\nu);\nonumber \\ L_{18}&:& k_{\mu \gamma }= 4 \omega
g_{\mu \gamma}-(k_\mu \omega +a_\mu -d_\mu)(k_\gamma \omega
+a_\gamma -d_\gamma), \nonumber \\ & & l_{\mu\gamma} = 6g_{\mu
\gamma}+4(a_\mu d_\gamma -a_\gamma d_\mu) -3 k_\gamma(k_\mu \omega
+a_\mu -d_\mu), \nonumber \\ & & m_{\mu\gamma} = -k_\mu
k_\gamma,\quad g_{\mu\nu\gamma}
=\epsilon(g_{\mu\gamma}(k_{\nu}\omega +a_\nu - d_{\nu})\nonumber
\\ & &\quad + g_{\nu\gamma}( k_{\mu}\omega +a_\mu - d_{\mu}) -
2g_{\mu\nu}(k_{\gamma}\omega +a_\gamma -d_\gamma)), \nonumber \\ &
& h_{\mu\nu\gamma} = \epsilon(g_{\mu\gamma}k_{\nu} -
g_{\mu\nu}k_{\gamma});\nonumber \\ L_{19} &:& k_{\mu\gamma} = -
g_{\mu\gamma} - c_{\mu}c_{\gamma},\quad l_{\mu\gamma} = 2 \epsilon
k_\gamma c_\mu, \quad  m_{\mu\gamma} =
-k_{\mu}k_{\gamma},\nonumber \\ & &  g_{\mu\nu\gamma} =
g_{\mu\gamma}c_{\nu} + g_{\nu\gamma}c_{\mu} -
2g_{\mu\nu}c_{\gamma},\quad
 h_{\mu\nu\gamma} = \epsilon(g_{\mu\gamma}k_{\nu}
- g_{\mu\nu} k_{\gamma} ;\nonumber \\
L_{20} &:& k_{\mu\gamma} = - g_{\mu \gamma} -(c_\mu -\epsilon
k_\mu)(c_\gamma -\epsilon k_\gamma),\quad \nonumber \\ & &
l_{\mu\gamma} = 2\epsilon k_{\gamma} c_\mu -2 k_\mu k_\gamma, \ \
m_{\mu\gamma} = -k_\mu k_\gamma,\nonumber \\ & &  g_{\mu\nu\gamma}
= g_{\mu\gamma}(\epsilon k_{\nu} -c_\nu) + g_{\nu\gamma}(\epsilon
k_{\mu} - c_\mu) -2g_{\mu\nu}(\epsilon
k_{\gamma}-c_\gamma),\nonumber \\ && h_{\mu\nu\gamma} =
\epsilon(g_{\mu\gamma}k_{\nu} - g_{\mu\nu}k_{\gamma}) ;\nonumber
\\ L_{21} &:& k_{\mu\gamma} = - g_{\mu \gamma} -(c_\mu -\alpha
\epsilon k_\mu)(c_\gamma -\alpha \epsilon
k_{\gamma}),\quad\nonumber \\ & & l_{\mu\gamma} = 2(\epsilon
k_{\gamma} c_\mu -\alpha k_\mu k_\gamma), \quad
  m_{\mu\gamma} = - k_{\mu}k_{\gamma},\nonumber \\
& &  g_{\mu\nu\gamma} = -g_{\mu\gamma}(c_\nu -\alpha \epsilon
k_{\nu})- g_{\nu\gamma}(c_\mu -\alpha \epsilon k_{\mu}) \nonumber
\\ & &\quad +2g_{\mu\nu}(c_\gamma -\alpha \epsilon k_{\gamma}),\quad
 h_{\mu\nu\gamma} = \epsilon( g_{\mu\gamma}k_{\nu}-
g_{\mu\nu}k_{\gamma});\nonumber \\ L_{22} &:& k_{\mu\gamma} = - 4
\omega g_{\mu\gamma} -(a_\mu -d_\mu+ k_{\mu}\omega)(a_\gamma
-d_{\gamma}+k_\gamma \omega),\nonumber \\ & & l_{\mu \gamma} =
4[2g_{\mu \gamma} +\alpha (b_\mu c_\gamma -c_\mu b_\gamma) -a_\mu
a_\gamma +d_\mu d_\gamma -\omega k_\mu k_\gamma],\nonumber \\ & &
m_{\mu\gamma} = -2 k_{\mu}k_{\gamma}, \quad g_{\mu\nu\gamma} =
\epsilon(g_{\mu\gamma}(a_\nu -d_\nu +k_\nu \omega) \nonumber \\ &
&\quad + g_{\nu\gamma}(a_\mu -d_\mu + k_\mu \omega) -2g_{\mu\nu}(a_\gamma
-d_\gamma +k_{\gamma}\omega)),\nonumber \\ & & h_{\mu\nu\gamma} =
\frac{3 \epsilon}{2}(g_{\mu\gamma}k_{\nu} -
g_{\mu\nu}k_{\gamma})-\epsilon \alpha [k_\gamma(b_\mu c_\nu -c_\mu
b_\nu)\nonumber \\ &&\quad +(k_{\mu}(b_{\nu}c_\gamma
-c_{\nu}b_{\gamma})+k_\nu(b_\gamma c_\mu -c_\gamma
b_\mu)].\nonumber
\end{eqnarray}
In the above formulae (\ref{3.31}) we have\ $\epsilon =1$ for $kx>0$
\ and\ $\epsilon=-1$ for $kx<0$. Furthermore, $\alpha$ is an arbitrary
parameter.

\subsection{Exact solutions}

Clearly, efficiency of the symmetry reduction procedure is subject
to our ability to integrate the reduced systems of ordinary
differential equations. Since the reduced equations are nonlinear,
it is not at all clear that it will be possible to construct their
particular or general solutions. That it why, we devote the first
part of this subsection to describing our technique for
integrating the reduced systems of nonlinear ordinary differential
equations (the further details can be found in \cite{m33}).

Note that in contrast to the case of the nonlinear Dirac equation,
it is not possible to construct the {\it general} solutions of the
reduced systems (\ref{3.13})--(\ref{3.15}). By this very reason,
we give whenever possible their {\it particular} solutions,
obtained by reduction of systems of equations in question by the
number of components of the dependent function. Let us emphasize
that the miraculous efficiency of the t'Hooft's ansatz \cite{m5}
for the Yang-Mills equations is a consequence of the fact that it
reduces the system of twelve differential equations to a single
conformally-invariant wave equation.

Consider system (\ref{3.13})--(\ref{3.15}), that corresponds to
the subalgebra $L_8$. We adopt the following ansatz
\be \label{3.32} {\bf B}_\mu = a_\mu {\bf e}_1 f(\omega) +d_\mu
{\bf e}_2 g(\omega) + b_\mu {\bf e}_3 h(\omega)
\ee
for the vector-function ${\bf B}_\mu$, where $f(\omega),\
g(\omega),\ h(\omega)$ are new unknown smooth functions of
$\omega$ and
$$
{\bf e}_1= (1,0,0)^T, \quad {\bf e}_2= (0,1,0)^T, \quad
{\bf e}_3= (0,0,1)^T.
$$

Now inserting (\ref{3.32}) into (\ref{3.13}), where the
coefficients (\ref{3.14}) are given in the list (\ref{3.31}) for
the case of the subalgebra $L_8$, we arrive at the system
of relations
\begin{eqnarray*}
&& a_\mu {\bf e}_1[-4 \omega \ddot f -4 \dot f
-\frac{\alpha^2}{\omega} f+ \frac{2 \alpha e}{\sqrt{\omega}} g h
+ e^2 (h^2 +g^2) f] \\ & &\quad + d_\mu{\bf e}_2[-4 \omega \ddot g -4
\dot g -\frac{\alpha^2}{\omega} g-\frac{2 \alpha e}{\sqrt{\omega}}
f h + e^2 (h^2 -f^2) g] \\ & &\quad + b_\mu{\bf e}_3[-4 \omega \ddot
h -4 \dot h +\omega^{-1} h-\frac{2 \alpha e}{\sqrt{\omega}} f g
+ e^2 (g^2 - f^2) h] =0.
\end{eqnarray*}
It is equivalent to the following system of three ordinary
differential equations:
\begin{eqnarray} \label{3.33}
&& 4 \omega \ddot f + 4\dot f +\frac{\alpha^2}{\omega} f- \frac{2
\alpha e}{\sqrt{\omega}} g\, h - e^2 (h^2 +g^2) f =0, \nonumber \\
& & 4 \omega \ddot g +4 \dot g +\frac{\alpha^2}{\omega} g + \frac{2
\alpha e}{\sqrt{\omega}} f\, h - e^2 (h^2 -f^2) g =0, \\ & & 4 \omega
\ddot h +4 \dot h -\omega^{-1} h + \frac{2 \alpha e}{\sqrt{\omega}}
f\, g - e^2 (g^2 -f^2) h =0.\nonumber
\end{eqnarray}
So that we reduce system of twelve ordinary differential
equations (\ref{3.13}) to the one containing three
equations only.

Next, choosing
\be \label{3.34}
{\bf B}_\mu =k_\mu {\bf e}_1 f(\omega) +b_\mu {\bf e}_2 g(\omega)
\ee
and inserting this expression into (\ref{3.13}) with coefficients
given by formulae (\ref{3.31}) for the case of the subalgebra
$L_8$ under $\alpha=0$ yield the system of two ordinary
differential equations
$$
4 \omega \ddot f +4 \dot f -e^2 g^2 f =0,\quad
4 \omega \ddot g + 4\dot g -\omega^{-1} g =0.
$$
Note that the second equation of the above system is linear.

In a similar way we have reduced some other systems of ordinary
differential equations (\ref{3.13}) to systems of two or three
equations. Below we list the substitutions for ${\bf B}_\mu
(\omega)$ and corresponding systems of ordinary differential
equations. Numbering of the systems below corresponds to
numbering of the corresponding subalgebras $L_j$ of the algebra
$p(1,3)$.
\begin{eqnarray} \label{3.35}
&1.& {\bf B}_\mu = a_\mu {\bf e}_1 f(\omega) + b_\mu {\bf e}_2
g(\omega) +c_\mu {\bf e}_3 h(\omega),\nonumber \\ && \ddot f
-e^2(g^2+h^2)f=0,\quad \ddot g+e^2(f^2-h^2) g =0,
\nonumber \\ & & \ddot h+e^2(f^2-g^2) h=0. \nonumber \\
&2.& {\bf B}_\mu = b_\mu {\bf e}_1 f(\omega) + c_\mu {\bf e}_2
g(\omega) + d_\mu {\bf e}_3 h(\omega),\nonumber \\ && \ddot f
+e^2(g^2+h^2)f=0,\quad \ddot g + e^2(f^2+h^2) g =0,
\nonumber \\ & & \ddot h+e^2(f^2+g^2) h=0. \nonumber \\
&5.& {\bf B}_\mu = k_\mu {\bf e}_1 f(\omega) + b_\mu {\bf e}_2
g(\omega),\nonumber \\ && \ddot f -e^2 g^2 f=0, \quad
\ddot g =0. \nonumber \\
&8.1.& (\alpha=0)\ \ {\bf B}_\mu = k_\mu {\bf e}_1 f(\omega) +
b_\mu {\bf e}_2 g(\omega), \nonumber \\ & & 4 \omega \ddot f + 4
\dot f -e^2 g^2 f =0, \quad 4 \omega \ddot g + 4 \dot
g -\omega^{-1} g =0. \nonumber \\
&8.2.& {\bf B}_\mu = a_\mu {\bf e}_1 f(\omega) + d_\mu {\bf e}_2
g(\omega) + b_\mu {\bf e}_3 h(\omega), \nonumber \\ & & 4 \omega
\ddot f + 4\dot f -\frac{\alpha^2}{\omega} f -\frac{ 2\alpha
e}{\sqrt{\omega}}g h -e^2(h^2+g^2)f=0, \nonumber \\ & & 4 \omega
\ddot g + 4\dot g +\frac{\alpha^2}{\omega} g +\frac{ 2\alpha
e}{\sqrt{\omega}} f h +e^2(f^2-h^2) g=0, \nonumber \\ & & 4
\omega \ddot h + 4\dot h -\omega^{-1} h +\frac{ 2\alpha
e}{\sqrt{\omega}}f g +e^2(f^2-g^2) h=0. \nonumber \\
&14.1.& {\bf B}_\mu = a_\mu {\bf e}_1 f(\omega) +d_\mu {\bf e}_2
g(\omega) +c_\mu {\bf e}_3 h(\omega), \nonumber \\ & & 16 \ddot f
-e^2(h^2+g^2) f=0,\quad 16 \ddot g +e^2 (f^2-h^2)g =0,
\\ & & 16 \ddot h+e^2 (f^2 -g^2) h =0. \nonumber \\
&14.2.& {\bf B}_\mu = k_\mu {\bf e}_1 f(\omega) +c_\mu {\bf e}_2
g(\omega),\nonumber \\ && 16 \ddot f -e^2 g^2 f=0,\quad \ddot g
=0. \nonumber \\ &15.1.& {\bf B}_\mu = a_\mu {\bf e}_1 f(\omega)
+d_\mu {\bf e}_2 g(\omega)+
(1+\alpha^2)^{\textstyle{-\frac{1}{2}}}(\alpha c_\mu + b_\mu) {\bf
e}_3 h(\omega), \nonumber \\ & & 16(1+\alpha^2) \ddot f -e^2
(h^2+g^2) f=0, \nonumber \\ & & 16(1+\alpha^2) \ddot g +e^2
(f^2-h^2) g=0, \nonumber \\ & & 16(1+\alpha^2) \ddot h +e^2
(f^2-g^2) h=0. \nonumber \\ &15.2.& {\bf B}_\mu = k_\mu {\bf e}_1
f(\omega) + (1+\alpha^2)^{\textstyle{-\frac{1}{2}}}(\alpha c_\mu
+b_\mu) {\bf e}_2 g(\omega), \nonumber \\ & & 16(1+\alpha^2) \ddot
f -e^2 f g^2=0,\quad \ddot g=0. \nonumber \\ &16.& {\bf
B}_\mu = k_\mu {\bf e}_1 f(\omega) + b_\mu {\bf e}_2 g(\omega),
\nonumber \\ & & 4\omega \ddot f +4\dot f - e^2  g^2 f=0,\quad
4\omega \ddot g +4\dot g -\omega^{-1} g=0. \nonumber \\ &18.& {\bf
B}_\mu = b_\mu {\bf e}_1 f(\omega) + c_\mu {\bf e}_2 g(\omega),
\nonumber \\ & & 4\omega \ddot f +6\dot f +e^2  g^2 f=0,\quad
4\omega \ddot g +6\dot g +e^2 f^2 g=0.
\nonumber \\
&19.& {\bf B}_\mu = k_\mu {\bf e}_1 f(\omega) + b_\mu
{\bf e}_2 g(\omega), \nonumber \\ & & \ddot f -e^2  g^2 f=0,
\quad \ddot g=0. \nonumber \\
&20.& {\bf B}_\mu = k_\mu {\bf e}_1 f(\omega) + b_\mu {\bf e}_2
g(\omega), \nonumber \\ & &  \ddot f -e^2 g^2 f=0,\quad
\ddot g=0. \nonumber \\
&21.& {\bf B}_\mu = k_\mu {\bf e}_1 f(\omega) + b_\mu {\bf e}_2
g(\omega), \nonumber \\ & &  \ddot f -e^2 g^2 f=0,\quad
\ddot g=0. \nonumber \\
&22.& (\alpha =0)\quad {\bf B}_\mu = b_\mu {\bf e}_1 f(\omega) + c_\mu
{\bf e}_2 g(\omega), \nonumber \\ & & 4\omega \ddot f +8\dot f
+ e^2 g^2 f = 0,\quad 4\omega \ddot g + 8\dot g + e^2
f^2 g = 0. \nonumber
\end{eqnarray}

So, combining symmetry reduction by the number of independent
variables and direct reduction by the number of the components of
the function to be found we have reduced the $SU(2)$ Yang-Mills
equations (\ref{3.1}) to comparatively simple systems of ordinary
differential equations (\ref{3.35}).

As a next step, we briefly review the procedure of integration
of equations (\ref{3.35}).

Choosing $f=0,\ g=h=u(\omega)$ reduces system 1 to the equation
\be \label{3.36} \ddot u =e^2 u^3,
\ee
that is integrated in terms of the elliptic functions. Note, that
this equation has the solution that is expressed in terms of
elementary functions, namely,
$$
u =\sqrt{2} (e \omega-C)^{-1},\quad C \in {\bf R}.
$$

System 2 with $f=g=h=u(\omega)$ reduces to the equation
$$
\ddot u + 2 e^2 u^3 =0,
$$
that is also integrated in terms of the elliptic functions.

Upon integrating the second equation of system 5 we get
$$
g = C_1 \omega + C_2,\quad C_1, C_2 \in {\bf R}.
$$
Provided $C_1 \not =0$, the constant $C_2$ is negligible
and we may put $C_2=0$. With this condition the first equation
of system 5 reads
\be
\label{3.37} \ddot f -e^2 C^{2}_{1}\omega^2 f=0. \ee
The general solution of equation (\ref{3.37}), which is equivalent
to the Bessel equation, is given by the formula
$$
f= \sqrt{\omega} Z_{\textstyle{\frac{1}{4}}} (\frac{\textstyle{i\,
e}}{\textstyle{2}} C_1 \omega^2).
$$
Here we use the designations\ $Z_\nu(\omega) = C_3 J_\nu (\omega)
+C_4 Y_\nu(\omega)$,\ where\ $J_\nu, Y_\nu$ are the Bessel
functions and $C_3, C_4$ are arbitrary constants.

Given the condition $C_1=0,\ C_2\not =0$,\ the general solution
of the first equation of system 5 reads as
$$
f = C_3 \cosh (C_2 e\, \omega) + C_4 \sinh (C_2 e\, \omega),
$$
where\ $C_3, \ C_4$ are arbitrary real constants.

Finally, if $C_1 =C_2=0$,\ then the general solution of the first
equation of system 5 is given by formula\ $f=C_3 \omega + C_4$,\
$C_3,\ C_4\in {\bf R}$.

Next, we integrate the second equation of system 8.1 to obtain
$$
g = C_1 \sqrt{\omega} +C_2 (\sqrt{\omega})^{-1},
$$
where $C_1,\ C_2$ are arbitrary integration constants. Inserting
the function $g$ into the first equation of system 8.1 yields the
linear differential equation
\be \label{3.38}
4 \omega^2 \ddot f + 4 \omega \dot f -e^2 (C_1 \omega + C_2)^2 f
=0.
\ee
For the case\ $C_1 C_2 \not=0$, equation (\ref{3.38}) is related
to the Whitteker equation. Here we restrict our considerations to
the case\ $C_1 C_2 = 0$, thus getting
\begin{eqnarray*}
&a)& C_1 \not =0,\quad C_2 =0,\quad f =Z_0\Bigl[ \frac{i\, e}{2}
C_1 \omega\Bigr ];\\
&b)& C_1 =0,\quad C_2 \not =0,\quad f=C_3
\omega^{\textstyle{\frac{e C_2}{2}}} + C_4
\omega^{\textstyle{-\frac{e C_2}{2}}};\\
&c)& C_1 = C_2 =0,\quad f =C_3 \ln \omega + C_4,
\end{eqnarray*}
where\ $C_3,\ C_4$ are arbitrary integration constants.

Analyzing equations 14.1 and  14.2 we arrive at the conclusion
that they reduce to equations 1 and 5, correspondingly, if we
replace $e$ by $\frac{e}{4}$. Analogously, replacing in systems 1,
5 the parameter $e$ by $\frac{e}{4}(1+\alpha^2)^{-\frac{1}{2}}$
yields systems 15.1 and 15.2, respectively.

Finally, system 22 with $\alpha =0$ is reduced by the change of
the dependent variable\ $f = g =u(\omega)$ to the Emden-Fauler
equation
$$
\omega \ddot u + 2 \dot u + \frac{\textstyle{e^2}}{\textstyle{4}}
u^3 =0,
$$
that has the following particular solution $u=e^{-1}
\omega^{\textstyle{-\frac{1}{2}}}$.

We have not succeeded in integrating systems of ordinary
differential equations 8.2 and 18. Furthermore, systems 19, 20, 21
coincide with system 5 and system 16 coincides with system 8.1.

Inserting the obtained forms of the  functions\ $f, g, h$\ into
(\ref{3.35}) with the subsequent substitution of the latter
expression into the corresponding ansatz (\ref{3.7})--(\ref{3.9})
yield invariant solutions of the $SU(2)$ Yang-Mills equations
(\ref{3.1}). Note that solutions of systems 5, 8.1, 14.2, 15.2,
16, 19, 20, 21 with\ $g = 0$,\ give rise to Abelian solutions of
the Yang-Mills equation, i.e., to solutions satisfying the
additional restriction ${{\bf A}_\mu \times {\bf A}_\nu} ={\bf 0}$. Such
solutions are of low interest for physical applications and are
not considered here. Below we give the full list of non-Abelian
invariant solutions of equations (\ref{3.1})
\begin{eqnarray} \label{3.39}
&1.& {\bf A}_{\mu} = ({\bf e}_{2}b_{\mu}+{\bf e}_3 c_\mu)
\sqrt{2}(e dx -\lambda)^{-1} ; \nonumber \\
&2.&  {\bf A}_{\mu} = ({\bf e}_{2} b_{\mu}+{\bf e}_3 c_\mu)
[\lambda {\rm sn}\, ({\sqrt{2}\over 2} e \lambda dx) {\rm dn}\,
({\sqrt{2}\over 2} e \lambda dx)] [{\rm cn}\, {\sqrt{2}\over 2} e
\lambda dx)]^{-1} ; \nonumber \\
&3.& {\bf A}_{\mu} = ({\bf e}_{2}b_{\mu}+{\bf e}_3 c_\mu)\lambda
[{\rm cn}\, (e \lambda dx)]^{-1} ; \nonumber \\
&4.& {\bf A}_{\mu} = ({\bf e}_{1}b_{\mu}+{\bf e}_2 c_\mu +{\bf
e}_3 c_\mu)\lambda {\rm cn}\, (e \lambda ax) ; \nonumber \\
&5.& {\bf A}_{\mu} = {\bf e}_{1}k_{\mu}|kx|^{-1} \sqrt{cx}
Z_{\textstyle{1\over 4}}[{i\over 2} e \lambda (cx)^2] + {\bf
e}_{2}b_{\mu}\lambda cx ; \nonumber \\
&6.& {\bf A}_{\mu} = {\bf e}_{1}k_{\mu}|kx|^{-1} [\lambda_1 \cosh
(e \lambda cx)+\lambda_2 \sinh (e \lambda cx)]
+ {\bf e}_{2}b_{\mu} \lambda ; \nonumber \\
&7.& {\bf A}_{\mu} = {\bf e}_{1}k_{\mu}Z_0[ {i\over 2} e \lambda
((bx)^2 +(cx)^2)] + {\bf e}_{2}(b_{\mu} cx-c_\mu
bx)\lambda ; \nonumber \\
&8.& {\bf A}_{\mu} ={\bf e}_{1}k_{\mu}[\lambda_1 ((bx)^2
+(cx)^2))^{\textstyle{\frac{e \lambda}{2}}}
+\lambda_2((bx)^2+(cx)^2)^{-{e\lambda \over 2}}\nonumber \\ & &\quad +
{\bf e}_{2}(b_{\mu} cx-c_\mu bx)\lambda ((bx)^2 +(cx)^2)^{-1};
\nonumber \\
&9.& {\bf A}_{\mu} = \lbrack{\bf e}_{2}(\frac{1}{8}(d_{\mu} -
k_\mu(kx)^2) + \frac{1}{2} b_\mu kx )+ {\bf e}_{3}c_{\mu} \rbrack
\lambda {\rm sn}\,({ e\sqrt{2}\over 8} \lambda (4 bx \nonumber \\
& &\quad +(kx)^2)) {\rm dn}\, ({e \sqrt{2}\over 8}\lambda (4 bx
+(kx)^2)) ({\rm cn}\,({e \sqrt{2} \over 8} \lambda (4 bx
+(kx)^2)))^{-1}; \nonumber \\
&10.& {\bf A}_\mu = [{\bf e}_2 (\frac{1}{8} (d_\mu -k_\mu
(kx)^2)+\frac{1}{2} b_\mu kx) +{\bf e}_3 c_\mu ] \nonumber
\\ & &\quad \times \lambda [{\rm cn}\, ({e \sqrt{2} \lambda \over 8} (4
bx +(kx)^2))]^{-1}; \nonumber \\
&11.& {\bf A}_{\mu} = [{\bf e}_{2}(\frac{1}{8} (d_\mu
-k_\mu(kx)^2)+\frac{1}{2} b_\mu kx)+{\bf e}_3 c_\mu ]
\nonumber \\ & &\quad \times 4 \sqrt{2}(e(4 bx +(kx)^2) -\lambda)^{-1};
\nonumber \\
&12.& {\bf A}_{\mu} = {\bf e}_{1}k_{\mu}\sqrt{4 bx+ (kx)^2}
Z_{\textstyle{1 \over 4}} ({i e\lambda \over 8}(4 bx
+(kx)^2)^2)\nonumber \\ & &\quad +{\bf e}_2 c_\mu \lambda (4 bx
+(kx)^2); \nonumber \\
&13.& {\bf A}_{\mu} = {\bf e}_{1}
k_{\mu}(\lambda \cosh (\frac{\textstyle{e
\lambda}}{\textstyle{4}}(4 bx +(kx)^2)) \\ & &\quad + \lambda_2 \sinh
(\frac{\textstyle{e \lambda}}{\textstyle{4}}(4 bx +(kx)^2))) +
{\bf e}_{2} c_{\mu}\lambda ; \nonumber \\
&14.& {\bf A}_{\mu} = \{ {\bf e}_{2}(d_{\mu} - \frac{1}{8}
k_{\mu}(kx)^{2} - {1\over 2} b_\mu kx )\nonumber \\ & &\quad +{\bf
e}_{3}(\alpha c_\mu +b_{\mu}+\frac{1}{2} k_\mu
kx)(1+\alpha^2)^{\textstyle{-{1\over 2}}} \}\nonumber \\ & &\quad
\times\lambda {\rm sn}\, [\frac{\textstyle{e \lambda
\sqrt{2}}}{\textstyle{8}} (4 (\alpha bx - cx)+\alpha
(kx)^2)(1+\alpha^2)^{\textstyle{-{1\over 2}}}] \nonumber \\
& &\quad \times {\rm dn}\, [\frac{\textstyle{e \lambda
\sqrt{2}}}{\textstyle{8}} (4 (\alpha bx -cx)+\alpha
(kx)^2)(1+\alpha^2)^{\textstyle{-{1\over 2}}}]\nonumber \\
& &\quad \times \{ {\rm cn}\, [\frac{\textstyle{e \lambda
\sqrt{2}}}{\textstyle{8}} ((4 \alpha bx - cx)+\alpha
(kx)^2)(1+\alpha^2)^{\textstyle{-{1\over 2}}}] \}^{-1}; \nonumber
\\
&15.& {\bf A}_{\mu} = \{ {\bf e}_{2}( d_{\mu} - \frac{1}{8}
k_{\mu}(kx)^{2})-{1\over 2} b_\mu kx)\nonumber \\ & &\quad +{\bf e}_3
(\alpha c_\mu + b_\mu+ \frac{1}{2} k_{\mu} kx
)(1+\alpha^2)^{\textstyle{-{1 \over 2}}} \}\nonumber \\ & &\quad
\times \{ {\rm cn}\, [\frac{\textstyle{e \lambda}}{\textstyle{4}}
(4 \alpha bx -cx)+\alpha (kx)^2 (1+\alpha^2)^{\textstyle{-{1\over
2}}} ]\}^{-1}; \nonumber \\
&16.& {\bf A}_{\mu} = \{ {\bf e}_{2} (d_{\mu} - \frac{1}{8}
k_{\mu}(kx)^{2} - \frac{1}{2} b_{\mu}kx)\nonumber \\ & &\quad + {\bf
e}_3 (\alpha c_\mu +b_\mu +\frac{1}{2} k_\mu
kx)(1+\alpha^2)^{\textstyle{-{1\over 2}}} \}\nonumber \\ &
&\quad \times 4 \sqrt{2} (1+\alpha^2)^{\textstyle{1\over 2}} [e(4(
\alpha bx -cx)+\alpha (kx)^2)]^{-1}; \nonumber \\
&17.& {\bf A}_\mu = {\bf e}_1 k_\mu \{ \sqrt{4(\alpha bx -cx)
+\alpha (kx)^2} Z_{\textstyle{1\over 4}}(\frac{\textstyle{i e
\lambda}}{\textstyle{8}}(4(\alpha bx -cx)\nonumber \\ &&\quad +\alpha
(kx)^2)^2)(1+\alpha^2)^{\textstyle{-{1\over 2}}}\}\nonumber \\ &
&\quad +{\bf e}_2 (\alpha c_\mu +b_\mu +\frac{1}{2} k_\mu kx) \lambda
(4 (\alpha bx -cx)+\alpha
(kx)^2)(1+\alpha^2)^{\textstyle{-\frac{1}{2}}}; \nonumber \\
&18.& {\bf A}_{\mu} = {\bf e}_{1} k_{\mu}\{ {\rm cn}\, [\frac{e
\lambda}{4} (1+\alpha^2)^{\textstyle{-{1\over 2}}}(4 (\alpha
bx-cx)+\alpha(kx)^2)]\nonumber \\ & &\quad +\lambda_2 \sinh
[\frac{\textstyle{e
\lambda}}{\textstyle{4}}(1+\alpha^2)^{\textstyle{-{1\over
2}}}(4(\alpha bx -cx)+\alpha(kx)^2]\}\nonumber \\ & &\quad + {\bf
e}_2(\alpha c_\mu +b_\mu+\frac{1}{2} k_\mu kx)\lambda
(1+\alpha^2)^{\textstyle{-{1\over 2}}};\nonumber \\
&19.& {\bf A}_{\mu} = {\bf
e}_{1}k_{\mu}|kx|^{-1}Z_0[\frac{\textstyle{i e
\lambda}}{\textstyle{2}}((bx)^2+(cx)^2)] +{\bf e}_2(b_\mu cx-c_\mu
bx) \lambda ;\nonumber \\
&20.& {\bf A}_{\mu} = {\bf e}_{1}
k_{\mu}|kx|^{-1}[\lambda_1((bx)^2+(cx)^2)^{\textstyle{e \lambda
\over 2}}+\lambda((bx)^2+(cx)^2)^{\textstyle-{e \lambda \over 2}}
]\nonumber \\ & &\quad +{\bf e}_2(b_\mu cx -c_\mu bx)\lambda ((bx)^2
+(cx)^2)^{-1}; \nonumber \\
&21.& {\bf A}_{\mu} ={\bf e}_1 k_\mu |kx|^{-1}\sqrt{cx}
Z_{\textstyle{1\over 4}}(\frac{\textstyle{i e
\lambda}}{\textstyle{2}}(cx)^2)+ {\bf e}_2 (b_\mu -k_\mu
bx(kx)^{-1})\lambda cx; \nonumber \\
&22.& {\bf A}_{\mu} = {\bf e}_{1}k_{\mu}|kx|^{-1}\lbrack
\lambda_{1}\cosh (\lambda e cx) +\lambda_2 \sinh(\lambda e
cx)\rbrack\nonumber \\ &&\quad + {\bf e}_{2}(b_{\mu}-
k_{\mu}bx(kx)^{-1}) \lambda; \nonumber \\
&23.&  {\bf A}_{\mu} = {\bf
e}_{1}k_{\mu}|kx|^{-1}\sqrt{\ln|kx|-cx}Z_{\textstyle{1\over 4}}({i
e \lambda\over 2} (\ln|kx|-cx)^2)\nonumber \\ & &\quad + {\bf e}_{2}(
b_{\mu} - k_{\mu}bx (kx)^{-1})\lambda(\ln|kx|-cx); \nonumber \\
&24.& {\bf A}_{\mu} ={\bf e}_{1}k_{\mu}|kx|^{-1}
\lbrack\lambda_{1}\cosh(\lambda e (
\ln|kx|-cx)) + \lambda_2 \sinh (\lambda e(\ln
|kx|-cx))]\nonumber \\ & &\quad + {\bf e}_{2}\lbrack b_{\mu} -
k_{\mu}bx (kx)^{-1}\rbrack \lambda ; \nonumber \\
&25.& {\bf A}_{\mu} = {\bf e}_{1}k_{\mu} |kx|^{-1}\sqrt{\alpha
\ln |kx| -cx} Z_{\textstyle{1\over 4}}({i e \lambda \over
2}(\alpha \ln|kx|-cx)^2)\nonumber \\ & &\quad +{\bf e}_{2}(b_{\mu} -
k_{\mu} bx -\ln|kx|)(kx)^{-1})\lambda(\alpha \ln|kx| -cx);
\nonumber \\
&26.& {\bf A}_{\mu} = {\bf e}_{1}
k_{\mu}|kx|^{-1}[\lambda_1 \cosh (\lambda e(\alpha \ln
|kx|-cx))\nonumber \\ & &\quad +\lambda_2 \sinh (\lambda e(\alpha
\ln|kx|-cx))]\nonumber \\ & &\quad +{\bf e}_{2}(b_{\mu} -
k_{\mu}(bx-\ln |kx|^{-1})(kx)^{-1})\lambda; \nonumber \\
&27.& {\bf A}_{\mu} = \lbrace {\bf e}_{1}(b_{\mu} -
k_{\mu}bx(kx)^{-1})\nonumber \\ &&\quad + {\bf e}_{2}(c_{\mu} -
k_{\mu}cx(kx)^{-1})\rbrace e^{-1}
(x_{\nu}x^{\nu})^{\textstyle{-{1\over 2}}};\nonumber\\
&28.& {\bf A}_{\mu} = \lbrace {\bf e}_{1}(b_{\mu} -
k_{\mu}bx(kx)^{-1}) + {\bf e}_{2}(c_{\mu} -
k_{\mu}cx(kx)^{-1})\rbrace f (x_{\nu}x^{\nu}).\nonumber
\end{eqnarray}
In the above formulae the symbol $Z_\alpha (\omega)$ stands for
the Bessel function, ${\rm sn}\,(\omega), {\rm dn}\,(\omega), {\rm
cn}\,(\omega)$ are the Jacobi elliptic functions having the module
${\sqrt{2}\over 2}$;\ $f(x_\nu x^\nu)$ is the general solution of
the ordinary differential equation
$$
\omega^2 \ddot{f} + 2\omega \dot{f} + \frac{e^2} {4}f^3 = 0
$$
and $\lambda, \lambda_1, \lambda_2$ are arbitrary real constants.

\section{Conditional symmetry and new solutions of the Yang-Mills
equations.}
\setcounter{equation}{0}
\setcounter{tver}{0}
\setcounter{lema}{0}

With all the wealth of exact solutions obtainable through Lie
symmetries of the Yang-Mills equations, it is possible to construct
solutions, that cannot be derived by the symmetry reduction
method. The source of these solutions is {\it conditional} or {\it
non-classical} symmetry of the Yang-Mills equations.

The first paper devoted to non-classical symmetry of partial
differential equations was published by Bluman and Cole \cite{2a}.
However, the real importance of these symmetries was understood
much later after appearing the papers \cite{15}--\cite{13},
\cite{m31,m32}, where the method of conditional symmetries had
been used in order to construct new exact solutions of a number of
nonlinear partial differential equations.

The methods for dimensional reduction of partial differential
equations based on their conditional symmetry can be
conventionally classified into two principal groups. The first
group is formed by the direct methods (the ansatz method by
Fushchych and the direct method by Clarkson \& Kruskal), relying
upon a special {\it ad-hoc} representation of the solution to be
found in the form of the ansatz containing some arbitrary elements
(functions) $f_1, f_2,\ldots, f_n$ and unknown functions
$\varphi_1, \varphi_2,\ldots, \varphi_m$ with fewer number of
dependent variables. Inserting the ansatz in question into
equation under study and requiring for the obtained relation to be
equivalent to a system of partial differential equations for the
functions $\varphi_1, \varphi_2,\ldots, \varphi_m$ yield nonlinear
determining equations for the functions $f_1, f_2,\ldots, f_n$.
Having solved the latter yields a number of ansatzes reducing a
given partial differential equation to one having fewer number of
dependent variables. The second group of methods (the
non-classical method by Bluman \& Cole, the method of conditional
symmetries by Fushchych and the method of side conditions by Olver
\& Rosenau) may be regarded as infinitesimal ones. They are in
line with the traditional Lie approach to the reduction of partial
differential equations, since they exploit symmetry properties of
the equation under study in order to construct its invariant
solutions. And again any deviation from the standard Lie approach
requires solving over-determined system of nonlinear determining
equations. A more profound analysis of similarities and differences
between these approaches can be found in \cite{m33,9,zhd99}.

So the principal idea of the method of ansatzes, as well as, of
the direct method of reduction of partial differential equations
is a special choice of the class of functions to which a solution
to be found should belong. Within the framework of the above
methods a solution of system (\ref{3.1}) is looked for in the form
\begin{displaymath}
{\bf A}_\mu=H_\mu\Bigl(x,{\bf B}_\nu(\omega(x))\Bigr),\quad
\mu=0,1,2,3,
\end{displaymath}
where $H_\mu$ are smooth functions chosen in such a way that
substitution of the above expressions into the Yang-Mills
equations yields a system of ordinary differential equations for
new unknown vector-functions ${\bf B}_\nu$ of one variable
$\omega$. However being posed in this way, the problem of reduction
of the Yang-Mills equations seems to be hopeless. Indeed, even if we
restrict ourselves to the case of a linear dependence of the above
ansatz on $B_\nu$
\begin{equation}\label{cond2}
{\bf A}_{\mu}(x)= R_{\mu\nu}(x){\bf B}^{\nu}(\omega),
\end{equation}
where ${\bf B}_{\nu}(\omega)$ are new unknown vector-functions and
$\omega=\omega(x)$ is the new independent variable, then the
requirement of reduction of (\ref{3.1}) to a system of ordinary
differential equations by virtue of (\ref{cond2}) gives rise to
the system of nonlinear partial differential equations for 17
unknown functions $R_{\mu\nu},\ \omega$. And what is more, the
system obtained is not at all simpler than the initial Yang-Mills
equations (\ref{3.1}). Consequently, an additional information
about the structure of the matrix-function $R_{\mu\nu}$ should be
input into ansatz (\ref{cond2}). This can be done in various
ways. But the most natural one is to use the information about the
structure of solutions provided by the Lie symmetry of the
equation under study.

In \cite{m33} we suggest an effective approach to study of
conditional symmetry of the nonlinear Dirac equation based on its
Lie symmetry. We have observed that all the Poincar\'e-invariant
ansatzes for the Dirac field $\psi(x)$ can be represented in the
unified form by introducing several arbitrary elements (functions)
$u_1(x),\ u_2(x),\ldots, u_N(x)$.  As a result, we get an ansatz
for the field  $\psi(x)$ which reduces the nonlinear Dirac
equation to system of ordinary differential equations, provided
functions $u_i(x)$ satisfy some compatible over-determined system
of nonlinear partial differential equations. After integrating it
we have obtained a number of new ansatzes that cannot in principle
be obtained within the framework of the classical Lie approach.

Here, following \cite{m46} we will show that the same idea proves
to work efficiently for obtaining new (non-Lie) reductions of the
Yang-Mills equations and for constructing new exact solutions of
system (\ref{3.1}).

\subsection{Non-classical reductions of the Yang-Mills equations}

In the previous section we give the complete list of
$P$(1,3)-inequivalent ansat\-zes for the Yang-Mills field, which
are invariant under the three-para\-me\-ter subgroups of the
Poincar\'e group $P(1,3)$. These ansatzes can be represented in
the unified form (\ref{3.7}), where ${\bf B}_{\nu}(\omega)$ are
new unknown vector-functions, $\omega=\omega(x)$ is the new
independent variable and the functions $a_{\mu\nu}(x)$ are given
by (\ref{3.8}).

In (\ref{3.8}), $\theta_{\mu}(x)$ are some smooth functions, and
what is more $\theta_a$ $=\theta_a(\xi$,\, $b_{\mu}x^{\mu}$,\,
$c_{\mu}x^{\mu})$,\ $a=1,2$;\ $\xi=(1/2)k_{\mu}x^{\mu}=
(1/2)(a_{\mu}x^{\mu}+d_{\mu}x^{\mu})$;\ $a_{\mu},\ b_{\mu},\
c_{\mu},\ d_{\mu}$ are arbitrary constants satisfying relations
(\ref{comm}).

The choice of the functions  $\omega(x),\ \theta_{\mu}(x)$ is
determined by the requirement that substitution of ansatz
(\ref{3.7}) into the Yang-Mills equations yields a system of
ordinary differential equations for the vector function ${\bf
B}_{\mu}(\omega)$. By the direct check one can become convinced
of the validity of the following statement \cite{m33,m46}:

\begin{tver}\ Ansatz (\ref{3.7}), (\ref{3.8}) reduces the
Yang-Mills equations (\ref{3.1}) to system of ordinary
differential equations, if and only if, the functions $\omega(x)$,\
$\theta_{\mu}(x)$ satisfy the following system of partial
differential equations:
\begin{eqnarray}
&1)&\omega_{x_{\mu}}\omega_{x^{\mu}}=F_1(\omega),\nonumber\\
&2)&\square\omega=F_2(\omega),\nonumber\\
&3)&a_{\alpha\mu}\omega_{x_{\alpha}}=G_{\mu}(\omega),\nonumber\\
&4)&a_{\alpha\mu x_{\alpha}}=H_{\mu}(\omega),\label{cond4}\\
&5)&R^{\alpha}_{\mu}a_{\alpha\nu x_{\beta}}
\omega_{x^\beta}=Q_{\mu\nu}(\omega), \nonumber\\
&6)&R^{\alpha}_{\mu}\square
a_{\alpha\nu}=S_{\mu\nu}(\omega),\nonumber\\ &7)&R^{\alpha}_{\mu}
a_{\alpha\nu x_{\beta}}a_{\beta\gamma} +a_{\nu}^{\alpha}
a_{\alpha\gamma x_{\beta}}a_{\beta\mu}+
a_{\gamma}^{\alpha}a_{\alpha\mu x_{\beta}}a_{\beta\nu}
=T_{\mu\nu\gamma}(\omega),\nonumber
\end{eqnarray}
where $F_1,\ F_2,\ G_{\mu},\ldots,T_{\mu\nu\gamma}$ are some
smooth functions of $\omega$,\ $\mu$, $\nu$, $\gamma=0,1,2,3$. And
what is more, the reduced system has the form
\begin{equation}
\begin{array}{l}
k_{\mu\gamma}\ddot{{\bf B}^{\gamma}}+ l_{\mu\gamma}\dot{\bf
B^{\gamma}}+m_{\mu\gamma}{\bf B}^{\gamma}+
eq_{\mu\nu\gamma}\dot{{\bf B}^{\nu}}\times{\bf B}^{\gamma}+
eh_{\mu\nu\gamma}{\bf B}^{\nu}\times{\bf B}^{\gamma}\\[2mm] \quad
+e^2{\bf B}_{\gamma} \times ({\bf B}^{\gamma}\times{\bf
B}_{\mu})={\bf 0},
\end{array}
\label{cond5}
\end{equation}
where
\begin{eqnarray}
k_{\mu\gamma}&=&g_{\mu\gamma}F_1-G_{\mu}G_{\gamma},\nonumber\\
l_{\mu\gamma}&=&g_{\mu\gamma}F_2+2Q_{\mu\gamma}-
G_{\mu}H_{\gamma}-G_{\mu}\dot G_{\gamma},\nonumber\\
m_{\mu\gamma}&=&S_{\mu\gamma}- G_{\mu}\dot
H_{\gamma},\label{cond6}\\
q_{\mu\nu\gamma}&=&g_{\mu\gamma}G_{\nu}+g_{\nu\gamma}
G_{\mu}-2g_{\mu\nu}G_{\gamma},\nonumber\\
h_{\mu\nu\gamma}&=&(1/2)(g_{\mu\gamma}H_{\nu}-
g_{\mu\nu}H_{\gamma})-T_{\mu\nu\gamma}.\nonumber
\end{eqnarray}
\end{tver}

Consequently, to describe all the ansatzes of the form
(\ref{3.7}), (\ref{3.8}) reducing the Yang-Mills equations to
a system of ordinary differential equations one has to construct
the general solution of the over-determined system of partial
differential equations (\ref{3.8}), (\ref{cond4}). Let us
emphasize that system (\ref{3.8}), (\ref{cond4}) is compatible
since the ansatzes for the Yang-Mills field ${\bf A}_\mu(x)$
invariant under the three-parameter subgroups of the Poincar\'e
group satisfy equations (\ref{3.8}), (\ref{cond4}) with some
specific choice of the functions $F_1,\ F_2,\ldots,
T_{\mu\nu\gamma}$ \cite{m45}.

Integration of system of nonlinear partial differential equations
(\ref{3.8}), (\ref{cond4}) has been performed in \cite{m33,m46}.
Here we indicate the principal steps of the integration procedure.
While integrating (\ref{3.8}), (\ref{cond4}) we use essentially
the fact that the general solution of system of equations 1, 2
from (\ref{cond4}) is known \cite{10}. With already known $\omega
(x)$ in hand we proceed to integrating linear partial differential
equations 3, 4 from (\ref{cond4}). Next, we insert the results
obtained into the remaining equations and get the final forms of
the functions $\omega(x)$,\ $\theta_{\mu}(x)$.

Before presenting the results of integration of system of partial
differential equations (\ref{3.8}), (\ref{cond4}) we make the
following remark. As the direct check shows, the structure of
ansatz (\ref{3.7}), (\ref{3.8}) is not altered by the change of
variables
\begin{equation}
\begin{array}{l}
\omega\to\omega'=T(\omega),\quad
\theta_0\to\theta_0'=\theta_0+T_0(\omega),\\[2mm]
\theta_1\to\theta_1'=\theta_1+e^{\theta_0}
\Bigl(T_1(\omega)\cos\theta_3+T_2(\omega)\sin\theta_3\Bigr),\\[2mm]
\theta_2\to\theta_2'=\theta_2+e^{\theta_0}
\Bigl(T_2(\omega)\cos\theta_3-T_1(\omega)\sin\theta_3\Bigr),\\[2mm]
\theta_3\to\theta_3'=\theta_3+T_3(\omega),
\end{array}
\label{cond7}
\end{equation}
where $T(\omega),\ T_{\mu}(\omega)$ are arbitrary smooth
functions. That is why, solutions of system (\ref{3.8}),
(\ref{cond4}) connected by relations (\ref{cond7}) are considered
as equivalent.

Integrating the system of partial differential equations under
study within the above equivalence relations we obtain the set of
ansatzes containing the ones equivalent to the
Poincar\'e-invariant ansatzes, obtained in the previous section.
That is why, we concentrate on essentially new (non-Lie) ansatzes.
It so happens that our approach gives rise to non-Lie ansatzes,
provided the functions $\omega(x),\ \theta_{\mu} (x)$ within the
equivalence relations (\ref{cond7}) have the form
\begin{equation}
\theta_{\mu}=\theta_{\mu} (\xi,\, b x,\,
c x),\quad \omega = \omega (\xi,\, b x,\,
c x), \label{cond8}
\end{equation}
where, as earlier, $b x = b_\mu x^\mu, c x = c_\mu x^\mu$.

List of inequivalent solutions of system of partial differential
equations (\ref{3.8}), (\ref{cond4}) satisfying (\ref{cond8}) is
exhausted by the following solutions:
\begin{eqnarray}
&1)&\theta_0=\theta_3=0,\quad \omega=(1/2)k  x,\quad
\theta_1=w_0(\xi)b  x + w_1(\xi)c  x,\nonumber\\ &
&\theta_2=w_2(\xi)b  x + w_3(\xi)c  x;\nonumber\\[3mm]
&2)&\omega=b  x + w_1(\xi), \quad\theta_0=\alpha\Bigl (c
x + w_2(\xi)\Bigr ),\nonumber\\ & &\theta_a=-(1/4)\dot w_a(\xi),\
\ a=1,2,\quad \theta_3=0,\label{cond9}\\[3mm] &3)&\theta_0=T(\xi),
\quad \theta_3=w_1(\xi),\quad \omega=b  x\cos w_1+c  x\sin
w_1+ w_2(\xi),\nonumber\\ & &\theta_1=\Bigl ((1/4)(\varepsilon
e^T+\dot T) (b  x\sin w_1-c  x\cos w_1)+w_3(\xi)\Bigr )
\sin w_1\nonumber\\ & &\quad +(1/4)\Bigl ( \dot w_1(b  x \sin
w_1-c  x\cos w_1)-\dot w_2\Bigr )\cos w_1,\nonumber\\ &
&\theta_2=-\Bigl ((1/4)(\varepsilon e^T+\dot T) (b  x\sin
w_1-c  x\cos w_1)+w_3(\xi)\Bigr ) \cos w_1\nonumber\\ & &\quad
+(1/4)\Bigl ( \dot w_1(b  x \sin w_1-c  x\cos w_1)-\dot
w_2\Bigr )\sin w_1;\nonumber\\[3mm] &4)&\theta_0=0,\quad \theta_3=
\arctan \ \Bigl ([c  x + w_2 (\xi)][b  x +
w_1(\xi)]^{-1}\Bigr ),\nonumber\\ & &\theta_a=-(1/4)\dot w_a
(\xi),\ \ a=1,2,\nonumber\\ & &\omega = \Bigl ([b  x +
w_1(\xi)]^2+[c  x + w_2(\xi)]^2\Bigr )^{1/2}.\nonumber
\end{eqnarray}

Here $\alpha\ne 0$ is an arbitrary constant, $\varepsilon=\pm 1$,\
$w_0,\ w_1,\ w_2,\ w_3$ are arbitrary smooth functions on $\xi=
(1/2) k  x$,\ $T=T(\xi)$ is a solution of the nonlinear
ordinary differential equation
\begin{equation}
(\dot T+\varepsilon e^T)^2+\dot w_1^2=\varkappa e^{2T}, \ \
\varkappa \in {\mathbb R}^1, \label{cond10}
\end{equation}
a dot over the symbol denotes differentiation with respect to
$\xi$.

Inserting ansatz (\ref{3.7}), where $a_{\mu\nu}(x)$ are given by
formulae (\ref{3.8}), (\ref{cond9}), into the Yang-Mills
equations yields systems of nonlinear ordinary differential
equations of the form (\ref{cond5}), where
\begin{eqnarray}
&1)& k_{\mu\gamma}=-(1/4)k_{\mu}k_{\gamma},\quad
l_{\mu\gamma}=-(w_0+w_3)k_{\mu}k_{\gamma},\nonumber\\ & &
m_{\mu\gamma} = -4\ (w_0^2+w_1^2+w_2^2+w_3^2)
k_{\mu}k_{\gamma}-(\dot w_0+\dot w_3)k_{\mu}k_{\gamma},\nonumber\\
& & q_{\mu\nu\gamma}=
(1/2)(g_{\mu\gamma}k_{\nu}+g_{\nu\gamma}k_{\mu}-
2g_{\mu\nu}k_{\gamma}),\nonumber\\ & & h_{\mu\nu\gamma}=
(w_0+w_3)(g_{\mu\gamma} k_{\nu}-g_{\mu\nu} k_{\gamma})+
2(w_1-w_2)\Bigl( (k_{\mu}b_{\nu}-k_{\nu}b_{\mu})\
c_{\gamma}\nonumber\\ & &\quad
+(b_{\mu}c_{\nu}-b_{\nu}c_{\mu})k_{\gamma}+(c_{\mu}k_{\nu}-
c_{\nu}k_{\mu})b_{\gamma}\Bigr);\nonumber\\[3mm] &2)&
k_{\mu\gamma}=-g_{\mu\gamma}-b_{\mu}b_{\gamma}, \quad
l_{\mu\gamma}=0,\quad m_{\mu\gamma}=-\alpha^2 (a_{\mu}
a_{\gamma}-d_{\mu}d_{\gamma}),\nonumber\\ &
&q_{\mu\nu\gamma}=g_{\mu\gamma}b_{\nu}+
g_{\nu\gamma}b_{\mu}-2g_{\mu\nu}b_{\gamma},\nonumber\\ &
&h_{\mu\nu\gamma}=\alpha\Bigl ((
a_{\mu}d_{\nu}-a_{\nu}d_{\mu})c_{\gamma}+(d_{\mu}c_{\nu}-
d_{\nu}c_{\mu})
a_{\gamma}+(c_{\mu}a_{\nu}-c_{\nu}a_{\mu})d_{\gamma} \Bigr
);\nonumber\\[3mm] &3)&
k_{\mu\gamma}=-g_{\mu\gamma}-b_{\mu}b_{\gamma}, \quad
l_{\mu\gamma}=-(\varepsilon/2)b_{\mu} k_{\gamma},\label{cond11}\\
& &m_{\mu\gamma}=-(\varkappa/4)k_{\mu}k_{\gamma}, \quad
q_{\mu\nu\gamma}=g_{\mu\gamma} b_{\nu}+g_{\nu\gamma}b_{\mu}-
2g_{\mu\nu}b_{\gamma},\nonumber\\ &
&h_{\mu\nu\gamma}=(\varepsilon/4)(g_{\mu\gamma}k_{\nu}
-g_{\mu\nu}k_{\gamma});\nonumber\\[3mm] &4)&
k_{\mu\gamma}=-g_{\mu\gamma}-b_{\mu}b_{\gamma}, \quad
l_{\mu\gamma}=-\omega^{-1}(g_{\mu\gamma}+ b_{\mu}
b_{\gamma}),\nonumber\\ &
&m_{\mu\gamma}=-\omega^{-2}c_{\mu}c_{\gamma},\quad
q_{\mu\nu\gamma}=g_{\mu\gamma} b_{\nu}+g_{\nu\gamma} b_{\mu}-
2g_{\mu\nu} b_{\gamma},\nonumber\\ &
&h_{\mu\nu\gamma}=(1/2)\omega^{-1}(g_{\mu\gamma}b_{\nu}
-g_{\mu\nu} b_{\gamma}).\nonumber
\end{eqnarray}

\subsection{Exact solutions}

Systems (\ref{cond5}), (\ref{cond11}) contain twelve nonlinear
second-order ordinary differential equations with variable
coefficients. That is why, there is little hope to construct
their general solutions. Nevertheless, it is possible to obtain
particular solutions of system (\ref{cond5}), whose coefficients
are given by formulae 2--4 from (\ref{cond11}).

Consider, as an example, system of ordinary differential equations
(\ref{cond5}) with coefficients given by the formulae 2 from
(\ref{cond11}). We look for its solutions of the form
\begin{equation}
{\bf B}_{\mu}= k_{\mu} {\bf e}_1 f(\omega) + b_{\mu} {\bf e}_2
g(\omega), \ \ fg\ne 0, \label{cond12}
\end{equation}
where ${\bf e}_1=(1, 0, 0),\ {\bf e}_2=(0, 1, 0)$.

Substituting expression (\ref{cond12}) into the above
mentioned system we get
\begin{equation}
\ddot f + (\alpha^2-e^2g^2)f=0,\quad f\dot g + 2\dot fg=0.
\label{cond13}
\end{equation}

The second ordinary differential equation from (\ref{cond13}) is
easily integrated
\begin{equation}
g=\lambda f^{-2},\ \ \lambda \in{\mathbb R}^1, \ \ \lambda\ne 0.
\label{cond14}
\end{equation}

Inserting the result obtained into the first ordinary
differential equation from (\ref{cond13}) yields the Ermakov-type
equation for $f(\omega)$
\begin{displaymath}
\ddot f +\alpha^2 f-e^2\lambda^2f^{-3}=0,
\end{displaymath}
which is integrated in elementary functions \cite{11a}
\begin{equation}
f=\Bigl (\alpha^{-2}C^2+\alpha^{-2}
(C^4-\alpha^2e^2\lambda^2)^{1/2} \sin 2|\alpha|\omega\Bigr
)^{1/2}. \label{cond15}
\end{equation}
Here $C\ne 0$ is an arbitrary constant.

Substituting (\ref{cond12}), (\ref{cond14}), (\ref{cond15}) into
the corresponding ansatz for ${\bf A}_{\mu}(x)$ we get the
following class of exact solutions of the Yang-Mills equations
(\ref{3.1}):
\begin{eqnarray*}
{\bf A}_{\mu}&=&{\bf e}_1 k_{\mu}\exp\ (-\alpha c  x-\alpha
w_2 )\Bigl (\alpha^{-2}C^2+\alpha^{-2}(C^4- \alpha^2
e^2\lambda^2)^{1/2}\\ &&\times \sin 2 \vert\alpha\vert (b
x+w_1)\Bigr )^{1/2}+ {\bf e}_2\lambda \Bigl
(\alpha^{-2}C^2+\alpha^{-2} (C^4-\alpha^2 e^2
\lambda^2)^{1/2}\qquad\\ &&\times \sin 2 \vert\alpha\vert (b
x+w_1)\Bigr )^{-1} \Bigl(b_{\mu}+ (1/2) k_{\mu} \dot w_1\Bigr ).
\end{eqnarray*}

In a similar way we have obtained the five other classes of exact
solutions of the Yang-Mills equations
\begin{eqnarray*}
  {\bf A}_{\mu}&=&{\bf e}_1 k_{\mu}e^{-T} (b  x \cos w_1+
c  x \sin w_1+w_2)^{1/2}Z_{1/4}\Bigl ((ie\lambda/2) (b
x\cos w_1\\ &&+ c  x \sin w_1+w_2)^2\Bigr )+ {\bf e}_2\lambda\
(b  x \cos w_1+c  x\sin w_1+w_2)\\ &&\times \Bigl
(c_{\mu}\cos w_1-b_{\mu}\sin w_1+2 k_{\mu} [(1/4)(\varepsilon
e^T+\dot T)(b  x\sin w_1\\ &&-c  x\cos w_1)+w_3]\Bigr
);\\[3mm]
  {\bf A}_{\mu}&=&{\bf e}_1 k_{\mu}
e^{-T} \Bigl (C_1\cosh[e\lambda(b  x\cos w_1+ c  x\sin
w_1+w_2)]+C_2\sinh[e\lambda\\ &&\times(b  x\cos w_1+ c  x
\sin w_1+w_2)]\Bigr )+ {\bf e}_2\lambda \Bigl ( c_{\mu}\cos
w_1-b_{\mu}\sin w_1\\ &&+ 2k_{\mu}[(1/4)(\varepsilon e^T+ \dot
T)(b  x\sin w_1-c  x\cos w_1)+w_3]\Bigr );\\[3mm]
  {\bf A}_{\mu}&=&{\bf e}_1 k_{\mu}
e^{-T}\Bigl (C^2(b  x\cos w_1+ c  x\sin
w_1+w_2)^2+\lambda^2 e^2C^{-2}\Bigr )^{1/2}\\ &&+ {\bf e}_2\lambda
\Bigl (C^2(b  x\cos w_1+ c  x\sin w_1+w_2)^2+ \lambda^2
e^2 C^{-2}\Bigr )^{-1}\\ &&\times \Bigl (b_{\mu}\cos
w_1+c_{\mu}\sin w_1- (1/2)k_{\mu}[\dot w_1 (b  x\sin w_1\\
&&-c  x\cos w_1)-\dot w_2]\Bigr );\\[3mm]
  {\bf A}_{\mu}&=&{\bf e}_1 k_{\mu} Z_0
\Bigl ((ie\lambda/2)[(b  x+w_1)^2+(c  x+ w_2)^2]\Bigr
)+{\bf e}_2\lambda \Bigl ( c_{\mu}(b  x+w_1)\\
&&-b_{\mu}(c  x+w_2)- (1/2)k_{\mu}[\dot w_1(c  x+
w_2)-\dot w_2(b  x+w_1)]\Bigr );\\[3mm]
  {\bf A}_{\mu}&=&{\bf e}_1 k_{\mu}\Bigl (C_1[(b  x+w_1)^2+
(c  x+w_2)^2]^{e\lambda/2} + C_2[(b  x+w_1)^2\\ &&+(c
x+w_2)^2]^{-e\lambda/2}\Bigr) +{\bf e}_2\lambda[(b  x+
w_1)^2+(c  x+w_2)^2]^{-1}\\ &&\times \Bigl (c_{\mu}(b
x+w_1)-b_{\mu}(c  x+w_2)- (1/2)k_{\mu}[\dot w_1(c
x+w_2)\\ &&-\dot w_2(b  x+w_1)]\Bigr ).
\end{eqnarray*}
Here $C_1,\ C_2,\ C\ne 0, \lambda$ are arbitrary parameters;
$w_1,\ w_2,\ w_3$ are arbitrary smooth functions on
$\xi=(1/2)k  x;\ T=T(\xi)$ is a solution of ordinary
differential equation (\ref{cond10}). Besides that, we use the
following notations:
\begin{eqnarray*}
& &k  x=k_{\mu}x^{\mu},\quad b  x=b_{\mu}x^{\mu}, \quad
c  x= c_{\mu}x^{\mu},\qquad\\ & &Z_s(\omega)=C_1J_s
(\omega)+C_2Y_s (\omega),\\ & &{\bf e}_1=(1,0,0),\quad {\bf
e}_2=(0,1,0),
\end{eqnarray*}
where $J_s,\ Y_s$ are the Bessel functions.

Thus, we have obtained the broad families of exact non-Abelian
solutions of the Yang-Mills equations (\ref{3.1}). It can be
verified by direct and rather involved computation that the
solutions obtained are not self-dual, i.e., that they do not
satisfy the self-dual Yang-Mills equations.


\subsection{Conditional symmetry formalism}

Now we briefly discuss the problem of conditional symmetry
interpretation of ansatzes (\ref{3.7}), (\ref{3.8}),
(\ref{cond9}). Consider, as an example, the ansatz determined by
the formulae 1 from (\ref{cond9}). As the direct computation shows,
generators of a three-parameter Lie group $G$ leaving it invariant
are of the form
\begin{equation}
\begin{array}{l}
Q_1= k_{\alpha}\partial_{\alpha},\\[2mm]
Q_2=b_{\alpha}\partial_{\alpha}- 2[
w_0(k_{\mu}b_{\nu}-k_{\nu}b_{\mu})+
 w_2(k_{\mu}c_{\nu}-k_{\nu}c_{\mu})]
\, {\displaystyle\mathop\sum\limits_{a=1}^3} A^{a\nu}
\partial_{\displaystyle{A^{a\mu}}},\\[2mm]
Q_3= c_{\alpha}\partial_{\alpha}- 2[
w_1(k_{\mu}b_{\nu}-k_{\nu}b_{\mu})+
w_3(k_{\mu}c_{\nu}-k_{\nu}c_{\mu})] \,
{\displaystyle\mathop\sum\limits_{a=1}^3} A^{a\nu}
\partial_{\displaystyle{A^{a\mu}}}.
\end{array}
\label{cond16}
\end{equation}

Evidently, system of partial differential equations (\ref{3.1}) is
invariant under the one-parameter group $G_1$ having the generator
$Q_1$. However, it is not invariant under the one-parameter groups
$G_2, G_3$ having the generators $Q_2,\ Q_3$. Consider, as an
example, the generator $Q_2$. Acting by the second prolongation of
the operator $Q_2$ (which is constructed in the standard way, see,
for example, \cite{m19,m20}) on the system of partial differential
equations (\ref{3.1}) we see that the resulting expression does
not vanish on the solution set of the equations (\ref{3.1}).
However, if we consider the constrained Yang-Mills equations
\begin{displaymath}
{\bf L}_\mu ={\bf 0},\quad Q_a {\bf A}_\mu = {\bf 0},\quad
a=1,2,3,
\end{displaymath}
then we see that the system obtained is invariant under the group
$G_2$. In the above formulae we use the designations
\begin{eqnarray*}
{\bf L}_\mu& \equiv &\square {\bf
A}_{\mu}-\partial^{\mu}\partial_{\nu} {\bf A}_{\nu} + e\Bigl
((\partial_{\nu}{\bf A}_{\nu})\times {\bf
A}_{\mu}-2(\partial_{\nu}{\bf A}_{\mu})\times {\bf A}_{\nu}\\
&&+(\partial^{\mu}{\bf A}_{\nu})\times {\bf A}^{\nu}\Bigr )+
e^2{\bf A}_{\nu} \times ({\bf A}^{\nu}\times {\bf A}_{\mu}),\\
Q_1{\bf A}_{\mu}&\equiv& k_{\alpha}\partial_{\alpha}{\bf
A}_{\mu},\\ Q_2 {\bf A}_{\mu}&\equiv&
b_{\alpha}\partial_{\alpha}{\bf A}_{\mu} +2\Bigl(
w_0(k_{\mu}b_{\nu}-k_{\nu}b_{\mu})+
w_2(k_{\mu}c_{\nu}-k_{\nu}c_{\mu})\Bigr ){\bf A}^{\nu},\\ Q_3{\bf
A}_{\mu}&\equiv& c_{\alpha}\partial_{\alpha}{\bf A}_{\mu} + 2\Bigl
( w_1(k_{\mu}b_{\nu}-k_{\nu}b_{\mu})+
w_3(k_{\mu}c_{\nu}-k_{\nu}c_{\mu})\Bigr ){\bf A}^{\nu}.
\end{eqnarray*}
The same assertion holds for the Lie transformation group $G_3$
having the generator $Q_3$. Consequently, the Yang-Mills equations
are conditionally-invariant with respect to the three-parameter
Lie transformation group $G=G_1\otimes G_2\otimes G_3$. It means
that solutions of the Yang-Mills equations obtained with the help
of the ansatz invariant under the group with generators
(\ref{cond16}) cannot be found by means of the classical symmetry
reduction procedure. We refer the reader interested in further
details to the monographs \cite{m21,m33}.

As very cumbersome computations show, the ansatzes determined by
the formulae 2--4 from (\ref{cond9}) also correspond to
conditional symmetry of the Yang-Mills equations. Hence it
follows, in particular, that the Yang-Mills equations should be
included into the long list of mathematical and theoretical
physics equations possessing non-trivial conditional symmetry
\cite{m21}.

\section{Symmetry reduction and exact solutions of the Maxwell
equations}
\setcounter{equation}{0}
\setcounter{tver}{0}
\setcounter{lema}{0}

In this part of the paper we exploit symmetry properties of the
(vacuum) Maxwell equations in order to construct their exact
solutions.

It is well-known that the electro-magnetic field for the case
of the vanishing current is described by the Maxwell equations
in vacuum
\begin{eqnarray} \label{4.1}
& & {\rm rot}\, {\bf E} = -\frac{\partial {\bf H}}{\partial
x_0},\quad {\rm div}\, {\bf H}=0, \nonumber \\
& & {\rm rot}\, {\bf H} = \frac{\partial {\bf E}}{\partial
\,x_0},\quad {\rm div}\, {\bf E} = 0
\end{eqnarray}
for the vector fields ${\bf E} = {\bf E}(x_0, {\bf x})$ and
${\bf H} = {\bf H}(x_0, {\bf x})$ (in the sequel, we call them
the Maxwell fields).

First, we give a brief overview of symmetry properties of
equations (\ref{4.1}) following \cite{m42}.

\subsection{Symmetry of the Maxwell equations}

As we have mentioned in the introduction, the maximal symmetry
group admitted by equations (\ref{4.1}) is the sixteen-parameter
group $C(1,3) \bigotimes H$. This group is the direct product of
the conformal group $C(1,3)$ generated by the Lie vector fields
\begin{eqnarray} \label{4.2}
& & P_{\mu} = \partial_{x_{\mu}}, \ \ \ J_{0a} =
x_{0}\partial_{x_{a}} +x_{a} \partial_{x_{0}}
+\varepsilon_{abc}(E_{b}
\partial_{H_{c}} -H_{b} \partial_{E_{c}}), \nonumber \\
& & J_{ab} = x_{b} \partial_{x_{a}}-x_{a} \partial_{x_{b}} +E_{b}
\partial_{E_{a}} -E_{a} \partial_{E_{b}} +H_{b}
\partial_{H_{a}} -H_{a} \partial_{H_{b}}, \nonumber \\
& & D = x_{\mu} \partial_{x_{\mu}}-2(E_{a} \partial_{E_{a}} +H_{a}
\partial_{H_{a}}), \\
& & K_0 = 2 x_0 D-x_\mu x^\mu \partial_{x_0} + 2 x_a
\varepsilon_{abc}(E_b \partial_{H_c} -H_b \partial_{E_c}),
\nonumber \\ & & K_a = -2 x_a D-x_\mu x^\mu \partial_{x_a} -2 x_0
\varepsilon_{abc}(E_b \partial_{H_c} -H_b \partial_{E_c})
\nonumber \\ \quad & &- 2 H_a (x_b \partial_{H_b}) -2 E_a(x_b \partial_{
E_b})+2 (x_b H_b) \partial_{H_a} +2(x_b E_b)
\partial_{E_a},\nonumber
\end{eqnarray}
and of the one-parameter Heviside-Larmor-Rainich group $H$
having the generator
\be \label{4.3} Q=E_a \partial_{H_a} - H_a
\partial_{E_a},
\ee
where $\varepsilon_{abc}$ is the third-order anti-symmetric tensor
with $\varepsilon_{123}=1$. In this section the indices denoted by
the Latin alphabet letters $a, b, c$ take the values $1, 2, 3$, and
the ones denoted by the Greek alphabet letters take the values $0,
1, 2, 3$, and the summation convention is used.

It is readily seen from (\ref{4.2}), (\ref{4.3}) that the action
of the group $C(1,3) \bigotimes H$ in the space $R^{1,3} \times
R^6$,\ where $R^{1,3}$ is Minkowski space of the variables\ $
x_0$, ${\bf x} =(x_1, x_2, x_3)$ and \ $R^6 $ is the
six-dimensional space of the functions ${\bf E} = (E_1, E_2, E_3),
\ {\bf H}=(H_1, H_2, H_3)$, is projective. And furthermore, the
basis generators of this group can be represented in the form
(\ref{2.11}).

The matrices $S_{\mu \nu}$ read as
\begin{eqnarray} \label{4.4}
& & S_{01} = \left( \matrix{0& \tilde S_{23}\cr -\tilde S_{23}&
0\cr }\right ), \qquad S_{02} = \left( \matrix{0&  -\tilde
S_{13}\cr \tilde S_{13}& 0\cr }\right ),\nonumber \\ && \nonumber
\\ & & S_{03} = \left( \matrix{0& \tilde S_{12}\cr -\tilde S_{12}&
0\cr }\right) , \quad S_{12} = \left( \matrix{ \tilde S_{12}&0\cr
0& \tilde S_{12}& \cr }\right ), \\ && \nonumber \\ & &  S_{13} =
\left( \matrix{ \tilde S_{13}&0\cr 0& \tilde S_{13}& \cr }\right
), \qquad S_{23} = \left( \matrix{ \tilde S_{23}&0\cr 0& \tilde
S_{23}& \cr }\right ),\nonumber
\end{eqnarray}
where $0$ is the zero $3 \times 3$ matrix and
$$
{\tilde S}_{12} = \left( \matrix{ 0&-1&0\cr 1& 0&0 \cr
0&0&0\cr}\right ), \quad {\tilde S_{13}}\left( \matrix{ 0&0&-1\cr
0& 0&0 \cr1&0&0\cr }\right ), \quad {\tilde S_{23}} = \left(
\matrix{ 0&0&0\cr 0& 0&-1 \cr 0&1&0\cr}\right ),
$$
$E$ is the unit $6 \times 6$ matrix. The matrix $-A$ corresponding
to operator $Q$ (\ref{4.3}) is given by the formula
\be
\label{4.5} A=\left( \matrix{0& -I\cr I& 0\cr
}\right),
\ee
where $0$ and $I$ are zero and unit $3 \times 3$ matrices,
correspondingly.

Hence, it follows that $C(1,3) \bigotimes H$-invariant ansatzes
for the Maxwell fields, that reduce (\ref{4.1}) to systems of
ordinary differential equations, can be represented in the
form (\ref{2.18}), namely,
\be \label{4.6}
{\bf V} = \Lambda (x_0, {\bf x}) {\bf \tilde V}(\omega)
\ee
with
$$
{\bf V}=\left(\matrix{ E_1 \cr E_2 \cr E_3 \cr H_1\cr H_2\cr
H_3\cr}\right ), \quad {\bf \tilde V}=\left(\matrix{ \tilde
E_1 \cr \tilde E_2 \cr \tilde E_3 \cr \tilde H_1\cr \tilde H_2\cr
\tilde H_3\cr}\right ).
$$
Here $\Lambda(x_0, {\bf x})$ is the $6\times 6$ matrix, which is
non-singular in some open domain of the space $R^{1,3}$ and
$\tilde E_a = \tilde E_a (\omega)$,\ $ \tilde H_a = \tilde
H_a(\omega)$ are new unknown functions of the variable
$\omega=\omega(x_0, {\bf x})$.

In addition, the Maxwell equations admit the following discrete
symmetry group \cite{m42}:
\be \label{4.7}
\Psi: \overline{x}_{\mu} = -x_\mu, \quad \overline{{\bf E}} =
-{\bf E}, \quad \overline{{\bf H}} = -{\bf H}.
\ee
The transformation properties of operators (\ref{4.2}), (\ref{4.3})
with respect to the action of the group $\Psi$ read as:
$$
P_\mu \to -P_\mu,\quad J_{\mu \nu} \to J_{\mu \nu},\quad D
\to D,\quad K_\mu \to -K_\mu,\quad Q \to Q.
$$
So that actions of discrete symmetry groups\ $\Psi$ \
(\ref{4.7}) and\ $\Phi_1$\ from Table 2.1 on the basis
operators of the algebra $\tilde p(1,3)$ coincide. Therefore,
we can use Assertions \ref{t4}, \ref{t5} and choose the parameter
$j$ to be equal to 2, i.e., $(-1)^j=1$.

In what follows we exploit invariance of the Maxwell equations
under the conformal group $C(1,3)$ in order to construct
their invariant solutions.

\subsection{Conformally-invariant ansatzes for the Maxwell fi\-elds}

First we will give two assertions, that simplify substantially the
problem of the full description of invariant solutions of the
Maxwell equations.

\bt \label{t41}
If  ${\bf E} = {\bf E}(x_0, x_3), \ {\bf H} = {\bf H}(x_0, x_3)$,
then it is possible to construct the general solution of equations
(\ref{4.1}). Is has the form
\begin{eqnarray*}
E_{1}&=& \varphi_{1}(\xi) + \psi_{1}(\eta), \quad H_{1} =
-\varphi_{2}(\xi) + \psi_{2} (\eta),\\ E_{2}&=&
\varphi_{2}(\xi) + \psi_{2}(\eta),\quad H_{2} = \varphi_{1}(\xi)-
\psi_{1}(\eta),\\ E_{3}&=& C_{1},\quad H_{3} = C_{2},\nonumber
\end{eqnarray*}
where $\varphi_{1}, \varphi_{2}, \psi_{1}, \psi_{2}$ are arbitrary
smooth functions;\ $\xi = x_{0} - x_{3}$,\ $ \eta = x_{0} +
x_{3}$;\ $C_{1}, C_{2}\in{\bf R}$.
\et

\bt \label{t42}
If\ ${\bf E} = {\bf E}(x_1, x_2, \xi),\ {\bf H}
= {\bf H}(x_1, x_2, \xi)$, where $\xi=\frac{1}{2}(x_0-x_3)$, then
it is possible to construct the general solution of the Maxwell
equations (\ref{4.1}). It is given by the following formulae:
\begin{eqnarray*}
E_1 &=& \frac{1}{2}(R + R^* + T_1 + T^{*}_{1}),\quad H_1
= \frac{1}{2}(i R - i R^* - T_2 - T^{*}_{2}), \nonumber \\
E_2 &=& \frac{1}{2}(iR - iR^* + T_2 + T^{*}_{2}),\quad H_2
=\frac{1}{2}(R + R^* - T_1 - T^{*}_{1}),  \\ E_3&=& S + S^*,\quad
H_3=i S - i S^*, \nonumber
\end{eqnarray*}
where
\begin{eqnarray*}
&&T_j= \frac{\textstyle{\partial^2 \theta_j}}{\textstyle{\partial
\xi^2}},\quad (j=1,2), \quad S = \frac{\textstyle{\partial
\theta_1}}{\textstyle{\partial \xi}} + i\frac{\textstyle{\partial
\theta_2}}{\textstyle{\partial \xi}} + \lambda  (z), \\ && R= -2
\Bigl (\frac{\textstyle{\partial \theta_1}}{\textstyle{\partial
z}} + i \frac{\textstyle{\partial \theta_2}}{\textstyle{\partial
z}} \Bigr ) + \xi \frac{\textstyle{d \lambda}}{\textstyle{d z}}.
\end{eqnarray*}
Here $\theta_j = \theta_j(z, \xi),\ \lambda (z)$ are arbitrary
functions analytic by the variable $z=x_1 + i x_2$;\ $j=1,2$;
$i$ is the imaginary unit, i.e., $i^2=-1$.
\et
Proof of the above assertions can be found in
\cite{m49}--\cite{m54}.

It follows from Assertions \ref{t41}, \ref{t42} that we have to
exclude from the further considerations those subalgebras of the
conformal algebra that yield solutions of the form covered by
these assertions. It is straightforward to check that we have to
skip subalgebras $L$ of the rank $3$ fulfilling the conditions
$$
<P_0 + P_3> \not \subset L,\quad <P_0 - P_3> \not \subset
L,\quad <P_0, P_3> \not \subset L,\quad <P_1, P_2>\not
\subset L.
$$
Owing to this fact, to get the full description of
conformally-invariant solutions of the Maxwell equations it
suffices to consider the following subalgebras of
the conformal algebra $c(1,3)$ (note, that we have also made
use of the discrete symmetry group $\Psi$ in order to simplify
their basis elements):
\begin{eqnarray*}
&& M_1=\langle J_{03}, G_1, P_2\rangle ; \quad M_2=\langle
G_1, G_2, J_{03}+ \alpha J_{12} \rangle ,\quad \alpha \in {\bf
R};\\ && M_3=\langle J_{12}, D,P_0 \rangle ; \quad
M_4=\langle J_{12}, D,P_3 \rangle , \\ && M_5=\langle J_{03}, D,P_1
\rangle ;\quad M_6=\langle J_{03}, J_{12}, D \rangle ;\\ &&
M_7=\langle G_1, J_{03}+ \alpha D, P_2 \rangle \quad (0< |\alpha|
\le 1); \\ && M_8=\langle J_{03}-D+M,G_1,P_2\rangle ; \quad
M_9=\langle J_{03}+2D,G_1+2T,P_2\rangle ; \\ && M_{10}=\langle
J_{12}, S+T,Z \rangle ;\quad M_{11}=\langle S+T+J_{12},
Z,G_1+P_2\rangle ; \\ && M_{12}=\langle P_2+K_2+ \sqrt{3}
(P_1+K_1)+K_0-P_0, J_{02}-D- \sqrt{3} J_{01},\\ & & \quad
P_0+K_0-2(K_2-P_2) \rangle ;\\ && M_{13}=\langle P_0+K_0\rangle
\bigoplus  \langle J_{12},K_3-P_3 \rangle ;\\ && M_{14}=\langle
2J_{12}+K_3-P_3, 2 J_{13}-K_2+P_2, 2 J_{23}+K_1-P_1\rangle ;\\ &&
M_{15}=\langle P_1+K_1+2 J_{03}, P_2+K_2+K_0-P_0, 2
J_{12}+K_3-P_3 \rangle.
\end{eqnarray*}
Here we use the following designations:
\begin{eqnarray*}
&&M = P_0 + P_3,\quad G_{0j}=J_{0j} - J_{j3},\ (j=1,2),\\
&&Z=J_{03} + D,\quad S={1 \over 2}(K_0+K_3),\quad T = {1\over 2}
(P_0-P_3).
\end{eqnarray*}

In the sequel, we consider the first ten subalgebras from the
above list. For these subalgebras we can represent the matrix
$\Lambda$ from ansatz (\ref{4.6}) as follows
\begin{eqnarray*}
\Lambda &= &\exp \{(\ln \theta ) E \} \exp (2 \theta_1 H_1)\exp
(2\theta_2 H_2)   \exp (-\theta_0 S_{03}) \exp( \theta_3
S_{12}),
\end{eqnarray*}
the matrices $S_{\mu \nu}$ having the form (\ref{4.4}). So that,
we have
$$
\Lambda = \theta\left(\matrix{C&G\cr -G&C\cr} \right),
$$
where
\begin{eqnarray*}
C &=& \left(\matrix{\cosh \theta_0 \cos \theta_3-r_1 & -\cosh
\theta_0 \sin \theta_3+r_2& 2\theta_1\cr \cosh \theta_0 \sin
\theta_3+r_2& \cosh \theta_0 \cos \theta_3+r_1& 2\theta_2 \cr -2
s_1&2 s_2&1\cr}\right),\\
G &=& \left(\matrix{\sinh \theta_0 \sin \theta_3+r_2 & \sinh
\theta_0 \cos \theta_3+r_1& 2\theta_2\cr -\sinh \theta_0 \cos
\theta_3+r_1& \sinh \theta_0 \sin \theta_3-r_2& -2\theta_1\cr 2
s_2&2 s_1&0\cr}\right),
\end{eqnarray*}
and furthermore,
\begin{eqnarray*}
r_1&=& 2[(\theta^{2}_{1} -\theta^{2}_{2}) \cos \theta_3 +2
\theta_1 \theta_2 \sin \theta_3] e^{-\theta_0}, \\ r_2&=&
2[(\theta^{2}_{1} -\theta^{2}_{2}) \sin \theta_3 -2 \theta_1
\theta_2 \cos \theta_3] e^{-\theta_0}, \\ s_1&=& 2[\theta_{1}\cos
\theta_3 +\theta_2\sin \theta_3 ] e^{-\theta_0}, \\ s_2&=&
2[\theta_{1}\sin \theta_3 -\theta_2\cos \theta_3 ]e^{-\theta_0}.
\end{eqnarray*}

After some algebra we obtain the following form of the
conformally-invariant ansatz for the Maxwell fields:
\begin{eqnarray} \label{4.12}
E_1&=& \theta \{(\tilde E_1 \cos \theta_3 -\tilde E_2 \sin
\theta_3) \cosh \theta_0 \nonumber \\ & & +(\tilde H_1 \sin
\theta_ 3 +\tilde H_2 \cos\theta_3) \sinh \theta_0 \nonumber \\
& & +2 \theta_1 \tilde E_3 + 2 \theta_2 \tilde H_3 + 4 \theta_1
\theta_2 \Sigma_1 + 2( \theta^{2}_{1} - \theta^{2}_{2} )
\Sigma_2\},\nonumber \\ E_2&=& \theta \{ (\tilde E_2 \cos \theta_3
+\tilde E_1 \sin \theta_3) \cosh \theta_0 \nonumber \\ & &
+(\tilde H_2 \sin \theta_3 -\tilde H_1 \cos \theta_3 ) \sinh
\theta_0 \\ & & - 2\theta_1 \tilde H_3+ 2\theta_2 \tilde E_3 +4
\theta_1 \theta_2 \Sigma_2 -2 (\theta^{2}_{1} -\theta^{2}_{2})
\Sigma_1 \},\nonumber \\ E_3&= &\theta \{ \tilde E_3 + 2 \theta_1
\Sigma_2 + 2 \theta_2 \Sigma_1\},\nonumber \\ H_1 &=& \theta \{ (
\tilde H_1 \cos \theta_3-\tilde H_2 \sin \theta_3 ) \cosh \theta_0
\nonumber \\ & & -(\tilde E_1 \sin \theta_3 +\tilde E_2 \cos
\theta_3) \sinh \ \theta_0\nonumber \\ & & + 2 \theta_1 \tilde
H_3 - 2 \theta_2 \tilde E_3 - 4 \theta_1 \theta_2 \Sigma_2 + 2(
\theta^{2}_{1} - \theta^{2}_{2}) \Sigma_1\},\nonumber \\ H_2& =&
\theta \{ (\tilde H_2 \cos \theta_3 + \tilde H_1 \sin \theta_3)
\cosh \theta_0\nonumber \\ & & +(\tilde E_1 \cos \theta_3 -
\tilde E_2 \sin \theta_3) \sinh \theta_0\nonumber \\ & & +2
\theta_1 \tilde E_3 + 2 \theta_2 \tilde H_3 + 4 \theta_1 \theta_2
\Sigma_1 + 2 ( \theta^{2}_{1} - \theta^{2}_{2})
\Sigma_2\},\nonumber \\ H_3&=& \theta \{ \tilde H_3+ 2 \theta_1
\Sigma_1 - 2 \theta_2 \Sigma_2\}.\nonumber
\end{eqnarray}
Here
\begin{eqnarray*}
\Sigma_1& =& [( \tilde H_2 - \tilde E_1) \sin \theta_3 - ( \tilde
E_2 + \tilde H_1) \cos \theta_3] e^{-\theta_0},\\ \Sigma_2&=&[(
\tilde E_2 + \tilde H_1) \sin \theta_3 + ( \tilde H_2 - \tilde
E_1) \cos \theta_3] e^{-\theta_0}.
\end{eqnarray*}
The form of the functions $\theta, \ \theta_\mu, \omega$ for each
of the subalgebras $M_j,\ (j=1,2,\ldots,10)$ is obtained from
Assertions \ref{t4}--\ref{t6} with $k=2$.
\begin{eqnarray*}
M_1&:& \theta=1, \ \theta_0 =-\ln|x_0-x_3|, \ \theta_1=-{1 \over
2} x_1 (x_0 -x_3)^{-1},\\ & & \theta_2 = \theta_3= 0, \ \omega
=x^{2}_{0} -x^{2}_{1}-x^{2}_{3}; \\ M_2&:& \theta=1, \ \theta_0
=-\ln|x_0-x_3|, \ \theta_1=-{1 \over 2} x_1 (x_0 -x_3)^{-1},\\ & &
\theta_2 = -{1\over 2} x_2 (x_0-x_3)^{-1}, \theta_3= \alpha
\ln|x_0 -x_3|, \\ & & \omega =x^{2}_{0} -x^{2}_{1}-x^{2}_{2}-
x^{2}_{3},\ \ \alpha \in {\bf R}; \\ M_3&:& \theta=(x_3)^{-2}, \
\theta_0 =\theta_1=\theta_2=0,\\ & & \theta_3= \arctan
\frac{x_2}{x_1},\
 \omega =(x^{2}_{1} +x^{2}_{2})x^{-2}_{3}; \\
M_4&:& \theta=(x_0)^{-2}, \ \theta_0 =\theta_1=\theta_2=0,\\ & &
\theta_3= \arctan \frac{x_2}{x_1},\
 \omega =(x^{2}_{1} +x^{2}_{2})x^{-2}_{0}; \\
M_5&:& \theta=(x_2)^{-2}, \ \theta_0 =\ln |(x_0 +x_3) x^{-1}_{2}|,
\ \theta_1=\theta_2=\theta_3= 0,\\ & &  \omega =(x^{2}_{0}
-x^{2}_{3})x^{-2}_{2}; \\ M_6&:& (x^{2}_{1} +x^{2}_{2})^{-1}, \
\theta_0 = -{1\over 2} \ln|(x_0-x_3)(x_0+x_3)^{-1}|, \\ & &
\theta_1 = \theta_2=0, \ \theta_3 = \arctan \frac{x_2}{x_1}, \
\omega = (x^{2}_{1} +x^{2}_{2})(x^{2}_{0} -x^{2}_{3})^{-1}; \\
M_7&:& 1)\ \alpha=-1 \\ &&\theta=(x_0-x_3)^{-1}, \ \theta_0= -{1\over 2} \ln
|x_0-x_3|, \\ & & \ \theta_1 =-{1\over 2} x_1(x_0-x_3)^{-1}, \
\theta_2=\theta_3=0, \\ & & \ \omega = x_0 +x_3
-x^{2}_{1}(x_0-x_3)^{-1}; \\
& & 2)\  \alpha \ne -1 \\ &&
\theta = |x^{2}_{0} -x^{2}_{1} -x^{2}_{3}|^{-1}, \ \theta_0
={1\over 2 \alpha} \ln |x^{2}_{0} -x^{2}_{1}-x^{2}_{3}|, \\ & &
\theta_1 = -{1\over 2} x_1 (x_0 -x_3)^{-1}, \ \theta_2 = \theta_3
=0, \\ & & \omega =2 \alpha \ln |x_0 -x_3| +(1-\alpha)\ln
|x^{2}_{0} -x^{2}_{1} -x^{2}_{3}|;\\
M_8&:& \theta = |x_0-x_3|^{-1}, \ \theta_0 =-{1\over 2}
\ln|x_0 -x_3|, \ \theta_1 = -{1 \over 2} x_1 (x_0 -x_3)^{-1}, \\ &
& \theta_2 = \theta_3=0, \ \omega = x_0+x_3-x^{2}_{1}(x_0
-x_3)^{-1}+\ln|x_0 -x_3|; \\ M_9&:& \theta = [(x_0 -x_3)^2 -4
x_1]^{-2}, \ \theta_0= {1\over 2} \ln|(x_0 -x_3)^2 -4 x_1|, \\ & &
\theta_1 =-{1\over 4} (x_0 -x_3), \ \theta_2 = \theta_3=0, \\ & &
\omega=[x_0 +x_3-x_1(x_0-x_3) +{1\over 6}(x_0 -x_3)^3]^2
[(x_0-x_3)^2 -4 x_1]^{-3}; \\ M_{10} &:& \theta =[(x_1 -(x_0 -x_3)
x_2)^2 (1+(x_0-x_3)^2)^{-1}]^{-1}; \\ & &\theta_0 ={1\over 2}
\ln[(x_1-(x_0 -x_3)x_2)^2(1+(x_0-x_3)^2)^{-3}], \\ & & \theta_1 =
-{1\over 2} (x_2 +(x_0 -x_3) x_1)(1+(x_0 -x_3)^2)^{-1}, \\ & &
\theta_2 = {1\over 2} (x_1 -(x_0 -x_3) x_2)(1+(x_0-x_3)^2)^{-1},
\\ & & \theta_3 =-\arctan (x_0 -x_3), \ \omega =[(x_0
+ x_3)(1+(x_0-x_3)^2)^2 \\ & & \quad -2 x_1 (x_2 +(x_0 -x_3) x_1)
 -(x_0 -x_3)(x^{2}_{1}(x_0- x_3)^2
-x^{2}_{2})]\\ & &\quad \times [x_1-(x_0-x_3)x_2]^{-2}-x_0 +x_3.
\end{eqnarray*}

\subsection{Exact solutions of the Maxwell equations}

Now we have to insert ansatzes (\ref{4.12}) into (\ref{4.1}).
However, it is more convenient to rewrite the Maxwell equations
(\ref{4.1}) in the following equivalent form:
\begin{eqnarray}\label{4.8}
& & \partial_{\textstyle{x_1}} (E_1 +H_2)
+\partial_{\textstyle{x_2}}(E_2 -H_1) =
(\partial_{\textstyle{x_0}} -\partial_{\textstyle{x_3}}) E_3,
\nonumber
\\ & & \partial_{\textstyle{x_1}} (E_1 -H_2)
+\partial_{\textstyle{x_2}}(E_2
+H_1) = -(\partial_{\textstyle{x_0}} +\partial_{\textstyle{x_3}})
E_3, \nonumber \\ & & \partial_{\textstyle{x_1}} (E_2 -H_1)
-\partial_{\textstyle{x_2}}(E_1 +H_2) =
-(\partial_{\textstyle{x_0}} -\partial_{\textstyle{x_3}}) H_3,
\nonumber
\\ & & \partial_{\textstyle{x_1}} (E_2 +H_1)
-\partial_{\textstyle{x_2}}(E_1
-H_2) = -(\partial_{\textstyle{x_0}} +\partial_{\textstyle{x_3}})
H_3, \nonumber \\ & &
(\partial_{\textstyle{x_0}}+\partial_{\textstyle{x_3}})(E_1 +H_2)
= \partial_{\textstyle{x_1}}E_3
+\partial_{\textstyle{x_2}}H_3, \nonumber
\\ & & (\partial_{\textstyle{x_0}}-\partial_{\textstyle{x_3}})(E_1
-H_2) =- \partial_{\textstyle{x_1}}E_3
+\partial_{\textstyle{x_2}}H_3,\\ & &
(\partial_{\textstyle{x_0}}-\partial_{\textstyle{x_3}})(E_2 +H_1)
=- \partial_{\textstyle{x_2}}E_3
-\partial_{\textstyle{x_1}}H_3, \nonumber
\\ & & (\partial_{\textstyle{x_0}}+\partial_{\textstyle{x_3}})(E_2
-H_1) = \partial_{\textstyle{x_2}}E_3
-\partial_{\textstyle{x_1}}H_3. \nonumber
\end{eqnarray}

We will give the calculation details for the case of the
subalgebra $M_1$ only, since the remaining subalgebras are handled
in the similar way. For the case in hand, ansatz (\ref{4.12}) can
be written in the form
\begin{eqnarray}\label{4.13}
& & E_1 +H_2 =f e^{\theta_0} +4 \theta_1 \tilde E_3 - 4
\theta^{2}_{1} e^{-\theta_0} h, \nonumber \\ & & E_1 -H_2 = h
e^{-\theta_0}, \quad E_2 +H_1 = \rho e^{-\theta_0},
\nonumber \\ & & E_2 -H_1 =g e^{\theta_0} -4 \theta_1 \tilde H_3 +
4 \theta^{2}_{1} e^{-\theta_0} \rho,  \\ & & E_3 =\tilde E_3 -2
\theta_1 h e^{-\theta_0}, \quad H_3 =\tilde H_3 - 2 \theta_1
\rho e^{-\theta_0}, \nonumber
\end{eqnarray}
where $\theta_0 = -\ln |x_0 -x_3|, \ \theta_1 = -{1 \over 2}
x_1(x_0-x_3)^{-1}$. The functions ${\tilde E}_3,\ {\tilde H}_3$
and
\begin{eqnarray} \label{4.14}
f&=& f(\omega) = \tilde E_1+\tilde H_2, \quad g = g(\omega)
= \tilde E_2 -\tilde H_1, \nonumber \\ h&=& h(\omega) = \tilde E_1
-\tilde H_2, \quad \rho = \rho(\omega) = \tilde E_2 + \tilde
H_1
\end{eqnarray}
are arbitrary smooth functions of the variable\ $\omega= x^{2}_{0}
- x^{2}_{1} -x^{2}_{3}$.

Inserting (\ref{4.13}) into the second and fourth equations
from (\ref{4.8}) gives equations
\be \label{4.15}
\dot{\tilde E}_3 =0, \quad \dot{\tilde H}_3=0.
\ee
We remind that the dot over symbol stands for the derivative with
respect to the variable $\omega$.

Similarly, we get from the sixth and seventh equations of system
(\ref{4.8}) the following reduced equations:
\be \label{4.16}
2 \omega \dot h +3 h=0, \quad 2 \omega \dot \rho + 3 \rho
=0.
\ee
Next, the fifth and eighth equations give rise to ordinary
differential equations of the form
\be \label{4.17}
2 \dot f -h =0, \quad 2 \dot g + \rho = 0.
\ee
Finally, substituting ansatz (\ref{4.13}) into the first and third
equations from (\ref{4.8}) yields
\begin{eqnarray} \label{4.18}
& & 4 \varepsilon \theta_1 [\omega \dot h + h + \dot f] = 2 \xi^{-1}
\tilde E_3, \nonumber \\ & & 4 \varepsilon \theta_1 [ \dot g
-\omega\dot \rho-\rho] = -2 \xi^{-1} \tilde H_3,
\end{eqnarray}
where $\varepsilon=1$ for $\xi = x_0 -x_3>0$ and $\varepsilon=-1$
for $x_0 -x_3<0$.

Taking into account (\ref{4.16}), (\ref{4.17}) we see that
$$
\tilde E_3=0, \quad \tilde H_3 =0.
$$

Summing up we conclude that the ansatz invariant with respect
to the subalgebra $M_1$ reduce the Maxwell equations
to the following system of ordinary differential equations:
\begin{eqnarray} \label{4.19}
& &2 \omega \dot h + 3 h =0,\quad 2 \omega \dot \rho + 3
\rho =0,\quad 2 \dot f-h=0,\nonumber \\ && 2 \dot g + \rho
=0, \quad \tilde E_3 =0, \quad \tilde H_3 =0. \nonumber
\end{eqnarray}

Below we give the reduced systems for the ansatzes invariant with
respect to the remaining subalgebras $M_2$--$M_{10}$. Note that
the functions $f, g, h, \rho$ are of the form (\ref{4.14}).
\begin{eqnarray} \label{4.20}
1.& &  {\mbox{System (\ref{4.19}).}}\nonumber \\ 2.& & \dot f =0,
\ {\tilde E}_3 =0, \ \dot g=0, \ {\tilde H}_3 =0, \nonumber \\ & &
\omega \dot h+2 h +\alpha \rho =0, \ \omega \dot \rho + 2 \rho
-\alpha h =0, \ \alpha \in {\bf R}. \nonumber \\ 3. & & 2 \omega
(1+\omega) \ddot{\tilde E}_3 +(7 \omega +2) \dot{\tilde E}_3+3
{\tilde E}_3 =0, \nonumber \\ & & f =h= -2 \sqrt{\omega} (\tilde
E_3+(1+\omega) \dot{\tilde E_3}), \nonumber \\ & & 2 \omega(1+\omega)
\ddot{\tilde H}_3+(7 \omega +2) \dot{\tilde H}_3 +3{ \tilde
H}_3=0, \nonumber \\ & & g =-\rho =2 \sqrt{\omega} ({\tilde H}_3
+(1 +\omega) \dot{\tilde H}_3). \nonumber \\ 4. & & 2 \omega(\omega-1)
\ddot{\tilde E}_3 +(7 \omega -2) \dot{\tilde E}_3 +3 {\tilde
E}_3=0, \nonumber \\ & & f =-h =2 \sqrt{\omega}({\tilde E}_3
+(\omega-1) \dot{\tilde E}_3), \nonumber \\ & & 2 \omega(\omega-1)
\ddot{\tilde H}_3 +(7 \omega -2) \dot{\tilde H}_3+3 {\tilde H}_3
=0, \nonumber \\ & & g=\rho =-2 \sqrt{\omega}({\tilde H}_3
+(\omega-1) \dot{\tilde H}_3). \nonumber \\ 5.& & 2
\omega(\omega-1)\ddot{\tilde E}_3 +(7 \omega -2) \dot{\tilde E}_3 +3
{\tilde E}_3 =0, \nonumber
\\ & & g =-\omega^{-1} \rho= 2 \varepsilon[{\tilde E}_3 +(\omega-1)
\dot{\tilde E}_3], \nonumber \\ & & 2 \omega(\omega-1) \ddot{\tilde H}_3
+(7 \omega-2) \dot{\tilde H}_3 +3 {\tilde H}_3 =0, \nonumber \\ & &
f=\omega^{-1} h=2 \varepsilon[{\tilde H}_3+(\omega-1) \dot{\tilde
H}_3], \nonumber
\\ & & \varepsilon=1\ \mbox{for}\ (x_0+x_3) x^{-1}_{2}>0,
\nonumber \\ & & \varepsilon =-1\ \mbox{for}\ (x_0+x_3)
x^{-1}_{2}<0. \nonumber
\\ 6.& &(\omega -1) \dot{\tilde E}_3 +{\tilde E}_3 =0, \nonumber
\\
& & 2 \omega \dot f +f = - 2 \varepsilon_2 \sqrt{|\omega|}\dot{\tilde
E}_3, \nonumber \\ & & 2 \omega \dot h +h =  2 \varepsilon_1
\sqrt{|\omega|}\dot{\tilde E}_3,  \\ & &(\omega -1) \dot{\tilde H}_3
+{\tilde H}_3 =0, \nonumber \\ & & 2 \omega \dot \rho +\rho =2
\varepsilon_1 \sqrt{|\omega|} \dot{\tilde H}_3, \nonumber \\ & & 2
\omega \dot g +g =2 \varepsilon_2 \sqrt{|\omega|} \dot{\tilde H}_3,
\nonumber \\ & & \varepsilon_1=1\ \mbox{for}\ x_0+x_3>0, \
\varepsilon_1 =-1\ \mbox{for}\ x_0+x_3<0. \nonumber \\ & &
\varepsilon_2=1\ \mbox{for}\ x_0-x_3>0, \ \varepsilon_2 =-1\
\mbox{for}\ x_0-x_3<0. \nonumber \\ 7.& & 1) \ \dot{\tilde E}_3 =0,
\quad 2 \dot f = \varepsilon h, \nonumber \\ & & \
\dot{\tilde H}_3 =0, \quad 2 \dot g = -\varepsilon \rho,
\nonumber \\ & & \ \varepsilon=1\ \mbox{for}\ x_0 -x_3>0,\nonumber
\\ & & \ \varepsilon=-1\ \mbox{for}\ x_0 - x_3<0.\nonumber \\ & &
2) {\tilde E}_3 =0, \ \ 2(1+\alpha)\dot h -(1+{1\over \alpha})h
=0,\nonumber \\ & & ({1\over \alpha}-2)f +2(1-\alpha) \dot f=
\varepsilon e^{\textstyle{-{1\over \alpha} \omega}}h, \nonumber \\
& & \ {\tilde H}_3 =0, \ \ 2(1+\alpha) \dot \rho -(1+{1\over
\alpha}) \rho=0,\nonumber \\ & & ({1\over \alpha}-2) g
+2(1-\alpha) \dot g=-\varepsilon e^{\textstyle{-{1\over \alpha}
\omega}} \rho,\nonumber \\ & & 0<|\alpha|\le 1, \ \alpha \not =-1
, \ \varepsilon = 1\ \mbox{for}\ x^{2}_{0} -x^{2}_{1}
-x^{2}_{3}>0, \nonumber \\ & & \varepsilon =-1\ \mbox{for}\
x^{2}_{0} -x^{2}_{1} -x^{2}_{3}<0. \nonumber \\ 8. & & \dot h =0,
\ \ \dot{\tilde E}_3 =0, \ \ \dot \rho =0, \ \ \dot{\tilde H}_3
=0, \nonumber \\ & & 2 \varepsilon \dot f-h =0, \quad 2
\varepsilon \dot g+\rho =0, \nonumber \\ & & \varepsilon=1\
\mbox{for}\ x_{0} -x_{3}>0, \ \varepsilon =-1\ \mbox{for}\ x_{0}
-x_{3}<0. \nonumber \\ 9. & & h = 4 \varepsilon f, \ \ \rho =-4
\varepsilon g, \nonumber \\ & & {\tilde E}_3 = -(9
\omega^2+{\varepsilon \over 4}) \dot f-15 \omega f=0, \nonumber \\
& & (36 \omega^2+\varepsilon) \ddot f+180 \omega \dot f+140 f=0,
\nonumber
\\ & & {\tilde H}_3 = 15 \omega g+(9 \omega^2+{\varepsilon \over 4})
\dot g, \nonumber \\ & & (36 \omega^2+\varepsilon) \ddot g +180
\omega \dot g+140 g=0, \nonumber \\ & & \varepsilon=1\ \mbox{for}\
\sigma>0,\ \varepsilon =-1\ \mbox{for}\ \sigma<0,\nonumber \\ & &
\sigma= 4x_{1} -(x_0-x_{3})^2. \nonumber \\ 10.& & \dot f = \dot
h, \ \ h=(\omega^2+1) \dot{\tilde E}_3 +\omega {\tilde E}_3,
\nonumber
\\ & & (\omega^2+1) \ddot{\tilde E}_3 +4 \omega \dot{\tilde E}_3 +
2{\tilde E}_3=0, \nonumber \\ & & \dot g= -\dot \rho, \ \ \rho
=(\omega^2+1) \dot{\tilde H}_3 +\omega {\tilde H}_3, \nonumber \\
& & (\omega^2+1) \ddot{\tilde H}_3 + 4 \omega \dot{\tilde H}_3 +
2{\tilde H}_3=0. \nonumber
\end{eqnarray}

The above systems are linear and therefore are easily integrated
(the integration details can be found in \cite{m49}--\cite{m54}).
Below we give the final result. Namely, we present the families of
exact solutions of the Maxwell equations (\ref{4.1}) invariant
with respect to the subalgebras $M_1$--$M_{10}$.
\begin{eqnarray*}
M_1&:& E_1 = C_2(x_0-x_3)^{-1}-2 x_3 C_1|x^{2}_{0}-x^{2}_{1}-
x^{2}_{3}|^{-{3\over 2}}, \\ & & E_2 = C_4(x_0-x_3)^{-1} +2 x_0
C_3 |x^{2}_{0}-x^{2}_{1}- x^{2}_{3}|^{-{3\over 2}}, \\ & &E_3=2
x_1 C_1 |x^{2}_{0}-x^{2}_{1}- x^{2}_{3}|^{-{3\over 2}}, \\ & & H_1
= -C_4 (x_0-x_3)^{-1} -2 x_3 C_3 |x^{2}_{0}-x^{2}_{1}-
x^{2}_{3}|^{-{3\over 2}}, \\ & & H_2 = C_2 (x_0-x_3)^{-1} -2 x_0
C_1 |x^{2}_{0}-x^{2}_{1}- x^{2}_{3}|^{-{3\over 2}}, \\ & & H_3 = 2
x_1 C_3 |x^{2}_{0}-x^{2}_{1}- x^{2}_{3}|^{-{3\over 2}}. \\ & & \\
M_2&:& E_1 = |\xi|^{-1} \{C_1 \cos (\alpha \ln|\xi|)-C_2
\sin(\alpha \ln|\xi|)\\ & & \quad -x_1 x_2 [h \sin (\alpha \ln |\xi|)
+\rho \cos(\alpha \ln|\xi|)]\\ & &\quad +{1\over
2}(\xi^{2}-x^{2}_{1}+x^{2}_{2})[h \cos (\alpha \ln|\xi|) -\rho
\sin (\alpha \ln |\xi|)]\}, \\ & & E_2 = |\xi|^{-1} \{C_2 \cos
(\alpha \ln|\xi|)+C_1 \sin(\alpha \ln|\xi|)\\ & &\quad +x_1 x_2 [\rho
\sin (\alpha \ln |\xi|) -h \cos(\alpha \ln|\xi|)]\\ & &\quad +{1\over
2}(\xi^{2}+x^{2}_{1}-x^{2}_{2})[h \sin (\alpha \ln|\xi|) +\rho
\cos (\alpha \ln |\xi|)]\}, \\ & & E_3=\varepsilon \{ h[x_1 \cos
(\alpha \ln |\xi|) +x_2 \sin (\alpha \ln |\xi|)]\\ & &\quad +\rho[x_2
\cos(\alpha \ln |\xi|)-x_1 \sin(\alpha \ln|\xi|)], \\ & & H_1=
|\xi|^{-1}\{ -C_2 \cos(\alpha \ln |\xi|) -C_1 \sin (\alpha \ln
|\xi|)\\
 & &\quad -x_1 x_2 [\rho \sin (\alpha \ln|\xi|)- h \cos (\alpha \ln
 |\xi|)]\\
& &\quad + {1\over 2} (\xi^2-x^{2}_{1} +x^{2}_{2})[h \sin (\alpha \ln
|\xi|) +\rho \cos (\alpha \ln |\xi|)]\}, \\ & & H_2 = |\xi|^{-1}\{
C_1 \cos(\alpha \ln |\xi|) -C_2 \sin (\alpha \ln |\xi|)\\ & &
\quad -x_1 x_2 [h \sin (\alpha \ln|\xi|)+ \rho \cos (\alpha \ln
|\xi|)]\\ & &\quad -{1\over 2} (\xi^2+x^{2}_{1} -x^{2}_{2})[h \cos
(\alpha \ln |\xi|) -\rho \sin (\alpha \ln |\xi|)]\}, \\ & &
H_3=\varepsilon \{ h[x_1 \sin (\alpha \ln |\xi|) -x_2 \cos (\alpha
\ln |\xi|)]\\ & &\quad +\rho[x_1 \cos(\alpha \ln |\xi|)+x_2
\sin(\alpha \ln|\xi|)],\\
& & \mbox{where}\ \xi=x_0 -x_3, \ h = \omega^{-2}[C_4 \cos (\alpha
\ln |\omega|) -C_3 \sin (\alpha \ln|\omega|)], \\ & & \rho=
\omega^{-2}[ C_3 \cos (\alpha \ln |\omega|)+C_4 \sin (\alpha \ln
|\omega|)], \ \omega = x_\mu x^\mu, \\ & & \alpha \in {\bf R}, \
\varepsilon=1,\ \mbox{for}\ \xi>0\ \mbox{and}\ \varepsilon=-1\
\mbox{for}\ \xi<0. \\ & & \\
M_3&:& E_a = -\frac{2 C_1 x_a}{x_3(x^{2}_{1} +x^{2}_{2})} +x_a
\sigma^{-{3\over 2}} A_{12}, \ \ E_3 = x_3 \sigma^{-{3\over 2}}
A_{12},\\ & & H_a = -\frac{2 C_3 x_a}{x_3(x^{2}_{1} +x^{2}_{2})}
+x_a \sigma^{-{3\over 2}} A_{34}, \ \ H_3 = x_3 \sigma^{-{3\over
2}} A_{34},\\ & & \mbox{where}\ A_{ij} = C_i (\ln \Bigl \vert
\frac{\sqrt{\sigma}-x_3}{\sqrt{\sigma}+x_3}\Bigr \vert +2
x^{-1}_{3} \sqrt{\sigma}) +C_j, \\ & & \sigma = x^{2}_{1}
+x^{2}_{2} +x^{2}_{3}, \ a=1,2. \\ & & \\
M_4&:& 1)\ E_{a} = \varepsilon_{ab} x_{b} \Bigl\lbrace
\frac{\textstyle{2C_{4}}}{\textstyle{x_{0}(x^{2}_{1}+x^{2}_{2}) }}
-\sigma^{\textstyle{-\frac{3}{2}}}A_{34} \Bigr \rbrace, \quad
 E_{3} = x_{0}\sigma^{\textstyle{-\frac{3}{2}}}A_{12};\\
& & H_{a} = -\varepsilon_{ab} x_{b} \Bigl\lbrace
\frac{\textstyle{2 C_{2}}}{\textstyle{x_{0}(x^{2}_{1}+x^{2}_{2})
}} -\sigma^{\textstyle{-\frac{3}{2}}}A_{12}\Bigr \rbrace,\quad
 H_{3} = x_{0}\sigma^{\textstyle{-\frac{3}{2}}}A_{34}, \\ & &
{\rm where}\ A_{ij} = C_{i} +C_{j}\Bigl(\ln \Bigl\vert
\frac{\textstyle{\sqrt{\sigma}-x_{0}}}{\textstyle{\sqrt{\sigma}+x_{0}}}
\Bigr \vert +2 x^{-1}_{0} \sqrt{\sigma}\Bigr), \\ & & \sigma =
x^{2}_{0} -x^{2}_{1} -x^{2}_{2} >0, \ a,b=1,2; \\ & &
2)\  E_{a} =-\varepsilon_{ab} x_{b} \Bigl\lbrace
\frac{\textstyle{C_{4}}}{\textstyle{x_{0}(x^{2}_{1}+x^{2}_{2})}}
-\sigma^{\textstyle{-\frac{3}{2}}}B_{34} \Bigr \rbrace, \quad
E_{3} = x_{0} \sigma^{\textstyle{-\frac{3}{2}}}B_{12};\\ &&
H_{a} = -\varepsilon_{ab} x_{b} \Bigl\lbrace
\frac{\textstyle{C_{2}}}{\textstyle{x_{0}(x^{2}_{1}+x^{2}_{2}) }}
-\sigma^{\textstyle{-\frac{3}{2}}}B_{12}\Bigr \rbrace,\quad
H_{3} = x_{0} \sigma^{\textstyle{-\frac{3}{2}}}B_{34}, \\ & & {\rm
where}\ B_{ij} = C_{i} +C_{j}(x^{-1}_{0}\sqrt{\sigma}
 -\arctan \frac{\textstyle{\sqrt{\sigma}}}{\textstyle{x_{0}}}), \
  \sigma =
x^{2}_{1} +x^{2}_{2} -x^{2}_{0} >0, \\ & &  a,b=1,2.\\ && {\rm Here
}\ \varepsilon_{ab},\ (a,b=1,2)\ {\rm is\ the\ anti-symmetric\
tensor\ of\ the\ second}\\ & & {\rm order\ with}\
\varepsilon_{12} = 1.\\ && \\
M_5&:& 1)\ E_{1} =\frac{\textstyle{2 x_{0}
C_{4}}}{\textstyle{x_{2}(x^{2}_{0} -x^{2}_{3})}} -x_{0}
\sigma^{\textstyle{-\frac{3}{2}}} A_{34},\ E_{2} =
\frac{\textstyle{2 x_{3} C_{2}}}{\textstyle{x_{2}(x^{2}_{0}
-x^{2}_{3})}}-x_{3} \sigma^{\textstyle{-\frac{3}{2}}} A_{12},\\ &&
H_{1} = -\frac{\textstyle{2 x_{0}
C_{2}}}{\textstyle{x_{2}(x^{2}_{0} -x^{2}_{3})}}+ x_{0}
\sigma^{\textstyle{-\frac{3}{2}}} A_{12},\ H_{2}
=\frac{\textstyle{2 x_{3} C_{4}}}{\textstyle{x_{2}(x^{2}_{0}
-x^{2}_{3})}} -x_{3} \sigma^{\textstyle{-\frac{3}{2}}} A_{34}, \\
& & E_{3} = x_{2} \sigma^{\textstyle{-\frac{3}{2}}} A_{12},\quad
H_{3} = x_{2} \sigma^{\textstyle{-\frac{3}{2}}} A_{34}, \\ &
& {\rm where}\ A_{ij} = C_{i} +C_{j} \Bigl(2
\frac{\textstyle{\sqrt{\sigma}}}{\textstyle{x_{2}}} -\ln \Bigl
\vert
\frac{\textstyle{\sqrt{\sigma}-x_{2}}}{\textstyle{\sqrt{\sigma} +
x_{2}}}\Bigr \vert \Bigr),\ \sigma =  x^{2}_{2}
+x^{2}_{3}-x^{2}_{0} >0; \\ & &
2)\ E_{1} = \frac{\textstyle{x_{0}
C_{4}}}{\textstyle{x_{2} (x^{2}_{0} -x^{2}_{3})}} -x_{0}
\sigma^{\textstyle{-\frac{3}{2}}} B_{34},\ E_{2} =
\frac{\textstyle{x_{3} C_{2}}}{\textstyle{x_{2} (x^{2}_{0}
-x^{2}_{3})}} -x_{3} \sigma^{\textstyle{-\frac{3}{2}}} B_{12},
\\ & & H_{1} = -\frac{\textstyle{x_{0} C_{2}}}{\textstyle{x_{2}
(x^{2}_{0}
-x^{2}_{3})}} +x_{0} \sigma^{\textstyle{-\frac{3}{2}}} B_{12},\
H_{2} = \frac{\textstyle{x_{3} C_{4}}}{\textstyle{x_{2}
(x^{2}_{0} -x^{2}_{3})}} -x_{3} \sigma^{\textstyle{-\frac{3}{2}}}
B_{34},\\ & & E_{3} = x_{2} \sigma^{\textstyle{-\frac{3}{2}}}
B_{12},\quad H_{3}= x_{2} \sigma^{\textstyle{-\frac{3}{2}}}
B_{34},\\ & & {\rm where}\ B_{ij} = C_{i} +C_{j}\Bigl (
\frac{\textstyle{\sqrt{\sigma}}}{\textstyle{x_{2}}} -\arctan
\frac{\textstyle{\sqrt{\sigma}}}{\textstyle{x_{2}}}\Bigr ),\ \
\sigma = x^{2}_{0} -x^{2}_{2} -x^{2}_{3}>0.\\ && \\
M_6&:& E_{1} = \frac{\textstyle{1}}{\textstyle{2}} \Bigl \lbrack
\frac{\textstyle{\xi (x_{1} C_{2} -x_{2} C_{5}) +\eta(x_{1} C_{3}
-x_{3} C_{6})}}{\textstyle{\xi \eta (x^{2}_{1} +x^{2}_{2})}}\\ &
&\quad -\frac{\textstyle{\varepsilon_{1} \xi (x_{1} C_{1} +x_{2} C_{4})
-\varepsilon_{2} \eta (x_{1} C_{1} -x_{2}
C_{4})}}{\textstyle{\sigma(x^{2}_{1} +x^{2}_{2})}}\Bigr \rbrack,\\
& & E_{2} = \frac{\textstyle{1}}{\textstyle{2}} \Bigl \lbrack
\frac{\textstyle{\xi (x_{1} C_{5} +x_{2} C_{2}) +\eta(x_{1} C_{6}
+x_{2} C_{3})}}{\textstyle{\xi \eta (x^{2}_{1} +x^{2}_{2})}}\\ &
&\quad +\frac{\textstyle{\varepsilon_{1} \xi (x_{1} C_{4} -x_{2} C_{1})
+\varepsilon_{2} \eta (x_{1} C_{4} +x_{2}
C_{1})}}{\textstyle{\sigma(x^{2}_{1} +x^{2}_{2})}}\Bigr \rbrack,
\\ & & H_{1} =\frac{\textstyle{1}}{\textstyle{2}} \Bigl \lbrack
\frac{\textstyle{\eta (x_{1} C_{6} +x_{2} C_{3})- \xi (x_{1} C_{5}
+x_{2} C_{2})}}{\textstyle{\xi \eta (x^{2}_{1} +x^{2}_{2})}} \\ &
& \quad +\frac{\textstyle{\varepsilon_{1} \xi (x_{2} C_{1} -x_{1} C_{4})
+\varepsilon_{2} \eta (x_{1} C_{4} +x_{2}
C_{1})}}{\textstyle{\sigma(x^{2}_{1} +x^{2}_{2})}}\Bigr \rbrack,\\
&& H_{2} = \frac{\textstyle{1}}{\textstyle{2}} \Bigl \lbrack
\frac{\textstyle{\xi (x_{1} C_{2} -x_{2} C_{5})- \eta (x_{1} C_{3}
-x_{2} C_{6})}}{\textstyle{\xi \eta (x^{2}_{1} +x^{2}_{2})}}\\ &
&\quad -\frac{\textstyle{\varepsilon_{1} \xi (x_{1} C_{1} +x_{2} C_{4})
+\varepsilon_{2} \eta (x_{1} C_{1} -x_{2}
C_{4})}}{\textstyle{\sigma(x^{2}_{1} +x^{2}_{2})}}\Bigr \rbrack,\\
& & E_{3} = C_{1} \sigma^{-1}, \quad H_{3} =C_{4}
\sigma^{-1},
\\ & & {\rm where}\ \sigma =x^{2}_{1} +x^{2}_{2}
+x^{2}_{3}-x^{2}_{0}, \ \ \xi = x_{0} +x_{3}, \eta = x_{0}-x_{3},
\\ & & \varepsilon_{1} = \cases{\ \ 1,\ {\rm if}\ x_{0} +x_{3}>0, \cr
-1,\ {\rm if}\ x_{0} +x_{3}<0,\cr} \quad
\varepsilon_{2} = \cases{\ \ 1,\ {\rm if}\
x_{0} -x_{3}>0, \cr -1,\ {\rm if}\ x_{0} -x_{3}<0.\cr}\\ && \\
M_7&:& 1)\ \alpha=-1 \\ &&
E_{1} = |\eta|^{\textstyle{-\frac{3}{2}}}(C_{1}
+\frac{\textstyle{1}}{\textstyle{4}} F) -x_{1} \eta^{-2} C_{2}
-\frac{\textstyle{1}}{\textstyle{2}} \varepsilon
|\eta|^{\textstyle{-\frac{1}{2}}} f(x^{2}_{1} \eta^{-2} -1), \\ &
& E_{2} = |\eta|^{\textstyle{-\frac{3}{2}}} (C_{3}
-\frac{\textstyle{1}}{\textstyle{4}} G) +x_{1} \eta^{-2} C_{4}
+\frac{\textstyle{1}}{\textstyle{2}} \varepsilon
|\eta|^{\textstyle{-\frac{1}{2}}} g(x^{2}_{1} \eta^{-2} +1),\\ & &
H_{1} = -|\eta|^{\textstyle{-\frac{3}{2}}}(C_{3}
-\frac{\textstyle{1}}{\textstyle{4}} G) -x_{1} \eta^{-2} C_{4}
-\frac{\textstyle{1}}{\textstyle{2}} \varepsilon
|\eta|^{\textstyle{-\frac{1}{2}}} g(x^{2}_{1} \eta^{-2} -1), \\ &
& H_{2} = |\eta|^{\textstyle{-\frac{3}{2}}} (C_{1}
+\frac{\textstyle{1}}{\textstyle{4}} F) -x_{1} \eta^{-2} C_{3}
-\frac{\textstyle{1}}{\textstyle{2}} \varepsilon
|\eta|^{\textstyle{-\frac{1}{2}}} f(x^{2}_{1} \eta^{-2} +1),\\ & &
E_{3} = \eta^{-1} C_{2} +x_{1} |\eta|^{\textstyle{-\frac{3}{2}}}
f, \quad H_{3} = \eta^{-1} C_{4} + x_{1}
|\eta|^{\textstyle{-\frac{3}{2}}}g\\ & &
{\rm Here}\ f = f(\omega), \ g = g(\omega),\ F=F(\omega), \
G=G(\omega)\ {\rm are}
\\ & &  {\rm arbitrary\ smooth\ functions},\
 \frac{\textstyle{d F}}{\textstyle{d \omega}} =f, \
 \frac{\textstyle{d G}}{\textstyle{d
\omega}} =g, \ \omega = \xi-x^{2}_{1} \eta^{-1}, \\ & &  \xi =
x_{0} +x_{3}, \ \ \eta = x_{0} -x_{3},\\ & & \varepsilon =\cases{\
\ 1,\ {\rm if}\ x_{0} -x_{3}>0,\cr -1,\ {\rm if}\
x_{0} -x_{3} <0. \cr} \\ & &
2)\ 0< |\alpha| \le 1 \\ &&
E_{1} = x_{3} |\sigma|^{\textstyle{-\frac{3}{2}}}C_{1} +C_{2}
\eta^{\textstyle{\frac{2 \alpha-1}{1-\alpha}}}, \ E_{2} = x_{0}
|\sigma|^{\textstyle{-\frac{3}{2}}}C_{3} +C_{4}
\eta^{\textstyle{\frac{2 \alpha-1}{1-\alpha}}}, \\ & &  E_{3}
=-x_{1} |\sigma|^{\textstyle{-\frac{3}{2}}}C_{1}, \\ & &  H_{1}
=-x_{3} |\sigma|^{\textstyle{-\frac{3}{2}}} C_{3} -C_{4}
\eta^{\textstyle{\frac{2 \alpha-1}{1-\alpha}}},\
H_{2} =x_{0} |\sigma|^{\textstyle{-\frac{3}{2}}} C_{1} +C_{2}
\eta^{\textstyle{\frac{2 \alpha-1}{1-\alpha}}},  \\ & & H_{3}
=x_{1} |\sigma|^{\textstyle{-\frac{3}{2}}} C_{3}.
\\ && {\rm If}\ \alpha =1,\ {\rm then}\ C_{2} = C_{4}=0.\\ &&
{\rm Here}\ \sigma = x^{2}_{0} - x^{2}_{1} - x^{2}_{3},\
\eta=x_{0} -x_{3}.\\ && \\
M_8&:& E_{1} =
-x_{1} \eta^{-2} C_{1} +\frac{\textstyle{1}}{\textstyle{4}}
|\eta|^{\textstyle{-\frac{3}{2}}} C_{2}(\xi +2 \eta -3 x^{2}_{1}
\eta^{-1} +\ln|\eta|)+|\eta|^{\textstyle{-\frac{3}{2}}} C_{3}, \\
& & E_{2} = x_{1} \eta^{-2} C_{4}
-\frac{\textstyle{1}}{\textstyle{4}}
|\eta|^{\textstyle{-\frac{3}{2}}} C_{5}(\xi -2 \eta -3 x^{2}_{1}
\eta^{-1} +\ln|\eta|)+|\eta|^{\textstyle{-\frac{3}{2}}} C_{6},\\ &
& H_{1} = -x_{1} \eta^{-2} C_{4}
+\frac{\textstyle{1}}{\textstyle{4}}
|\eta|^{\textstyle{-\frac{3}{2}}} C_{5}(\xi +2 \eta -3 x^{2}_{1}
\eta^{-1}+\ln|\eta|)-|\eta|^{\textstyle{-\frac{3}{2}}} C_{6}, \\ &
& H_{2} = -x_{1} \eta^{-2} C_{1}
+\frac{\textstyle{1}}{\textstyle{4}}
|\eta|^{\textstyle{-\frac{3}{2}}} C_{2}(\xi - 2 \eta -3 x^{2}_{1}
\eta^{-1} +\ln|\eta|)+|\eta|^{\textstyle{-\frac{3}{2}}} C_{3}, \\
& & E_{3} = \eta^{-1} C_{1} +x_{1}
|\eta|^{\textstyle{-\frac{3}{2}}} C_{2}, \ H_{3} = \eta^{-1} C_{4}
+x_{1} |\eta|^{\textstyle{-\frac{3}{2}}} C_{5},
\\ & & {\rm where}\ \xi = x_{0} +x_{3}, \ \eta = x_{0}
-x_{3}.\\ && \\
M_9&:& 1)\ E_{1} = \varphi^{-2}
[A_{12}(\varphi^{\textstyle{\frac{1}{2}}}
-\varphi^{\textstyle{-\frac{1}{2}}}(\eta^{2} -4) -12 \eta \omega)
-\eta B_{12}], \\ & & E_{2} = \varphi^{-2}[A_{34}
(\varphi^{\textstyle{\frac{1}{2}}}
-\varphi^{\textstyle{-\frac{1}{2}}} (\eta^{2} +4) -12 \eta \omega
) -\eta B_{34}],\\ & & E_{3} = \varphi^{-2}[4 A_{12}(\eta
\varphi^{\textstyle{-\frac{1}{2}}} +6 \omega) +2 B_{12}], \\ & &
H_{1} =-\varphi^{-2}
[A_{34}(\varphi^{\textstyle{\frac{1}{2}}}-\varphi^{\textstyle
{-\frac{1}{2}}} (\eta^{2} -4) -12 \eta \omega) -\eta B_{34} ], \\
& & H_{2} = \varphi^{-2}[A_{12}(\varphi^{\textstyle{\frac{1}{2}}}
-\varphi^{\textstyle{-\frac{1}{2}}} (\eta^{2}+4)-12 \eta
\omega)-\eta B_{12}], \\ & &  H_{3} =-\varphi^{-2}[4 A_{34}(\eta
\varphi^{\textstyle{-\frac{1}{2}}}+6 \omega) +2 B_{34}],\\ & &
{\rm where}\ A_{ij} =(1+36 \omega^{2})^{\textstyle{-\frac{3}{2}}}
[C_{i} \sigma^{\textstyle{\frac{1}{3}}}(4 \sqrt{1+36
\omega^{2}}-72 \omega)\\ & &\quad + C_{j}
\sigma^{\textstyle{-\frac{1}{3}}}(4 \sqrt{1+36 \omega^{2}}+72
\omega)], \\ & & B_{ij} = 16(1+36
\omega^{2})^{\textstyle{-\frac{1}{2}}} (C_{i}
\sigma^{\textstyle{\frac{1}{3}}} -C_{j}
\sigma^{\textstyle{-\frac{1}{3}}}),\\ & &   \sigma = 6 \omega
+\sqrt{36 \omega^{2} +1},\ \omega = (\xi -x_{1} \eta
+\frac{\textstyle{1}}{\textstyle{6}}
\eta^{3})\varphi^{\textstyle{-\frac{3}{2}}},  \\
 & &   \varphi=4
x_{1} -\eta^{2} >0, \\ & &  \xi = x_{0} +x_{3}, \ \eta
=x_{0}-x_{3}; \\ & &
2)\ E_{1} = \varphi^{-2} [A_{12}
(\varphi^{\textstyle{\frac{1}{2}}}
-\varphi^{\textstyle{-\frac{1}{2}}}(\eta^{2} +4) +42 \eta \omega)
-\eta B_{12}], \\ & & E_{2} =
\varphi^{-2}[A_{34}(\varphi^{\textstyle{\frac{1}{2}}}-
\varphi^{\textstyle{-\frac{1}{2}}} (\eta^{2} -4)-42 \eta \omega)
-\eta B_{34}],  \\ & & E_{3} = -\varphi^{-2}[4 A_{12}(\eta
\varphi^{\textstyle{-\frac{1}{2}}}+21 \omega) -2 B_{12}], \\ & &
H_{1} -\varphi^{-2}[A_{34}(\varphi^{\textstyle{\frac{1}{2}}}-
\varphi^{\textstyle{-\frac{1}{2}}} (\eta^{2} +4) +42 \eta \omega)
-\eta B_{34}], \\ & &  H_{2} =
\varphi^{-2}[A_{12}(\varphi^{\textstyle{\frac{1}{2}}}-
\varphi^{\textstyle{-\frac{1}{2}}} (\eta^{2} -4) +42 \eta \omega)
-\eta B_{12}], \\ & & H_{3}= \varphi^{-2}[A_{34}(\eta
\varphi^{\textstyle{-\frac{1}{2}}}+21 \omega) -2 B_{34}], \\ & &
{\rm where}\ A_{ij} = (1-36 \omega^{2})^{\textstyle{-\frac{3}{2}}}
\lbrace \cos \sigma [72
\omega C_{j} -4 C_{i} \sqrt{1-36 \omega^{2}}]\\ &&\quad -\sin \sigma
[72 \omega C_{i} + 4 C_{j}\sqrt{1-36 \omega^{2}}]\rbrace, \\ & &
B_{ij} = 16(1-36 \omega^{2})^{\textstyle{-\frac{1}{2}}}[C_{i} \sin
\sigma -C_{j} \cos \sigma], \ \sigma
=\frac{\textstyle{1}}{\textstyle{3}} \arcsin 6 \omega, \\ & & |6
\omega| <1, \  \varphi = \eta^{2} -4 x_{1} >0, \ \omega = (\xi
-x_{1} \eta +\frac{\textstyle{1}}{\textstyle{6}} \eta^{3})
\varphi^{\textstyle{-\frac{3}{2}}}, \\ & &  \xi = x_{0}+x_{3}, \
\eta = x_{0} -x_{3}; \\ & &
3)\ E_{1} = \varphi^{-2}[A_{12}(\varphi^{\textstyle{\frac{1}{2}}}-
\varphi^{\textstyle{-\frac{1}{2}}}(\eta^{2} +4) -12 \eta \omega)
-\eta B_{12}], \\ & & E_{2} = \varphi^{-2}[A_{34}
(\varphi^{\textstyle{\frac{1}{2}}}-\varphi^{\textstyle
{-\frac{1}{2}}}(\eta^{2}-4)-12 \eta \omega) -\eta B_{34}], \\ & &
E_{3} = \varphi^{-2} [-4 A_{12}(\eta
\varphi^{\textstyle{-\frac{1}{2}}} -6 \omega) +2 B_{12}], \\ & &
H_{1} = -\varphi^{-2}[A_{34}(\varphi^{\textstyle{\frac{1}{2}}}-
\varphi^{\textstyle{-\frac{1}{2}}} (\eta^{2} +4) -12 \eta \omega)
-\eta B_{34}], \\ & & H_{2} = \varphi^{-2}[A_{12}
(\varphi^{\textstyle{\frac{1}{2}}}-\varphi^{\textstyle{-\frac{1}{2}}}
(\eta^{2}-4) -12 \eta \omega) -\eta B_{12}], \\ & & H_{3} =
\varphi^{-2}[ 4A_{34}(\eta \varphi^{\textstyle{-\frac{1}{2}}}-6
\omega )-2 B_{34}], \\ && {\rm where}\ A_{ij} = (36
\omega^{2}-1)^{\textstyle{-\frac{3}{2}}}[C_{i}
\sigma^{\textstyle{\frac{1}{3}}}(4 \sqrt{36 \omega^{2}-1} -72
\omega)\\ & &\quad +C_{j} \sigma^{\textstyle{-\frac{1}{3}}}(4
\sqrt{36 \omega^{2}-1}+72 \omega)], \\ & & B_{ij} = 16 (36
\omega^{2}-1)^{\textstyle{-\frac{3}{2}}}[C_{i}
\sigma^{\textstyle{\frac{1}{3}}}-C_{j}
\sigma^{\textstyle{-\frac{1}{3}}}], \\ & & \sigma = 6 \omega
+\sqrt{36 \omega^{2} -1}, \ \ |6 \omega|>1,\
\varphi = 4 x_{1} -\eta^{2} >0, \\
& &  \omega = (\xi -x_{1} \eta
+\frac{\textstyle{1}}{\textstyle{6}} \eta^{3})
\varphi^{\textstyle{-\frac{3}{2}}}, \ \xi = x_{0} +x_{3}, \ \eta =
x_{0}-x_{3}. \\ && \\
M_{10}&:& E_1 = \sigma^{-1}(1+\xi^2)^{-1}\{ x_1
C_5 -x_2 C_6 -(1+\omega^2)^{-1}[\xi x_1 (C_1 \omega \\ & &\quad +C_2)
+\xi x_2 (C_3 \omega +C_4) -{1\over 2}(1-\xi^2)(x_1 (C_1 -\omega
C_2)\\ & &\quad + x_2(C_3-\omega C_4))]\} + {1\over 2}
\sigma^{-2}(1+\xi^2)(1+\omega^2)^{-1}[x_1 (C_1 -\omega C_2) \\ &
&\quad -x_2 (C_3 -\omega C_4)],\\ && E_2 = \sigma^{-1}(1+\xi^2)^{-1}\{
x_1 C_6 +x_2 C_5 +(1+\omega^2)^{-1}[\xi x_1 (C_3 \omega \\ & &
\quad + C_4) -\xi x_2 (C_1 \omega +C_2) +{1\over 2}(1-\xi^2)(x_2 (C_1
-\omega C_2)\\ & &\quad - x_1(C_3-\omega C_4))]\}+ {1\over 2}
\sigma^{-2}(1+\xi^2)(1+\omega^2)^{-1}[x_1 (C_3 -\omega C_4) \\ &
&\quad + x_2 (C_1-\omega C_2)],\\ & & E_3 =
\sigma^{-1}(1+\omega^2)^{-1}[C_1(\omega +\xi) +C_2(1-\xi \omega)],
\\ && H_1 = -\sigma^{-1}(1+\xi^2)^{-1}\{ x_1 C_6 +x_2 C_5
+(1+\omega^2)^{-1}[\xi x_1 (C_3 \omega \\ & &\quad + C_4) -\xi x_2 (C_1
\omega +C_2) +{1\over 2}(1-\xi^2)(x_2 (C_1 -\omega C_2)\\ & &
\quad - x_1(C_3-\omega C_4))]\} +{1\over 2}
\sigma^{-2}(1+\xi^2)(1+\omega^2)^{-1}[x_1 (C_3 -\omega C_4) \\ &
&\quad + x_2 (C_1 -\omega C_2)],\\ && H_2 = \sigma^{-1}(1+\xi^2)^{-1}\{
x_1 C_5 -x_2 C_6 -(1+\omega^2)^{-1}[\xi x_1 (C_1 \omega \\ & &
\quad + C_2) +\xi x_2 (C_3 \omega +C_4) -{1\over 2}(1-\xi^2)(x_1 (C_1
-\omega C_2)\\ & &\quad + x_2(C_3-\omega C_4))]\}- {1\over 2}
\sigma^{-2}(1+\xi^2)(1+\omega^2)^{-1}[x_1 (C_1 -\omega C_2) \\ &
&\quad - x_2 (C_3-\omega C_4)],\\ & & H_3 =
\sigma^{-1}(1+\omega^2)^{-1}[C_3(\omega +\xi) +C_4(1-\xi \omega)],
\\ & & \mbox{where}\ \sigma = x^{2}_{1} +x^{2}_{2}, \
\omega=\eta(1+\xi^2) \sigma^{-1} -\xi, \ \eta = x_0 +x_3, \\ & &
\xi=x_0-x_3.
\end{eqnarray*}
In the above formulae $C_j,\ (j=1,2,\ldots, 6)$ are arbitrary
real constants.

Note that the constructed Maxwell fields are, generally speaking,
non-orthogonal. However, provided some additional restrictions on
the parameters $C_1,\ldots, C_6$ are imposed, they become
orthogonal. Consider, as an example, the last solution from the
above list. Imposing the orthogonality condition ${\bf E}\cdot
{\bf H} = 0$ yields the following restrictions on the choice of
$C_1,\ldots, C_6$:
$$
C_2 C_6 = C_4 C_5, \quad C_1 C_6 = C_1 C_3 + C_2 C_4 + C_3
C_5.
$$

Next, for the solution invariant under the subalgebra $M_1$ the
orthogonality condition leads to the following set of algebraic
equations to be satisfied by the parameters $C_1,\ldots, C_6$
$$
C_2 C_3 = C_1 C_4, \quad C_1 C_3 =0.
$$

\section{Concluding remarks}

The range of applications of the Lie group methods for solving
systems of linear and nonlinear partial differential equations is
so wide that it is simply impossible to give a detailed account of
all the available techniques, even, if we restrict our
considerations to some fixed group, like the conformal group
$C(1,3)$. However, the basic ideas and methods exposed in the
present review paper are easily adapted to the cases of other
groups of importance for the modern physics. In particular, it is
straightforward to modify the general reduction method suggested
here in order to make it applicable for solving equations of
non-relativistic physics, where the central role is played by the
Galilei and Schr\"odinger groups.

Furthermore, the general method exposed in the paper applies
directly to solving the full Maxwell equations with currents. It
can be used also to construct exact classical solutions of the
Yang-Mills equations with Higgs fields and of their
generalizations. Generically, the method developed in the paper
can be efficiently applied to any conformally-invariant wave
equation, on the solution set of which a covariant representation
of the conformal algebra (\ref{2.11}) is realized.

We do not consider here the solution techniques based on the
symmetry reduction of different versions of the self-dual
Yang-Mills equations to integrable models (we refer the interested
reader to the papers \cite{m13}--\cite{m15}, \cite{m22}--
\cite{m24}, \cite{tafel90} for a detailed exposition of the
results in this field available by now).

The results on exact solution of nonlinear generalizations of the
Maxwell equations are also beyond the scope of the present review.
The survey of these results, as well as, the extensive list of
references can be found in \cite{m21}.

\end{document}